\documentclass[twocolumn]{aastex631}
\usepackage{CJKutf8}
\usepackage{amsmath}
\usepackage{graphicx}
\usepackage{graphics}
\begin{document}
\begin{CJK}{UTF8}{gbsn}

\title{ A comparative study of ultraluminous infrared galaxies in the IRAS and SDSS Surveys}


\author[0000-0001-8485-2814]{Shaohua Zhang (张少华)}
\affiliation{Shanghai Key Lab for Astrophysics, Shanghai Normal University, Shanghai 200234, People's Republic of China}

\author{Zhijian Luo (罗智坚)}
\affiliation{Shanghai Key Lab for Astrophysics, Shanghai Normal University, Shanghai 200234, People's Republic of China}

\author[0000-0002-6710-7537]{Xiheng Shi (史习珩)}
\affiliation{MNR Key Laboratory for  Polar Science, Polar Research Institute of China, Shanghai 200136, People's Republic of China}

\author{Chenggang Shu (束成钢)}
\affiliation{Shanghai Key Lab for Astrophysics, Shanghai Normal University, Shanghai 200234, People's Republic of China}

\author[0000-0001-8244-1229]{Hubing Xiao (肖胡兵)}
\affiliation{Shanghai Key Lab for Astrophysics, Shanghai Normal University, Shanghai 200234, People's Republic of China}

\author[0000-0003-1956-9021]{Hongyan Zhou (周宏岩)}
\affiliation{MNR Key Laboratory for  Polar Science, Polar Research Institute of China, Shanghai 200136, People's Republic of China}
\affiliation{CAS Key Laboratory for Research in Galaxies and Cosmology, Department of Astronomy, University of Sciences and Technology of China, Hefei, Anhui 230026, People's Republic of China}

\begin{abstract}
We present a comprehensive study of Ultraluminous Infrared Galaxies (ULIRGs), leveraging data from the IRAS Faint Source Catalogue (FSC) and the spectroscopic catalog in the Sloan Digital Sky Survey (SDSS) DR16. Our meticulous cross-matching technique significantly enhances the reliability of ULIRG identification, resulting in the identification of 283 reliable ULIRGs, including 102 new detections, while discarding 120 previously reported false sources. Covering a redshift range of $z = 0.018 - 0.996$, with a median redshift of $\bar{z} = 0.259$, our uniform sample reveals apparent interaction features in approximately 40\% of ULIRGs, increasing  to 92\% for those with $z < 0.1$. Through optical spectra analysis, it is indicated that over 58\% of ULIRGs host an AGN,  which is twice as high as the detections based solely on infrared colors. Moreover, a pronounced excess of radio emissions associated with AGN activity results in a steeper radio-far-infrared correlation.  Notably, Type I ULIRGs exhibit properties similar to those of narrow-line Seyfert 1 galaxies (NLS1s), with an elevated incidence rate of \ion{Mg}{2} BALs (16.7\%), surpassing that of typical optically selected quasars by over tenfold, consistent with current evolutionary models. We anticipate that forthcoming telescopes such as the China Space Station Telescope (CSST) and Leighton Chajnantor Telescope (LCT) will provide deeper insights into ULIRG morphology, dust distribution, molecular gas, and AGN activity.
\end{abstract}


\section{Introduction}

Ultraluminous infrared galaxies (ULIRGs) stand out as remarkable celestial objects, characterized by their intense infrared luminosity in the 8 $\mu$m – 1000 $\mu$m wavelength band, classically exceeding $\rm 10^{12}~L_{\sun}$, comparable to the bolometric luminosity of quasars (see Sanders \& Mirabel 1996 for an in-depth review).  	
As one of the significant outcomes of the Infrared Astronomical Satellite (IRAS) mission (Neugebauer et al. 1984), ULIRGs were discovered in large numbers, and appeared as bright, pointlike sources at 60 $\mu$m, lacking visible counterparts (Houck et al. 1984; Soifer et al. 1984).

Efforts to unveil the mysteries surrounding ULIRGs have surged, exploring their origin, infrared power source, and evolutionary path (Lonsdale et al. 2006; Perez-Torres et al. 2021).  Prevailing consensus attributes ULIRG formation to intense galactic collisions and mergers, as demonstrated by the study of Soifer et al. (1984), which meticulously examined 86 infrared galaxies, revealing a substantial fraction as interacting spiral galaxies.  Additionally, New Technology Telescope (NTT) images of all 16 southern ULIRGs reveal strongly interacting systems showing double nuclei, wisps, and tails that are characteristic of advanced mergers (Melnick \& Mirabel 1990). 

Observations further reveal post-merger ULIRGs with active galactic nuclei (AGN) manifestations in the later stages of merging (Clements et al. 1996; Murphy et al. 1996; Zheng et al. 1999; Canalizo \& Stockton 2001; Cui et al. 2001; Bushouse et al. 2002; Veilleux et al. 2002). The enormous infrared radiation emitted by ULIRGs results from the re-emission of dust, heated by intense star formation and AGN activity triggered and fueled by these merger events (Leech et al. 1988; Sanders et al. 1988a; Veilleux et al. 1995; Genzel et al. 1998; Farrah et al. 2003).
In this dynamic astronomical panorama, ULIRGs have been observed in association with submillimeter galaxies (SMGs; Hughes et al. 1998; Barger et al. 1998; Efstathiou \& Rowan-Robinson 2003; Rowan-Robinson et al. 2018), quasars and hot dust-obscured galaxies (Hot DOGs; Eisenhardt et al. 2012; Bridge et al. 2013; Casey et al. 2014; Farrah et al. 2017; Efstathiou et al. 2021). 
ULIRGs play a pivotal role in the evolution path from galaxy mergers to elliptical galaxies and quasars, offering essential insights into the evolutionary connection between circumnuclear massive starbursts and AGNs.

Nevertheless, the complexity of the astronomical landscape is apparent. According to Jones et al. (2014), the investigation revealed that submillimeter-detected WISE-selected Hot DOGs qualify as hyperluminous infrared galaxies  (HyLIRGs; $L\rm_{IR} > 10^{13}~L_{\sun}$), harboring very powerful AGN.  Chen \& Liu (2024) observed that the majority of DOGs are characterized by AGN dominance in the mid-IR, far-IR, and submillimeter wavelengths, whereas ULIRGs are marked by a prevalence of star formation in the far-IR. Efstathiou et al. (2022) contributed to this narrative by discovering that all HERUS ULIRGs exhibit high rates of star formation, with bolometrically significant AGNs present in every galaxy of the sample, and dual AGNs identified in two sources. Importantly, there is no evidence that strong AGN appears either at the beginning or end of a starburst episode or that starbursts and AGN affect each other.

Despite the significant progress in understanding ULIRGs, previous investigations identified specific issues related to the use of IRAS data.  These issues mainly stem from the positional uncertainty of IRAS sources, leading to notable overestimation in the infrared luminosity of numerous sources, consequently misclassifying them as ULIRGs. This discrepancy arises from the presence of multiple optical and near-infrared counterparts within the positional uncertainty ellipse, contributing to challenges in precisely determining the true characteristics of these objects. 
This study aims to overcome these challenges by generating a comprehensive and reliable ULIRG sample, leveraging optical images, multi-band colors, and AGN activity analysis based on photometric and spectroscopic data.  Section 2 provides an overview of the data and the ULIRG identification method, followed by the presentation and discussion of results in Section 3, and concluding remarks in Section 4. Adopting standard cosmological parameters ($H\rm_0 = 70kms^{-1}~Mpc^{-1}$, $\Omega\rm_m = 0.3$, and $\Omega\rm_{\Lambda} = 0.7$), this work seeks to contribute to a deeper understanding of ULIRGs and their pivotal role in the cosmic tapestry.

\section{Identification of ULIRGs}
\subsection{Samples and cross-matching}

This work starts from the IRAS Faint Source Catalog (FSC92, $|b| > 10^\circ$, Version 2.0, Moshir et al. 1992) and the  Sloan Digital Sky Survey (SDSS)  Data Release 16 (DR16; Ahumada et al. 2020).
The IRAS survey, a remarkable feat in thermal infrared sky coverage, provided comprehensive data through multiple detectors. The cornerstone of IRAS, the FSC92 catalog, comprises information on 173,044 sources in unconfused regions, featuring flux densities typically exceeding 0.2 Jy at 12, 25, and 60 $\mu$m, and above 1.0 Jy at 100 $\mu$m. The catalog ensures reliabilities of at least 98.5\% at 12 and 25 $\mu$m and approximately 94\% at 60 $\mu$m. 
Notably, the FSC92 achieves a limiting magnitude approximately one magnitude deeper than the IRAS Point Source Catalog (PSC, Version 2.0, IPAC 1986), positioning it more favorably for distant sources like galaxies and quasars. However, FSC92 sources present substantial positional uncertainties, ranging from $1-13$ arcsec in the in-scan direction (typical value of 5 arcsecs) to $3-55$ arcsec in the cross-scan direction (typical value of 16 arcsecs), described as ``error ellipses."

The SDSS DR16 release is distinguished as the latest and final data release of optical spectra from the SDSS extended Baryon Oscillation Spectroscopic Survey (eBOSS; Dawson et al. 2016). 
Despite being the fourth data release of SDSS-IV, SDSS DR16 incorporates spectroscopically-confirmed sources from SDSS-I/II/III. 
Covering 5,789,200 optical spectra of stars, galaxies, and quasars, SDSS DR16 is a substantial resource. 
Post filtering out spectra marked as ``STAR"\footnote{https://www.sdss.org/dr16/spectro/catalogs/.} in the ``CLASS" parameter, over 4.5 million spectra with effective redshift measurements remain. After cross-referencing astrometric positions, 386,462 sources with repeated spectral observations are identified within the catalog, totaling 4,044,951 sources across 9,376 deg$^2$. The catalog serves as an extensive repository of galaxies and quasars within SDSS.
While the catalog may still contain sources classified as ``Unknown"\footnote{https://www.sdss.org/dr16/scope/\#Opticalspectroscopydatastatistics}; however, it is unnecessary to worry about those distracters, which will be gradually eliminated in the following analysis.

Previous studies have favored the positional uncertainty ellipse of each IRAS source to identify their SDSS counterparts (e.g., Cao et al. 2006; Hwang et al. 2007; Hou et al. 2009), considering an SDSS source within the $3\sigma$ error ellipse of an IRAS source as a match. This method is used in this work. 
Since the positional uncertainty of the SDSS is much smaller than that of IRAS, the positional uncertainty of the optical identification itself is negligible. 
The cross-matching reveals 17,428 IRAS sources with SDSS counterparts in DR16, including 3,996 with more than one counterpart within the 3$\sigma$ error ellipse. 
Notably, we abandon the conventional "likelihood ratio (LR)" method for confirming the most probable counterpart (e.g., Sutherland \& Saunders 1992; Cao et al. 2006; Hwang et al. 2007 and Hou et al. 2009), and instead opt for a robust approach—cross-matching with the Wide-field Infrared Survey Explorer (WISE; Wright et al. 2010) photometry database. This ensures reliable identification of counterparts and mitigates uncertainties associated with flux intensity and source brightness in the infrared waveband.

\subsection{Luminosity selection criteria}

\figurenum{1}
\begin{figure*}[tp]
\plotone{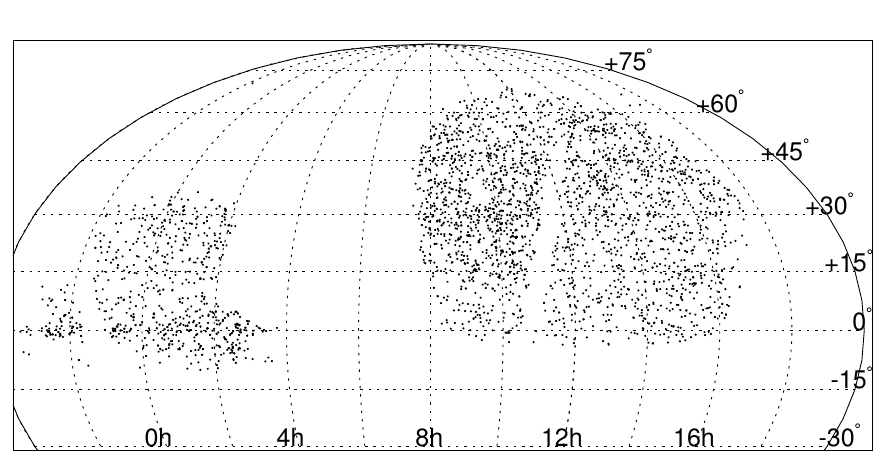}
\caption{Distribution of the primary ULIRG candidates in Equatorial coordinates (map centered at $\rm RA = 8h$). \label{fig:f1}}
\end{figure*}

To reduce the workload, we initially rejected the majority of IRAS sources using ULIRG luminosity selection criteria.
As the 12 and 25 $\mu$m flux qualities for the matched IRAS sources are primarily upper limits, our subsequent analysis focuses on sources with reliable 60 $\mu$m detections.
From the total list of 17,428 matched IRAS sources, we narrowed down our selection to 17,235 sources with ``high" or ``moderate" quality index for their 60 $\mu$m flux densities, following the FSC92 catalog conventions (where upper limits, moderate quality, and high quality are marked as numbers 1, 2, and 3, respectively).
 Following methods similar to Cao et al. (2006) and Hou et al. (2009), we calculate the far-infrared (FIR) luminosity ($L_{\rm FIR}$) using the 60 and 100 $\mu$m flux densities (Helou et al. 1988; Sanders \& Mirabel 1996), which is then converted to the total infrared luminosity ($L_{\rm IR}$; Calzetti et al. 2000).

The formulae are as follows:
\begin{equation}
F_{\rm FIR} = 1.26 \times 10^{-14} ~\left[2.58 ~f_{60~\mu m} + f_{100~\mu m} \right] ~ \rm (W~ m^{-2}),
\end{equation}
\begin{equation}
L_{\rm FIR} = 4 ~\pi~ D_{\rm L}^2 ~F_{\rm FIR} ~ \rm (L_{\sun}),
\end{equation}
\begin{equation}
L_{\rm IR} {\rm (1-1000~ \mu m)} = 1.75~ L_{\rm FIR},
\end{equation}
where $f_{\rm 60~\mu m}$, $f_{\rm 100~\mu m}$ are the IRAS flux densities in Jy at 60 and 100 $\mu$m,
$D_{\rm L}$ is the luminosity distance based on the counterpart's redshift, and the luminosities are in $\rm L_{\sun}$.
Notably, approximately 41\% of the sources have upper limits for their 100 $\mu$m flux qualities. Given that the 100 $\mu$m flux has a minimal impact on $L_{\rm IR}$ calculations, as pointed out by Cao et al. (2006) and corroborated by Kim \& Sanders (1998) and Hwang et al. (2007), we do not impose quality limits on the 100 $\mu$m flux density.

From the pool of 17,235 reliable sources, we identify the primary sample, comprising 2,715 ULIRG candidates with infrared luminosities exceeding $10^{12}~\rm L_{\sun}$. Of these, 254 still have multiple optical counterparts. At this stage, we refrain from designating the ``true" IRAS-SDSS association, retaining all associations for further analysis.

Figure \ref{fig:f1} displays the sky coverage of these ULIRG candidates in equatorial coordinates, illustrating comprehensive coverage across nearly the entire SDSS DR16 spectroscopic survey regions. 
A comparison with the ULIRG candidate sample from Hou et al. (2009), who identified 308 ULIRGs from the SDSS DR6 catalog, reveals a recovery of 306 out of their 308 ULIRGs. 
Two sources in their sample, F10212+2506 and F14336-0147, are excluded from our study due to their lower infrared luminosities, identified as $\rm 10^{11.53}~L_{\sun}$ and $\rm 10^{11.09} ~L_{\sun}$, respectively. 
Further examining the optical spectra of their counterparts, we find that the redshift of F10212+2506 was overestimated by Hou et al. (2009), while that of F14336-0147 is underestimated in the DR16 catalog. This primary sample serves as the basis for our Catalog, providing a foundation for detailed identification and subsequent in-depth analysis.

\figurenum{2}
\begin{figure}[tp]
\centering
\includegraphics[width=0.48\textwidth]{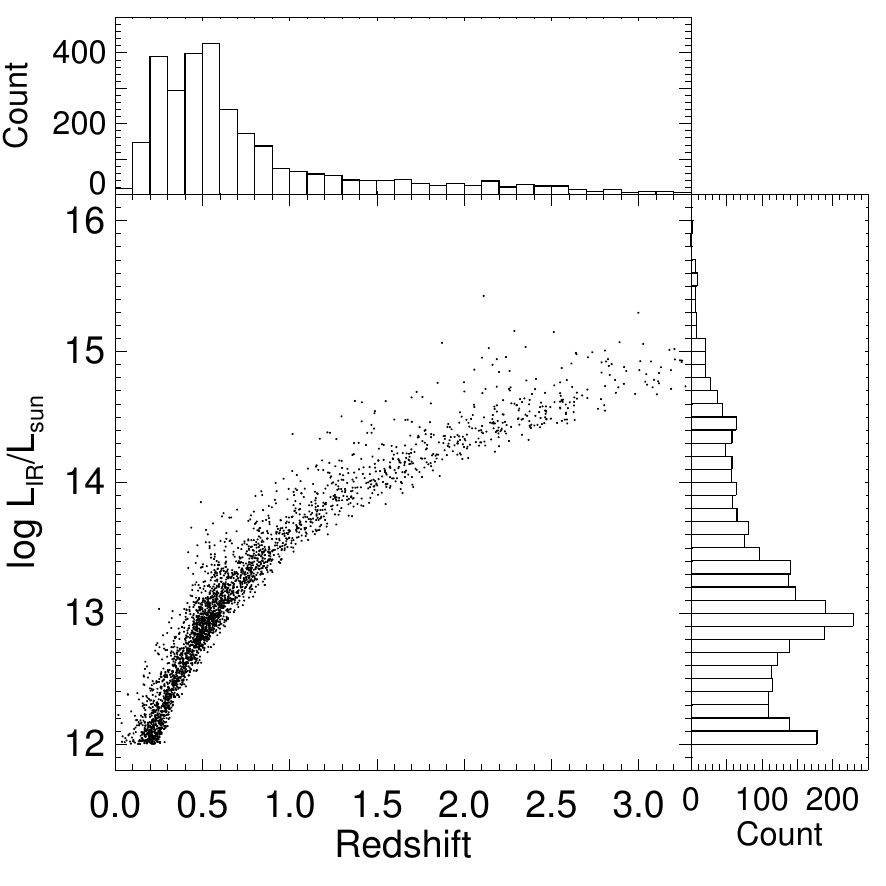}
\caption{The infrared luminosity of the primary ULIRG candidates versus redshift. The primary candidates' redshift and infrared luminosity distributions are shown in the upper and right-hand panels. As the SDSS-III/IV depth increases, many SDSS sources at high redshift are matched, and the estimated infrared luminosity also tends to be higher.}
\label{fig:f2}
\end{figure}

In Figure \ref{fig:f2}, we illustrate the infrared luminosities of the 2715 primary ULIRG candidates identified in this study plotted against their redshifts. The upper and right-hand panels display their redshift and luminosity distributions, respectively. 
The redshift range of the primary candidates spans from $z\sim 0.02$ to $4.0$ with a median value of approximately $0.56$. 
This range is notably broader and extends further than the redshift range of $0.02-0.6$ with a median redshift $z \simeq 0.2$ observed in previously known ULIRGs (e.g., Cao et al. 2006; Hwang et al. 2007; Hou et al. 2009).
This difference could be attributed to distinct selection criteria for spectroscopic targets. The SDSS-III/IV has probed up to two magnitudes deeper than SDSS-I/II and other bright galaxy surveys employed in previous studies. Consequently, the parent sample of our primary ULIRG candidates, namely the DR16 spectroscopic catalog, encompasses a substantial number of fainter and more distant (higher redshift) sources.
Moreover, thus, the primary ULIRG candidates in this study exhibit significantly larger infrared luminosities compared to those of previously known ULIRGs. Their median infrared luminosity is approximately $L\rm_{IR} \simeq 10^{13} ~L_{\sun}$, nearly approaching the maximum value observed in previously known ULIRGs.

\subsection{Examining candidates through WISE detections}

\figurenum{3}
\begin{figure}[tp]
\centering
\includegraphics[width=0.48\textwidth]{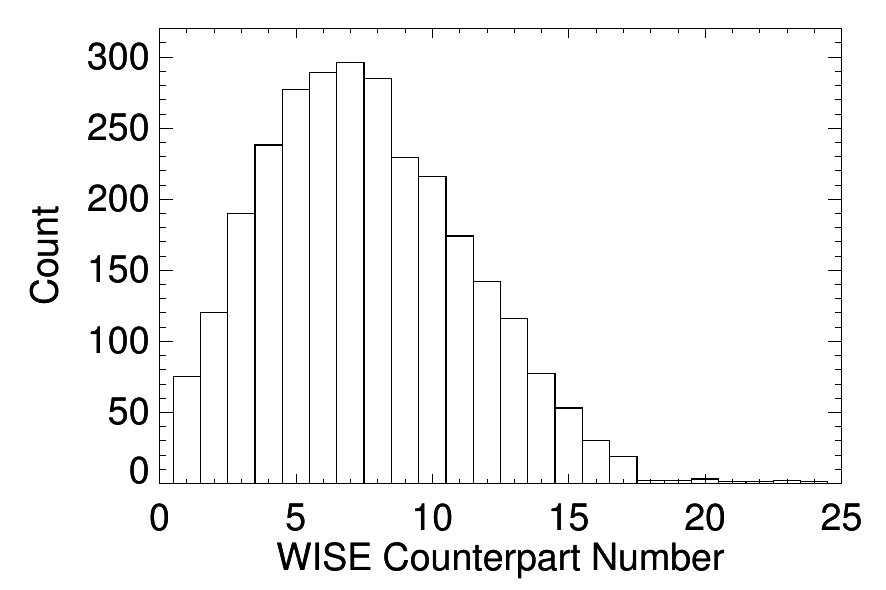}
\caption{The number distribution of the WISE counterparts for each primary ULIRG candidate, these primary candidates have 8 WISE counterparts within 3$\sigma$ error ellipse on average.
\label{fig:f3}}
\end{figure}

To verify the reliability of the IRAS-SDSS associations, we conducted an additional examination by scrutinizing WISE detections within the positional uncertainty ellipse of each IRAS source. 
In comparison to IRAS, WISE extensively mapped the entire sky at 3.4, 4.6, 12, and 22 $\mu$m, denoted as $W1$, $W2$, $W3$, and $W4$, with higher angular resolutions of 6.1, 6.4, 6.5, and 12.0 arcsec in the four bands, and higher photometric sensitivities of 0.068, 0.098, 0.86, and 5.4 mJy ($5\sigma$) for the four bands, respectively.

The comprehensive AllWISE source catalog\footnote{http://wise2.ipac.caltech.edu/docs/release/allwise/} encompasses attributes for over 747 million point-like and resolved objects detected on coadded Atlas Images (Cutri et al. 2013). Through the IRAS-WISE cross-matching process, we successfully identified 24,341 WISE counterparts in the AllWISE source catalog for the 2715 primary ULIRG candidates within the $3\sigma$ error ellipse. Figure \ref{fig:f3} illustrates the distribution of the number of WISE counterparts for each primary ULIRG candidate.
On average, the primary ULIRG candidates exhibit 8 WISE counterparts, with a maximum of 24 WISE counterparts. The far-infrared fluxes detected by IRAS are attributed to the contributions of all WISE counterparts within the 3$\sigma$ error ellipse.

\figurenum{4}
\begin{figure}[tp]
\centering
\includegraphics[width=0.48\textwidth]{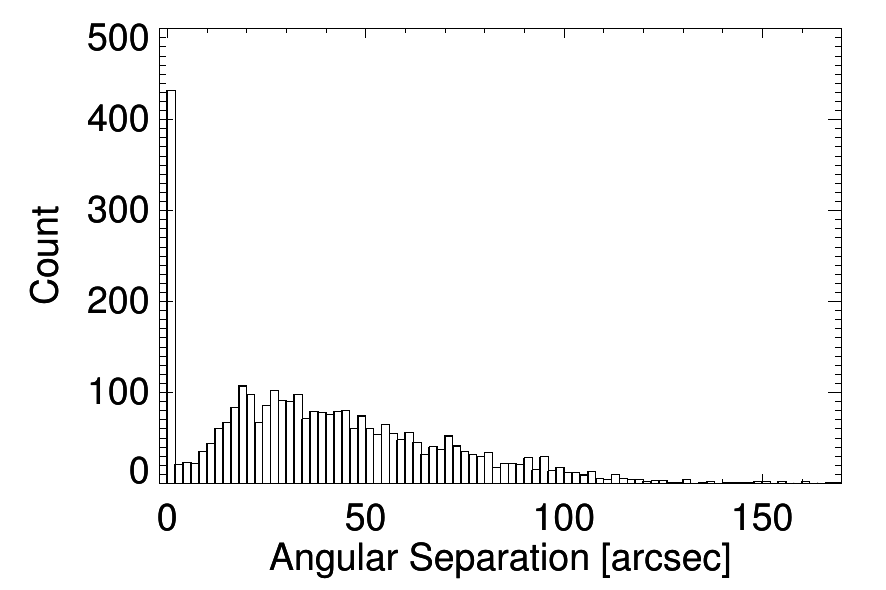}
\caption{The distribution of the angular separation between the brightest WISE sources at $W4$ band and the SDSS counterpart for each primary ULIRG candidate. The size of each bin of the histogram is 2 arcsec. \label{fig:f4}}
\end{figure}

We systematically organized the WISE counterparts for each ULIRG candidate based on their magnitudes in the $W4$ band, selecting the most luminous sources in the mid-infrared. 
The WISE $W4$ band centers at a wavelength of 22 $\mu$m, which is close to the IRAS 60 $\mu$m band.
Our working assumption is that the brightest WISE source in the W4 band is likely the WISE counterpart of the IRAS source.
We computed the distances between the brightest WISE source and the SDSS counterpart for each ULIRG candidate, illustrating the angular separation distribution in Figure \ref{fig:f4}. The distribution reveals two distinct bimodal patterns—a narrow, sharp peak at zero offsets and an exceptionally broad peak spanning from 6 to 170 arcsec.

For a valid WISE-SDSS association, the maximum angular separation is set to 2.0 arcsec, a commonly adopted criterion in SDSS quasar catalogs (e.g., P{\^a}ris et al. 2014, 2017, 2018). This choice helps minimize false-positive matches, maintaining the rate at a low level of approximately $1\%$ (Krawczyk et al. 2013; P{\^a}ris et al. 2018). 
This approach effectively addresses the issue of one IRAS source having multiple SDSS counterparts, except the source F08449+2332.

\begin{figure}[tp]
\centering
\includegraphics[width=0.3\textwidth]{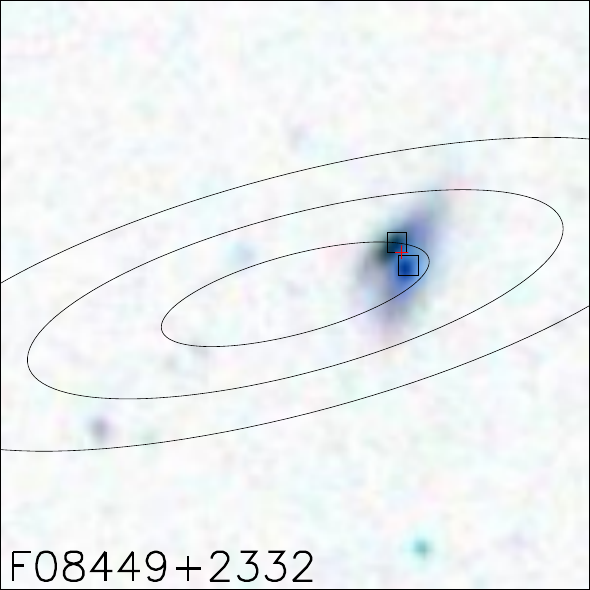}
\caption{The $1\times1$ arcmin$^2$ finding chart of F08449+2332 extracted from the SDSS $gri$ composite image. The IRAS object names are presented in the bottom left-hand corner.
The finding chart centered on the IRAS source position. The North is up, and the east is to the left-hand side. The ellipses represent 1, 2, and 3$\sigma$ positional uncertainty, and the red cross denotes the WISE counterpart.
Two SDSS counterparts (SDSS J084750.17+232108.6 and SDSS J084750.26+232110.9) in the SDSS are marked by the squares, which positional separation is only 2.5 arcsec, less than the angular resolutions of WISE.}\label{fig:f5}
\end{figure}

In the case of F08449+2332, we examined its $gri$ composite image using the SDSS DR16 Image List Tool (Figure \ref{fig:f5}). F08449+2332 exhibits two closely situated SDSS counterparts, namely, SDSS J084750.17+232108.6 and SDSS J084750.26+232110.9, with an angular separation of only 2.5 arcsecs (approximately half to one-third of the angular resolutions of WISE), thus, they cannot be resolved by WISE.
In conclusion, our WISE-SDSS association resulted in 432 matches, classifying the corresponding 431 IRAS sources as reliable ULIRG candidates.

\figurenum{5}

\figurenum{6}
\begin{figure}[tp]
\centering
\begin{minipage}[t]{0.48\textwidth}
\centering
\includegraphics[width=\textwidth]{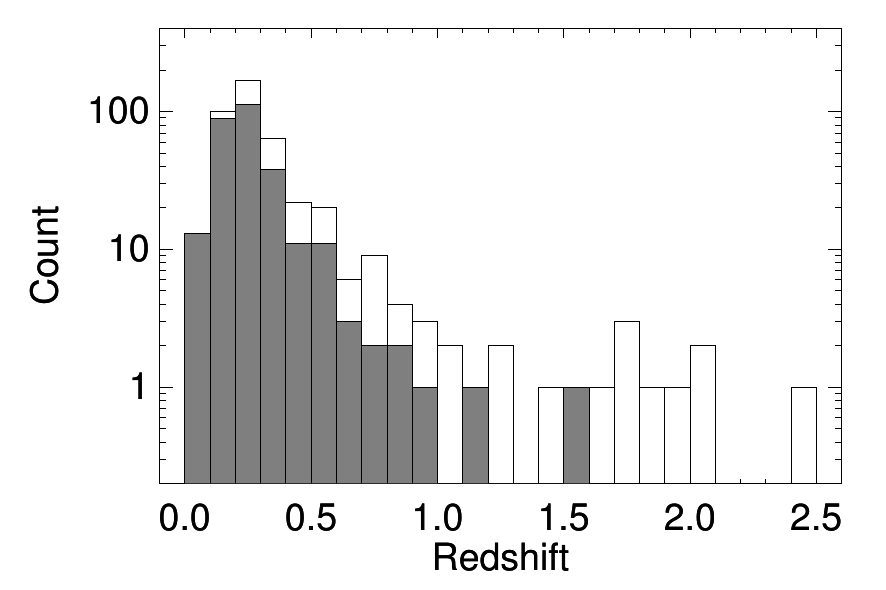}
\end{minipage}\\
\begin{minipage}[t]{0.48\textwidth}
\centering
\includegraphics[width=\textwidth]{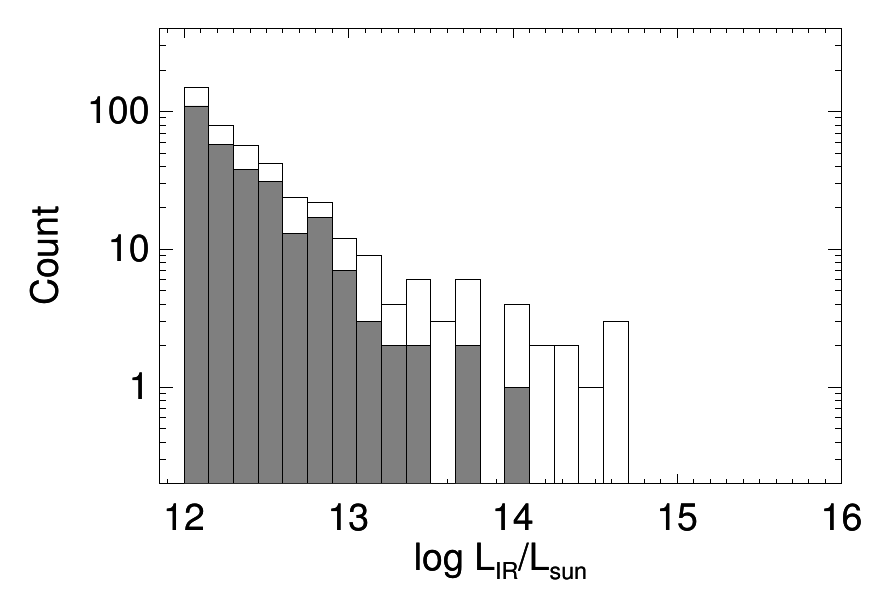}
\end{minipage}\\
\begin{minipage}[t]{0.48\textwidth}
\centering
\includegraphics[width=\textwidth]{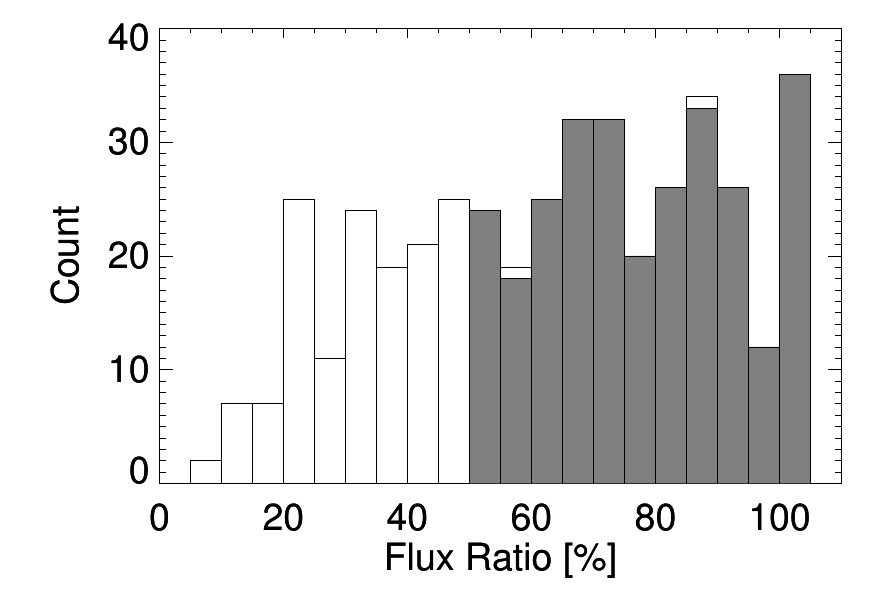}
\end{minipage}
\caption{Distributions of redshift (top panel), infrared luminosity (middle panel), and flux ratio (bottom panel) of the reliable ULIRG candidates. The shadow regions are the distributions of the final ULIRGs with a flux ratio larger than 50\%. \label{fig:f6}}
\end{figure}

Mirroring the presentation of primary ULIRG candidates, we depict the redshift and infrared luminosity distributions of reliable ULIRG candidates in Figure \ref{fig:f6}. Notably, a substantial number of IRAS sources with high-redshift SDSS counterparts have been excluded, resulting in the predominance of credible ULIRG candidates with redshifts below 1.0. Furthermore, their infrared luminosities generally fall below $\rm 10^{13}~ L_{\sun}$.

To quantitatively evaluate the contribution of SDSS counterparts in infrared, we computed the total flux from all matched WISE sources within the $3\sigma$ uncertainty ellipse. We derived the flux ratio (at the WISE $W4$ band) of the brightest source to all sources, presenting the flux ratio distribution of the 431 reliable ULIRG candidates in the bottom panel of Figure \ref{fig:f6}. Of these, 285 sources with a flux ratio exceeding 50\%, indicative of dominance in IRAS emission, were preliminarily selected to constitute our ULIRG sample.
Subsequently, a meticulous examination of the spectroscopic commissioning data of the SDSS counterpart for all sources was conducted to verify their redshifts. 
Notably, the redshifts of three high-redshift sources (F08409+4153, F08519+2017, and F15346+1409) were significantly overestimated in the SDSS DR16 release. Correct redshift values derived from their optical spectra are 0.0071, 0.3060, and 0.0002, respectively. Consequently, sources F08409+4153 and F15346+1409 were deemed invalid, and redshift and infrared luminosity values for F08519+2017 underwent subsequent revisions.
Our refined ULIRG sample comprises 283 sources, with their redshift, infrared luminosity, and flux ratio distributions displayed in the shaded regions of the panels in Figure \ref{fig:f6}.

Table \ref{tab:t1} provides a comprehensive list of our ULIRGs, encompassing details such as the names of IRAS sources, identifications of their optical counterparts, redshifts, IRAS flux densities and qualities, and infrared luminosities. To facilitate visual understanding, Figure \ref{fig:f7} displays $2.5\times2.5$ arcmin$^2$ finding charts for all sources within our ULIRG sample. These charts are extracted from the SDSS $gri$ composite images, with each image centered on the position of the corresponding IRAS source. The SDSS counterpart is denoted by a black square, and the red crosses represent WISE sources found within the $3\sigma$ error ellipse.

\figurenum{7}
\begin{figure*}[tp]
\centering
\includegraphics[width=\textwidth]{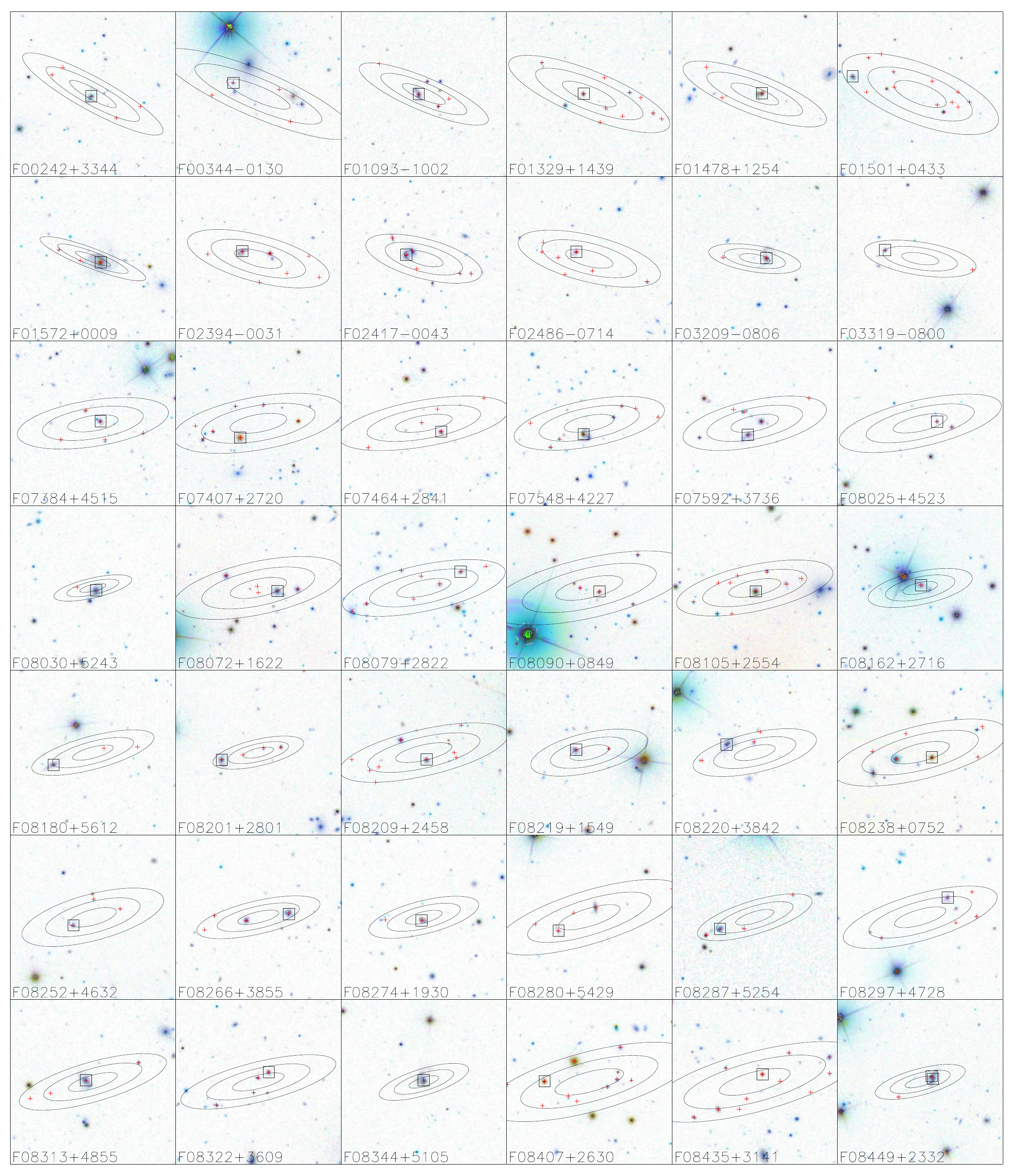}
\caption{Finding charts of the 285 ULIRGs. The $2.5\times2.5$ arcmin$^2$ finding charts extracted from the SDSS $gri$ composite images centered on IRAS sources' positions. Same as Figure \ref{fig:f5}, the north is up, and the east is to the left-hand side, and the ellipses represent 1, 2, and 3$\sigma$ positional uncertainty. The square and red crosses represent the SDSS counterpart and the $WISE$ sources within the $3\sigma$ error ellipse. For  F111388+0108,  the scale of the image is $\rm 5 \times 5~ arcmin^2$.\label{fig:f7}}
\end{figure*}

\figurenum{7}
\begin{figure*}[tp]
\centering
\includegraphics[width=\textwidth]{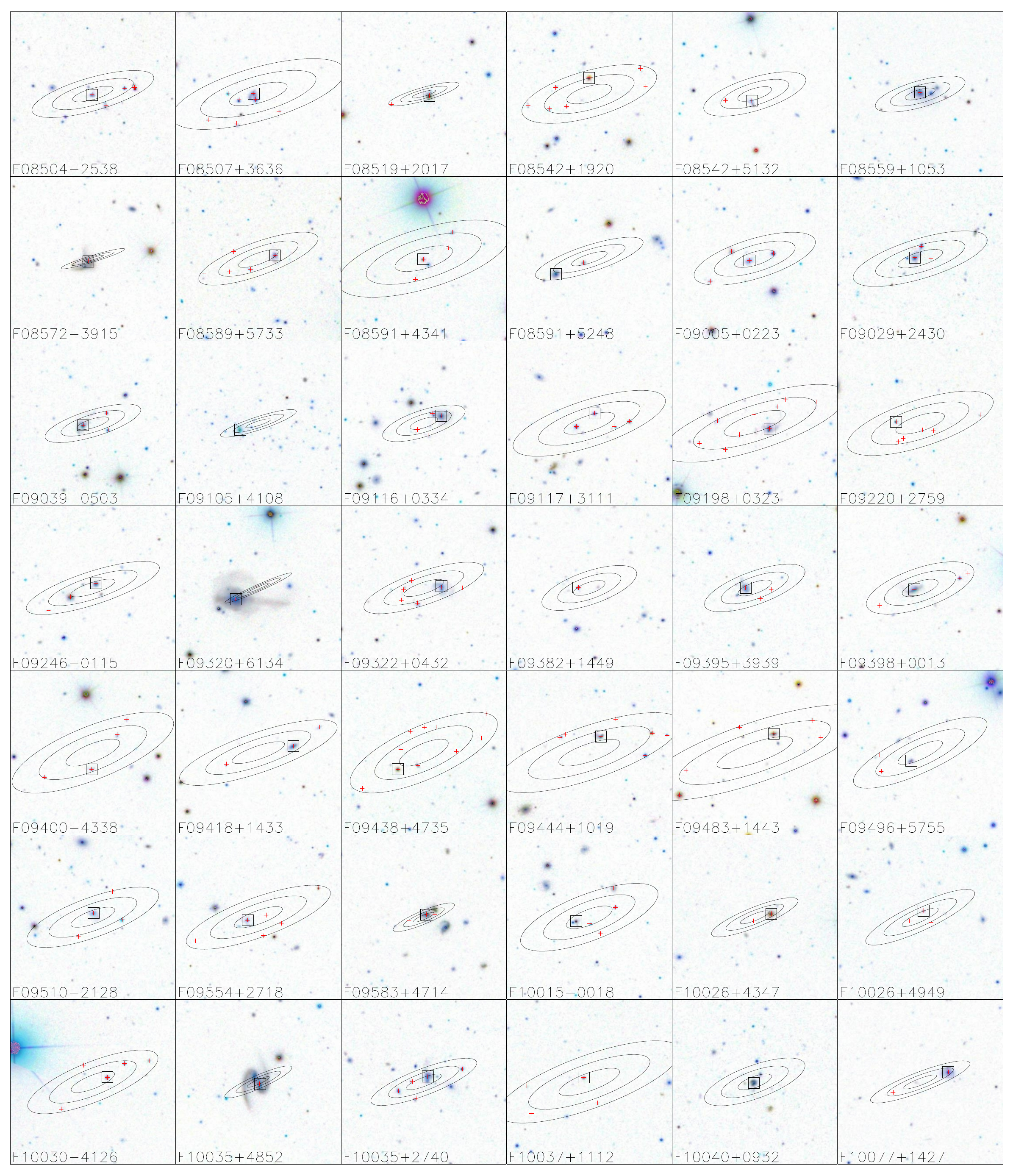}
\caption{Continued. \label{fig:f7}}
\end{figure*}

\figurenum{7}
\begin{figure*}[tp]
\centering
\includegraphics[width=\textwidth]{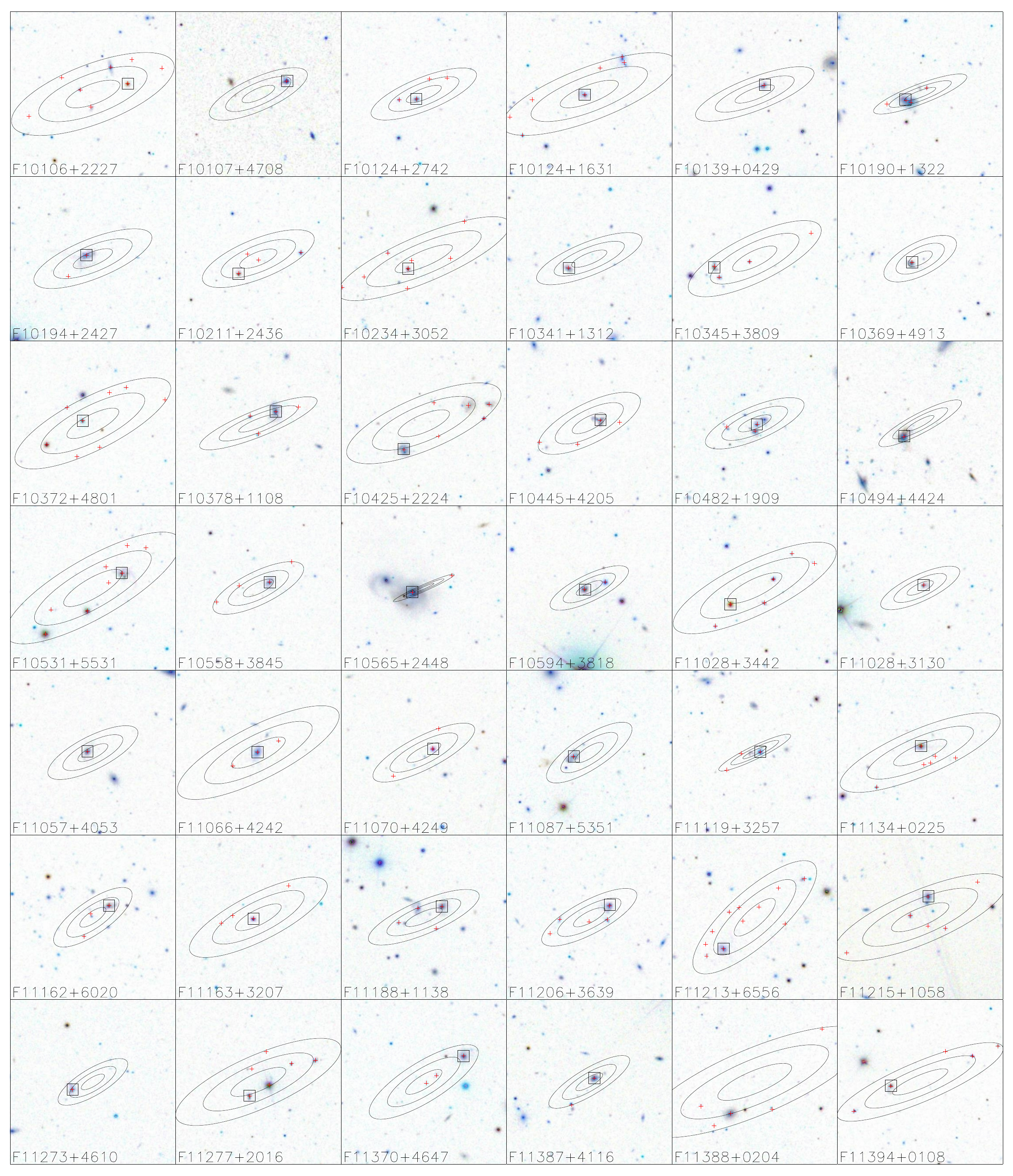}
\caption{Continued. \label{fig:f7}}
\end{figure*}

\figurenum{7}
\begin{figure*}[tp]
\centering
\includegraphics[width=\textwidth]{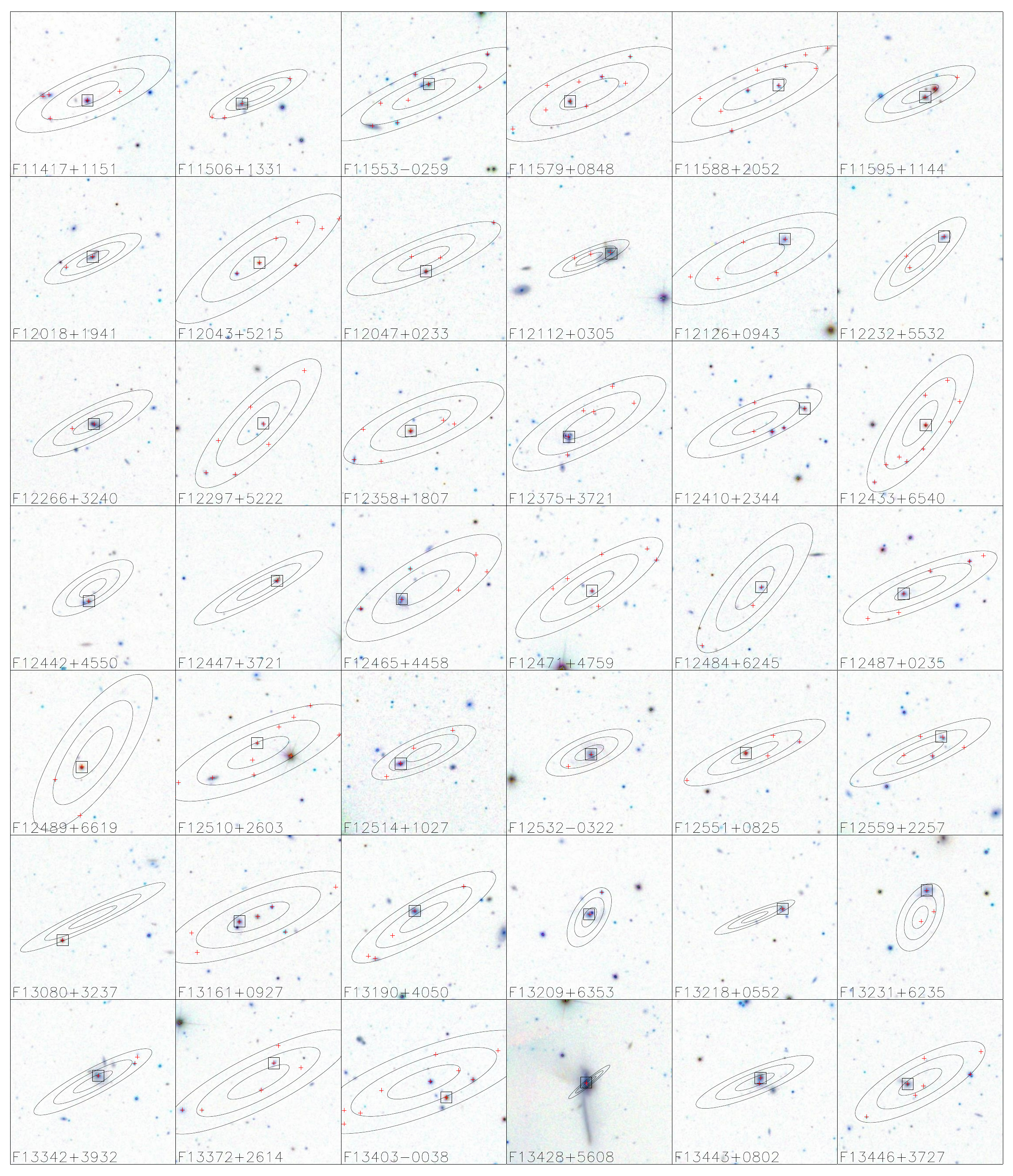}
\caption{Continued. \label{fig:f7}}
\end{figure*}

\figurenum{7}
\begin{figure*}[tp]
\centering
\includegraphics[width=\textwidth]{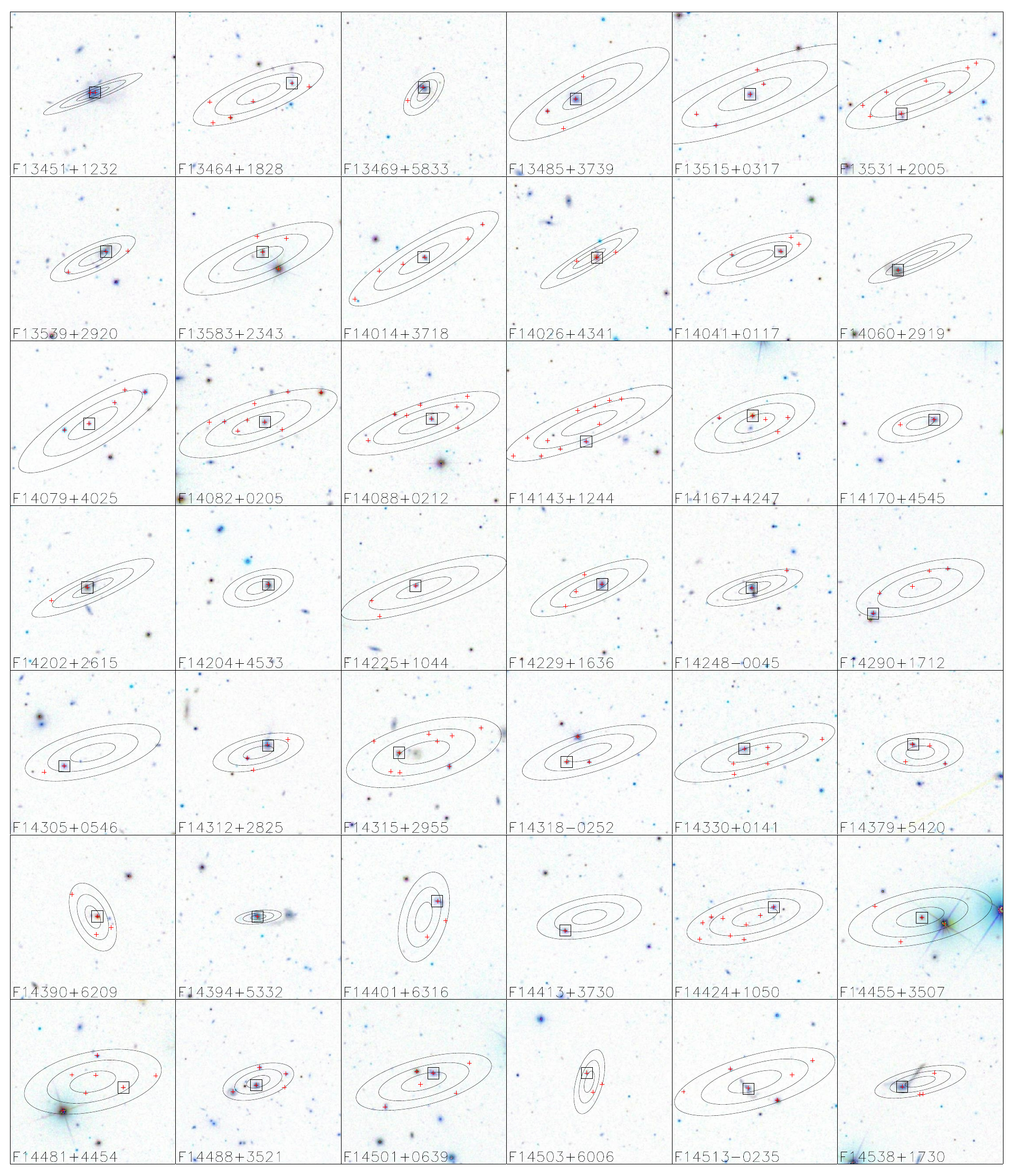}
\caption{Continued. \label{fig:f7}}
\end{figure*}

\figurenum{7}
\begin{figure*}[tp]
\centering
\includegraphics[width=\textwidth]{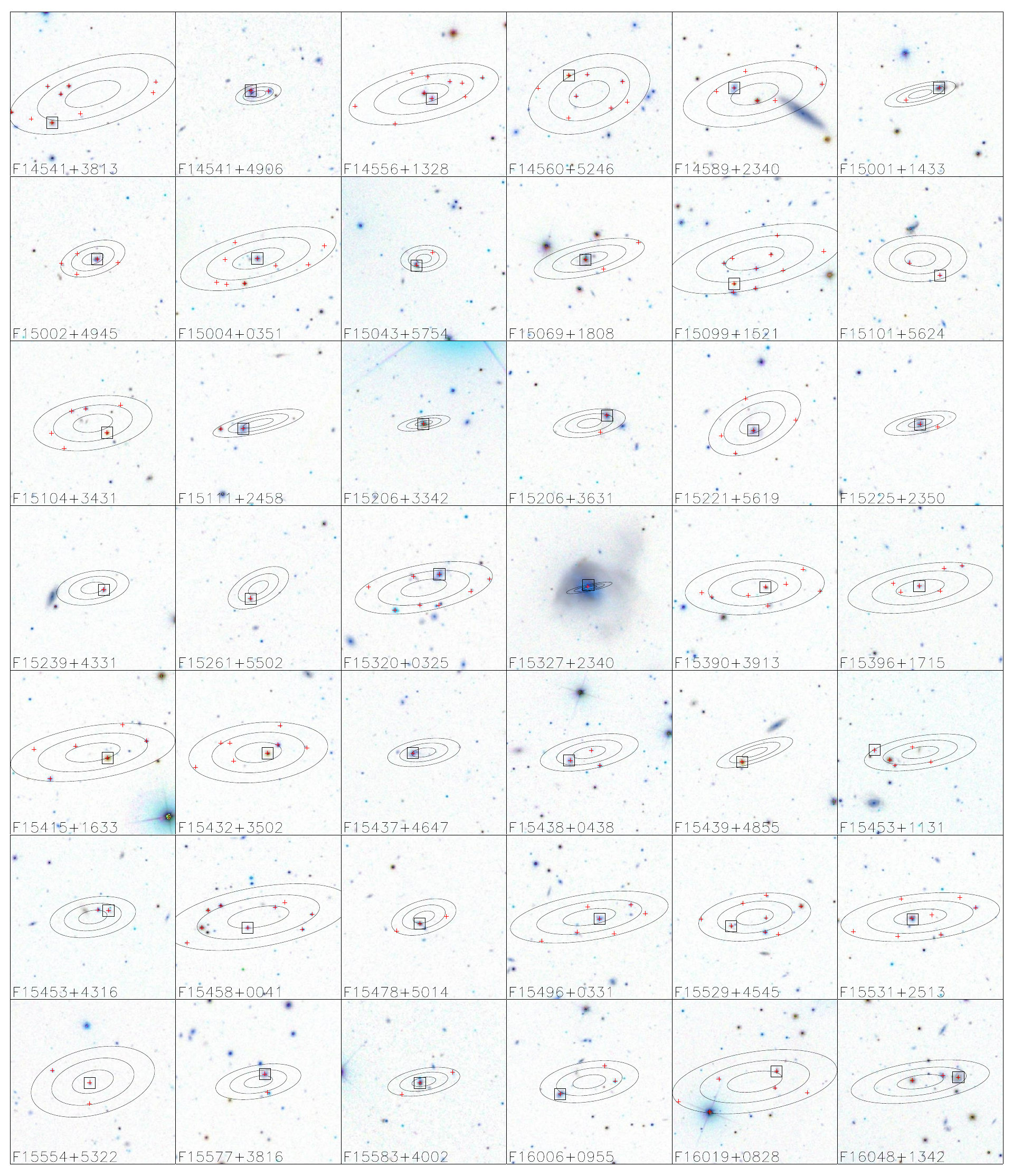}
\caption{Continued. \label{fig:f7}}
\end{figure*}

\figurenum{7}
\begin{figure*}[tp]
\centering
\includegraphics[width=\textwidth]{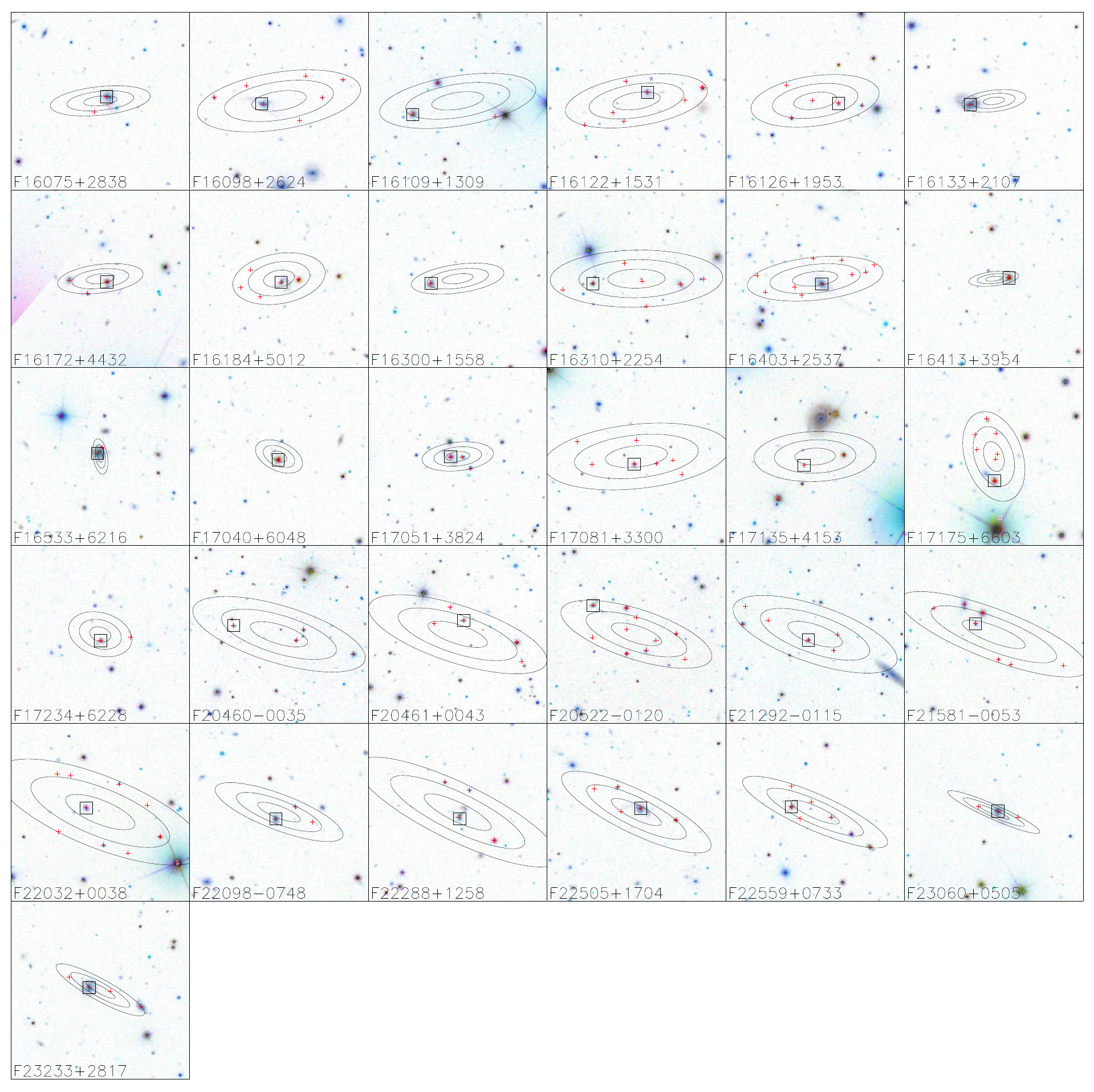}
\caption{Continued. \label{fig:f7}}
\end{figure*}

The criteria for ULIRG selection are stringent, potentially leading to the inadvertent exclusion of certain ``true" ULIRGs. Typically, ULIRGs are categorized as either ``cool" or ``warm" systems, based on the dominant power source of mid-/far-infrared radiation, such as starburst and AGN activity (Sanders et al. 1988b; Veilleux et al. 2002). 
In the context of pure type I AGN, the flux difference between the $W4$ and 60 $\mu$m bands shows only a marginal variation, constituting approximately $20-30\%$ of the total flux. 
Conversely, in star-forming templates, this factor could increase by a factor of $2\sim6$ at 60 $\mu$m compared to the $W4$ band (Polletta et al. 2007). 
It is crucial to recognize that an elevation in this factor also results in a proportional increase in the overall brightness of the system. This observation implies that AGNs are more likely to be identified as ULIRGs due to their greater luminosity in the mid-infrared waveband compared to star-forming galaxies.

\begin{figure}[tp]
\centering
\begin{minipage}[t]{0.48\textwidth}
\centering
\includegraphics[width=\textwidth]{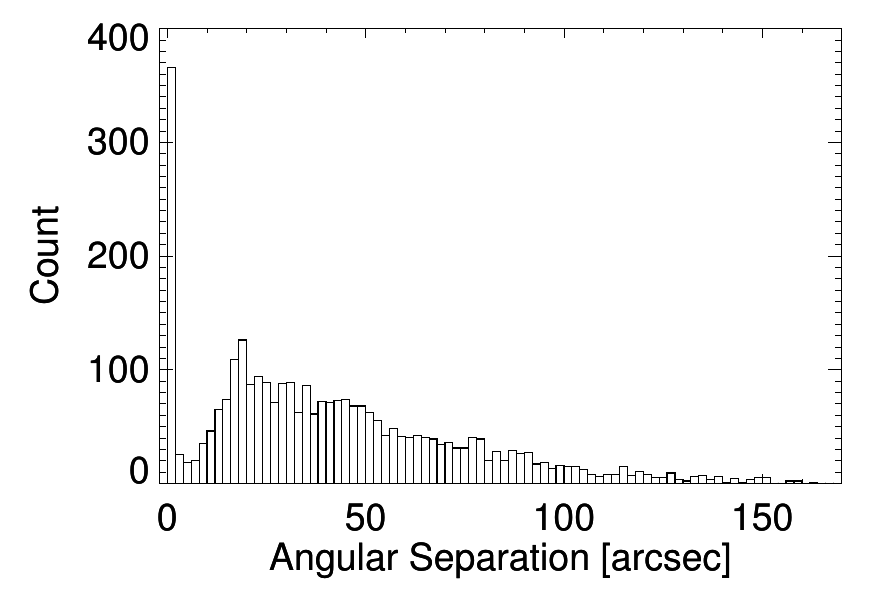}
\end{minipage}
\begin{minipage}[t]{0.48\textwidth}
\centering
\includegraphics[width=\textwidth]{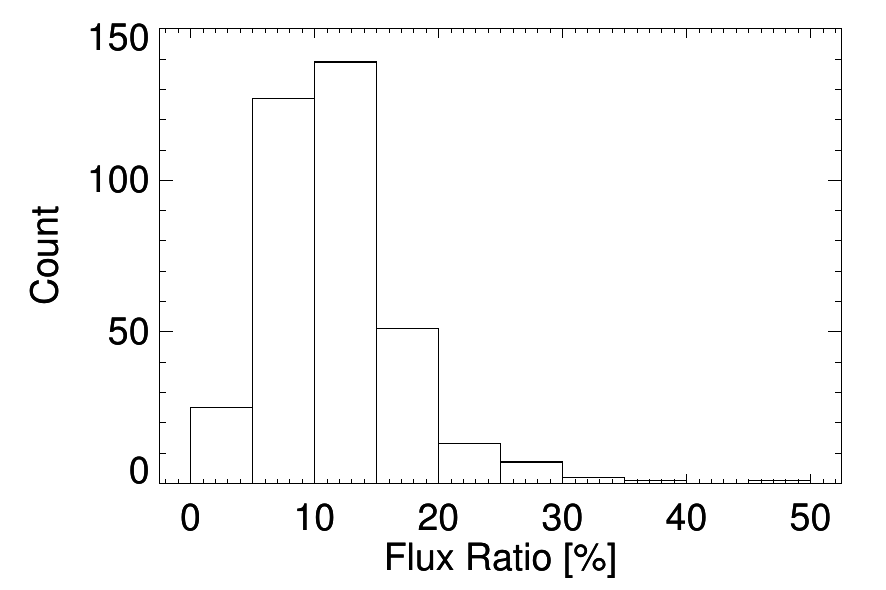}
\end{minipage}
\caption{Top panel: Distribution of the angular separation between the second bright WISE sources at the $W4$ band and the SDSS counterpart for each primary ULIRG candidate. The size of each bin of the histogram is 2 arcsec.
Bottom panel: Flux ratio distribution of the second bright WISE sources within 2 arcsec of the SDSS counterparts.
\label{fig:f8}}
\end{figure}

To explore potential selection bias, we opted for the second brightest source at the $W4$ band for each primary ULIRG candidate. In Figure \ref{fig:f8}'s top panel, we depict the separation distribution between the second brightest WISE sources at the $W4$ band and the SDSS counterpart for each primary ULIRG candidate. This distribution closely resembles that of the brightest WISE sources, with 366 effective matches within 2.0 arcsec.
We computed the flux ratio of the effective WISE source at the $W4$ band to all WISE sources within a 3$\sigma$ uncertainty ellipse for these 366 effective matches. The distribution of this flux ratio is presented in the bottom panel of Figure \ref{fig:f8}. 
It is worth noting that, despite being the second brightest source within the corresponding uncertainty ellipse, their fluxes remain relatively faint. The distribution of flux ratios peaks at approximately 10\%, with the majority of sources displaying ratios below 20\%. Only four sources exceed 30\%. This analysis suggests that the aforementioned concerns regarding selection bias may not be well-founded.

We have carried out  detailed  quantitative analyses to assess the impact of positional uncertainties on our method. To simulate potential inaccuracies, we artificially shifted the positions of the IRAS sources by 5 degrees in arbitrary directions, then executed cross-matching using the method we previously described.
This exercise reveals 3,640 IRAS sources have corresponding SDSS counterparts. Notably, 314 of these sources are associated with multiple counterparts within the $3\sigma$ error ellipses. 
Furthermore, we experimented with shifting the positions by various distances, yet the counts of IRAS-SDSS matches remained basically consistent.
The substantial number of matches for the artificially shifted IRAS positions can be mainly ascribed to the average number density of extragalactic sources within the SDSS and the large positional uncertainties inherent in the IRAS.
Upon refining the selection to IRAS sources with ``high" or ``moderate" quality for their 60  $\mu m$ flux densities, the count was reduced to 2,923.
Applying an infrared luminosity threshold, we selected 2,234 ``primary ULIRG  candidates" with infrared luminosities of  $L\rm_{IR} \ge 10^{12}~\rm L_{\sun}$.  
These ``primary ULIRG  candidates" show higher redshifts than the genuine ones,with a range from  $z = 0.05$ to $6.87$, and a median redshift  of $z = 0.67$. 
The corss-matching with the WISE data has exposed that  approximately a quarter of the SDSS counterparts  are missing corresponding entries in the AllWISE catalog.
After imposing the flux ratio threshold at W4, we have further refined the  IRAS-SDSS-WISE matches to 32, which could be considered potential error introduced by our method, mainly due to the uncertainties in the IRAS-SDSS matching process.
Similarly, considering the  $1\%$ false matching  rate in the SDSS-WISE data, we estimate that  there would be approximately 27 suprious matchs in our primary sample of 2,715 ULIRG candidates. Assuming that the same proportion of these false matches would pass the W4 flux ratio threshold, we anticipate that roughly 18 of these matches would incorrectly be classified as valid.

\figurenum{8}

\section{Results and Discussion}

\subsection{Interaction Features in Optical Images}

\figurenum{9}
\begin{figure*}[tp]
\centering
\includegraphics[width=\textwidth]{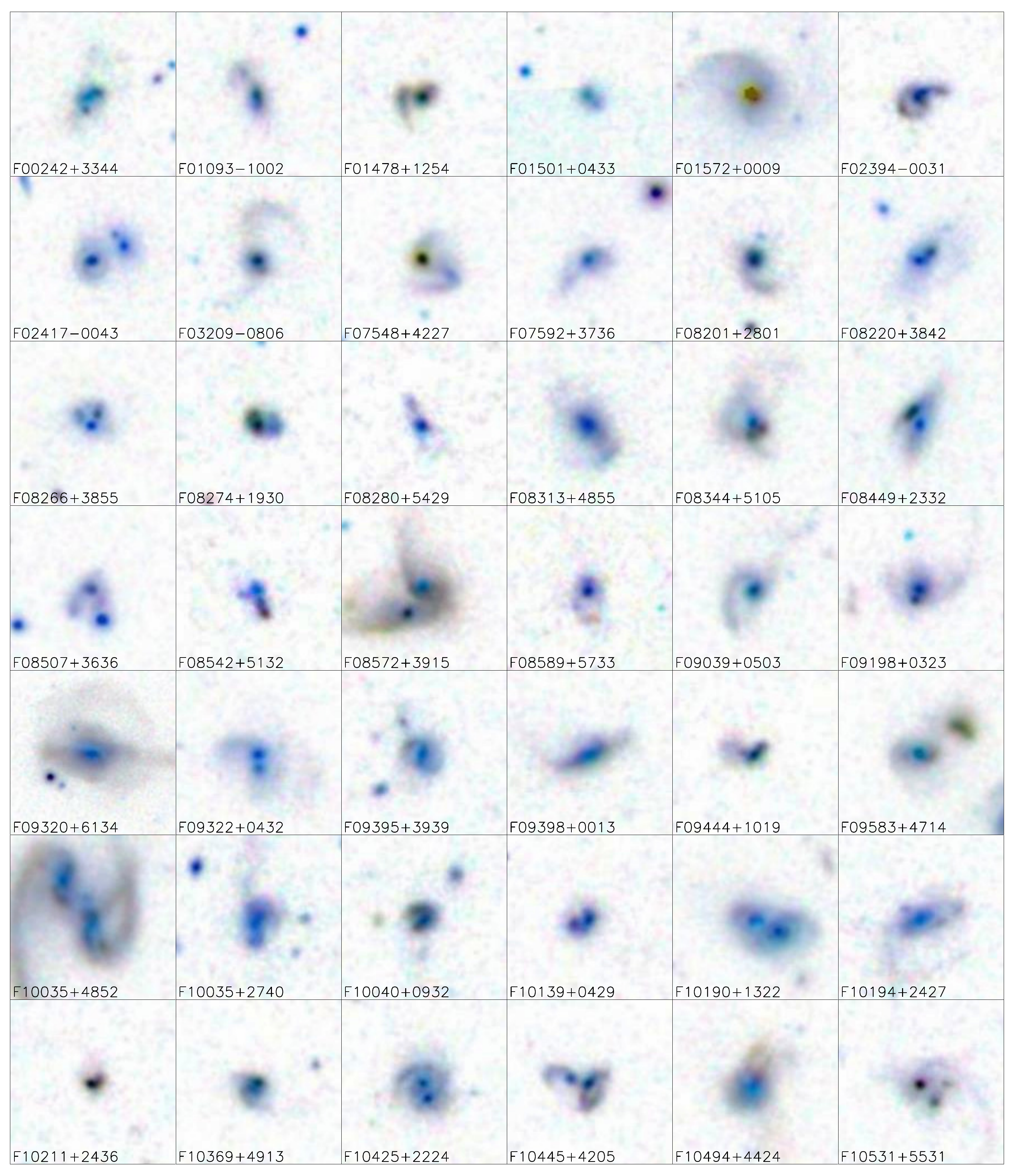}
\caption{The $30\times30$ arcsec$^2$ optical images of ULIRGs with obvious interacting features, which are centered on their SDSS counterparts' positions. For 5 cases of them (F09320+6134, F10565+2448, F13428+5608, F14394+5332, and F15327+2340), the scales of the images are $1\times1$ arcmin$^2$.	\label{fig:f9}}
\end{figure*}

\figurenum{9}
\begin{figure*}[tp]
\centering
\includegraphics[width=\textwidth]{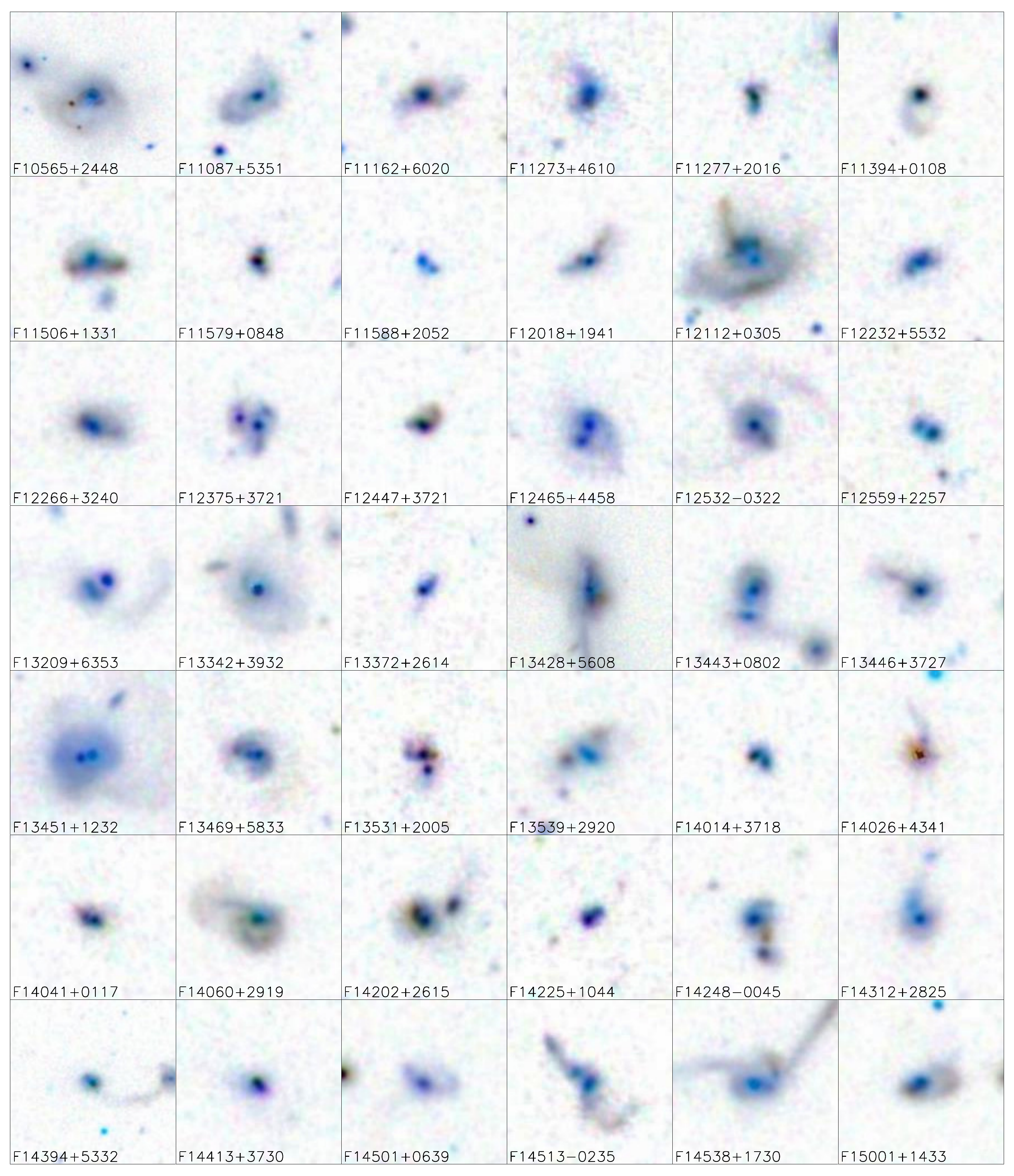}
\caption{Continued. \label{fig:f9}}
\end{figure*}

\figurenum{9}
\begin{figure*}[tp]
\centering
\includegraphics[width=\textwidth]{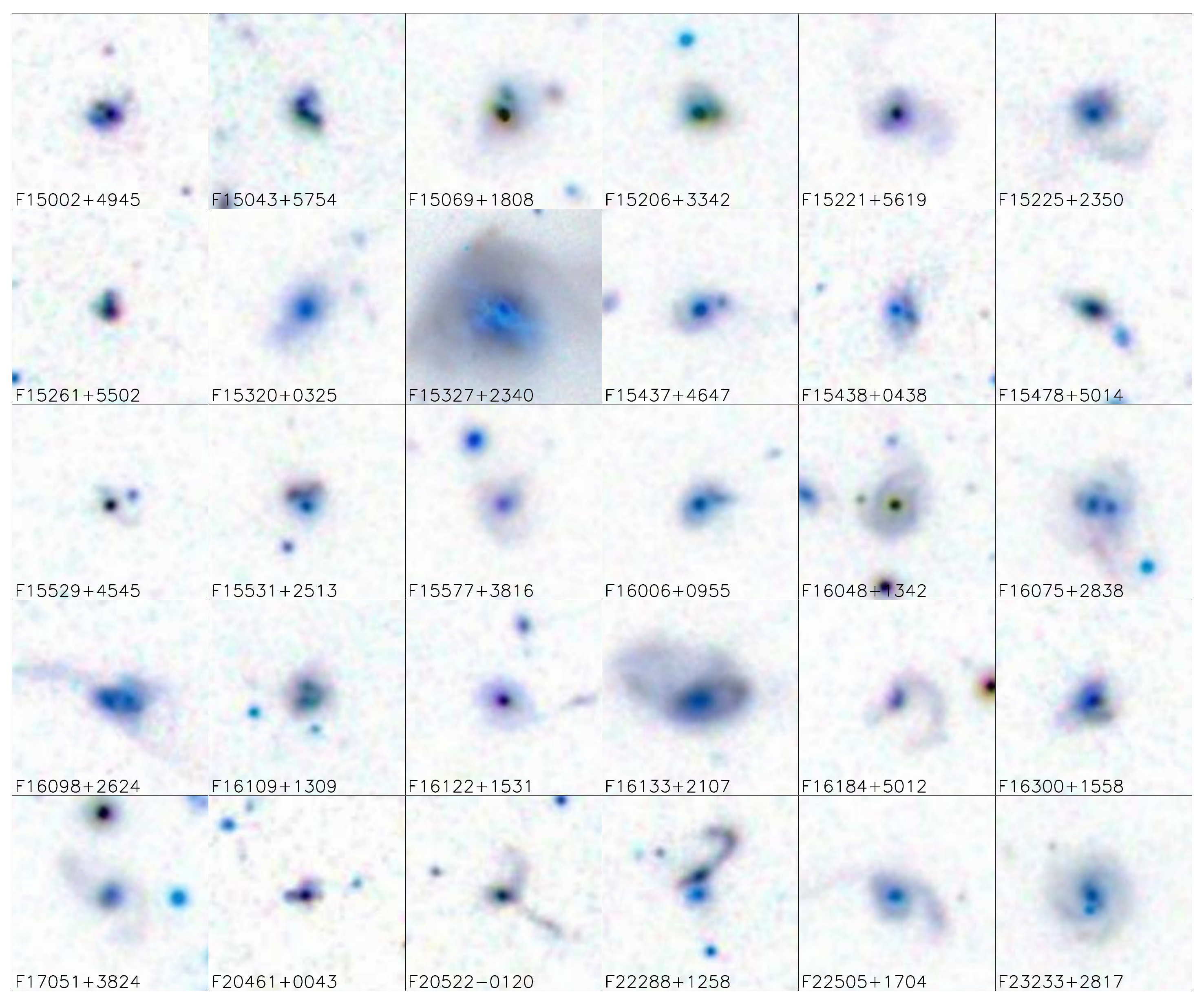}
\caption{Continued. \label{fig:f9}}
\end{figure*}

Previous works have revealed that ULIRGs have a wide morphological distribution, and their structures generally vary from ellipsoid to multiple nuclei. Some of them exhibit the appearance of a merger of two nuclei or the gas bridges and the tidal asymmetries structures in the optical and near-infrared images, and this indicates that there has been an interaction and merger process in the history of ULIRGs (e.g., Sanders et al. 1988a; Farrah et al. 2001; Veilleux et al. 2002; Xia et al. 2012).
In this study, we employed the SDSS DR16 Image List Tool\footnote{https://skyserver.sdss.org/dr16/en/tools/chart/list.aspx} to examine ULIRG images. 
Only 114 cases show apparent interaction features in their SDSS images, and the fraction is slightly lower than previous similar findings (e.g., Hou et al. 2009).
Due to the limited resolution of the SDSS image, we can only mark the ULIRGs with a noticeable interaction feature.

The relative fractions of interacting and merging systems varied across different subsamples of ULIRGs, ranging from as low as $10\%-20\%$ in some cases to as high as $80\%$ or more in others (e.g., Sanders et al. 1988a; Melnick \& Mirabel 1990; Zou et al. 1991; Leech et al. 1994; Clements et al. 1996; Murphy et al. 1996; Bushouse et al. 2002; Perna et al. 2021). Perna et al. (2021) observed 25 nearby ($z < 0.165$) ULIRGs using the integral field spectrograph MUSE, revealing systems with both AGN and Starburst (SB) nuclear activity in pre- and post-coalescence phases of major mergers. Five of their ULIRGs are included in our sample, with MUSE images providing detailed insights.
For instance, F01572+0009 exhibits ordered disk-like rotations and a merger classification with a single nucleus and small nearby companions. F10190+1322 displays two overlapping disks centered at the positions of the two nuclei, separated by $\sim 7$ kpc. F12112+0305 is identified as an interacting system, while F13451+1232 presents tidal tails and two nuclei separated by $\sim 4.18$ kpc. F15327+2340 shows a disturbing kpc-scale disk in the innermost nuclear regions and two nuclei separated by $\sim 0.37$ kpc.

Figure \ref{fig:f9} provides $30\times30$ arcsec$^2$ finding charts extracted from SDSS $gri$ composite images. Interaction types of ULIRGs were classified the interaction types of ULIRGs using zoom-in images, revealing that nearly half of ULIRGs show one nucleus with tail features, approximately 35\% have two identified nuclei with well-developed tidal tails or bridges, and the remaining 18\% have two close or even overlapped nuclei. It is unclear whether the latter are galaxy pairs.
In almost all cases, only one source in the system has effective redshift measurements. ULIRGs exhibit clearer interaction features, with a percentage of about 92\% for $z < 0.1$, 66\% for $0.1 \le z < 0.2$, 32\% for $0.2 \le z < 0.3$, 18\% for $0.3 \le z < 0.4$, and 11\% for $z \ge 0.4$. The fraction decreases with increasing redshift, suggesting a selection effect in classifying interactive features.
Notably, the ULIRGs identified in this study generally have higher redshifts than previous findings, resulting in a lower percentage of sources with interactive features.

Due to the limit of the spatial resolution and survey depth of most ground-based imaging surveys, most previous studies on galaxy mergers are often limited to the early stages of galactic gravitational convergence. 
In contrast, signs such as large-scale asymmetric structures or tidal tails and stellar streams produced by galaxy mergers tend to dissipate in the late stages. 
However, high-resolution images ($\sim 0.1''$) have revealed dual-core structures at the kpc scale in the galaxy's central region (e.g., Bushouse et al. 2002; Koss et al. 2018). This dual-core design is likely linked to the two phases of galaxy mergers and black hole mergers.
By comparing the ratio of AGN, galaxy morphology, and stellar population of dual-core galaxies, galaxy pairs, and isolated galaxies, a comprehensive understanding of galaxy merging into black hole pre-merging can be established. This approach may reveal a series of frontier problems, including the formation of nuclear spheres and the mechanism of black hole triggering. High-resolution imaging observations are imperative for clarifying these issues.

The upcoming China Space Station Telescope (CSST), a 2-meter space telescope boasting a large field of view ( $\rm \sim 1.1~deg^2$) and sharing the same orbit as the Chinese Space Station, is poised for launch. 
The CSST Optical Survey (CSS-OS; Zhan 2011, 2018, 2021; Cao et al. 2018; Gong et al. 2019) will undertake a comprehensive multi-band survey with high spatial resolution (approximately $0.15''$ for the survey camera). 
This survey encompasses the NUV, $u$, $g$, $r$, $i$, $z$, and $y$ bands, spanning the wavelength range of $255 - 1000$ nm, from the NUV to NIR.
Over the next approximately ten years, the CSS-OS is projected to cover an extensive sky area of $\rm 17,500 ~deg^2$ with deep limiting magnitudes, reaching around $26$ AB mag or higher in the $g$ and $r$ bands for $5\sigma$ point source detection.
The high-resolution, high-quality, and deep images captured by the CSST are expected to provide a more detailed understanding of the morphology and interactions of ULIRGs. 
The investigation of the interacting and merger processes of supermassive binary black holes and galaxies stands out as one of the pivotal scientific focuses for the CSST. 

\subsection{Infrared Colours}

\figurenum{10}
\begin{figure*}[tp]
\centering
\begin{minipage}[t]{0.48\textwidth}
\centering
\includegraphics[width=\textwidth]{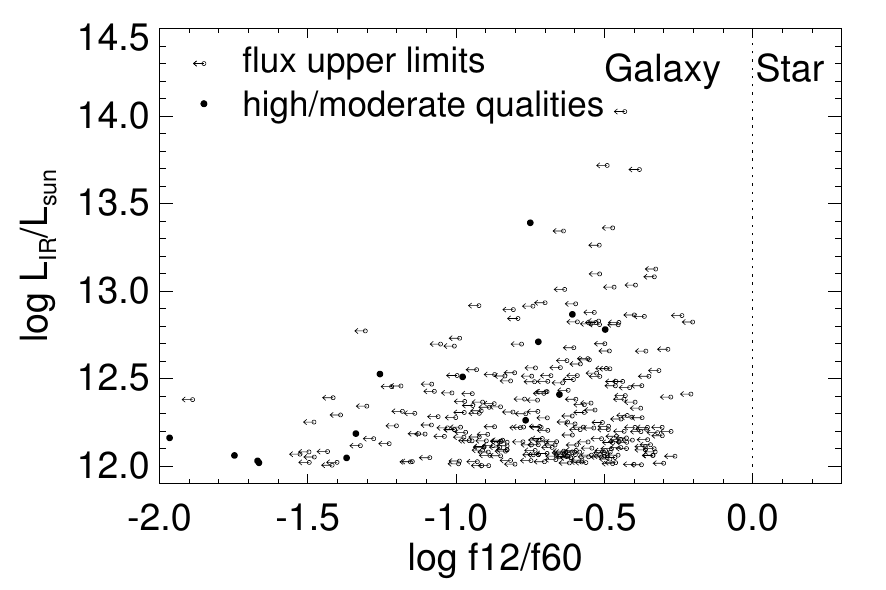}
\end{minipage}
\begin{minipage}[t]{0.48\textwidth}
\centering
\includegraphics[width=\textwidth]{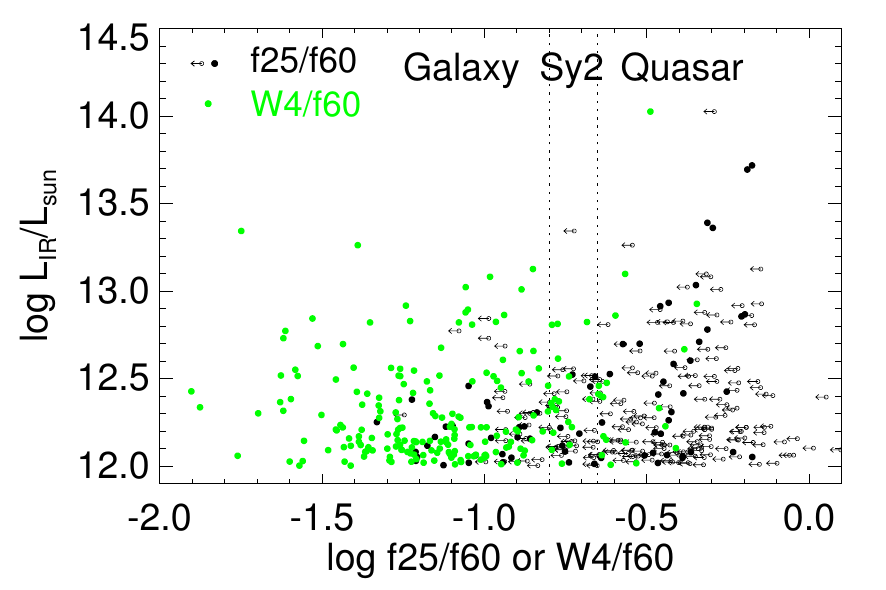}
\end{minipage}
\begin{minipage}[t]{0.48\textwidth}
\centering
\includegraphics[width=\textwidth]{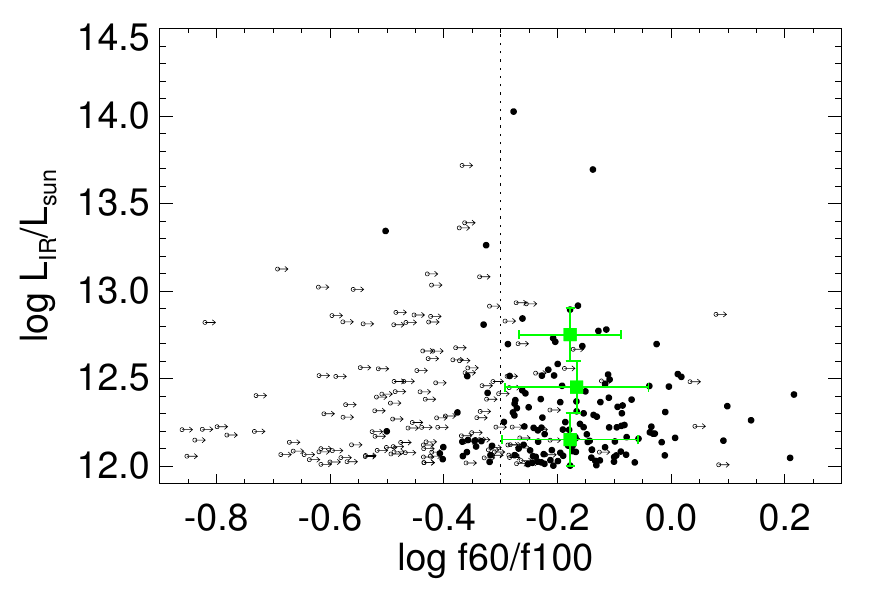}
\end{minipage}
\begin{minipage}[t]{0.48\textwidth}
\centering
\includegraphics[width=\textwidth]{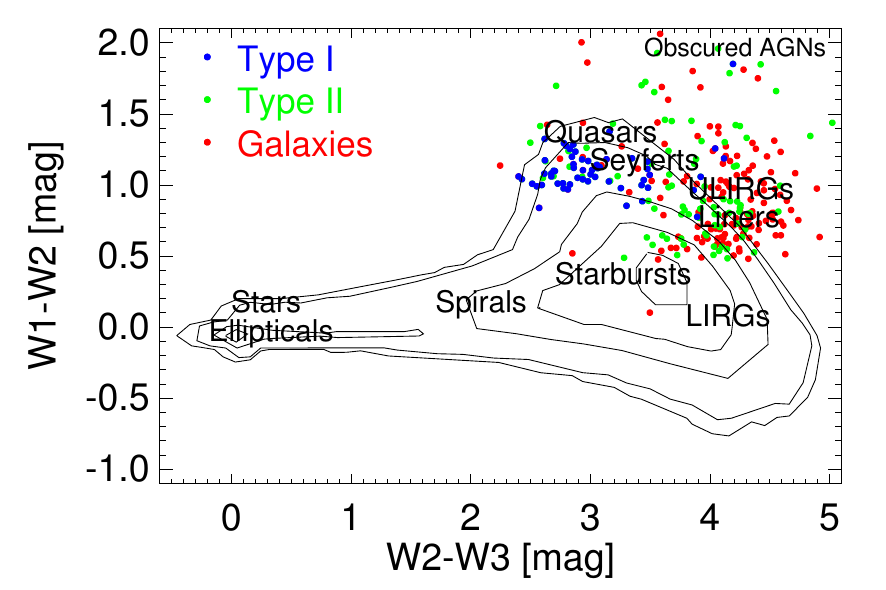}
\end{minipage}
\caption{The infrared luminosity and color diagrams for the ULIRGs. The open circles represent the sources with flux upper limits and high/moderate flux qualities with the open-circle-arrows and the filled circles, respectively.
Top-left panel: The dotted line represents the boundary between galaxies and stars (Kim \& Sanders 1998).
Top-right panel: For the sources with flux upper limits, we use the WISE $W4$ magnitudes (centered at 22 $\mu$m) to replace the upper limits at 25 $\mu$m; the green filled circles represent the colors of $W4/f60$. The dotted lines represent the boundaries between normal galaxies, Seyfert 2, and quasars (Neff \& Hutchings 1992).
Bottom-left panel: The dotted line represents the selection criteria (${\rm log}~ f60/f100> -0.3$) of the 1-Jy sample of ULIRGs (Kim \& Sanders 1998). The squares are at the mean color and the central luminosity of ULIRGs in the three luminosity bins. The vertical error bars define the limiting luminosities of the bins, and the horizontal error bars represent the standard deviation of ${\rm log}~ f60/f100$ colors in each bin.
Bottom-right panel:  WISE color-color diagram for the ULIRGs. The red, blue, and green filled circles symbolize the galaxies (not included in the Milliquas catalog), Type I and II  (distinguished by the presence or absence of broad emission lines as per Paliya et al. (2024) and Wu \&Shen (2022), see details in Section 4). The plotted isodensity contours refer to WISE thermal sources and locations of various source classes are also highlighted. The contour data are adopted from Massaro et al. (2011).
\label{fig:f10}}
\end{figure*} 

The classification of infrared sources has long relied on the infrared colors of $f12/f60$, $f25/f60$, and $f60/f100$. 
Following the approach of Hwang et al. (2007), we extended our examination to the infrared colors of ULIRGs. 
The top-left panel of Figure \ref{fig:f10} illustrates the relationship between infrared luminosity and the color of $f12/f60$ for ULIRGs. The $f12/f60$ color proves effective in distinguishing stars from galaxies (e.g., Cohen et al. 1987), 
particularly as infrared-bright stars typically exhibit spectral energy distributions peaking around 12 $\mu$m, characterized by a logarithmic slope of ${\rm log}~ f12/f60 > 0$ (e.g., Cohen et al. 1987; Kim \& Sanders 1998).
In this panel, open circles with arrows represent the 268 sources with flux upper limits at 12 $\mu$m, while the filled circles represent the remaining 15 ULIRGs with high or moderate flux qualities at 12 $\mu$m. 
Notably, none of the ULIRGs fall into the region of ${\rm log}~ f12/f60 > 0$, indicating that our ULIRG sample is free from contamination by infrared-bright stars.
 This result is consistent with our approach during the construction of the ULIRG sample, where all objects labeled as ``STAR" were initially removed from the SDSS DR16 spectral catalog (see Section 2.1).

In the bottom-left panel of Figure \ref{fig:f10},  we illustrate the relationship between infrared luminosity and the color of $f60/f100$ for ULIRGs. Open circles represent the sources with flux upper limits at 100 $\mu$m with the arrows, while filled circles represent sources with high or moderate flux qualities. For subsequent analysis, we focus on sources with high or moderate flux qualities.
The color of ${\rm log}~ f60/f100$ ranges from $-0.50$ to $+0.22$ with a mean value of $-0.18$. Notably, $86\%$ (123 out of 143) of ULIRGs exhibit colors of ${\rm log}~ f60/f100$ surpassing $-0.3$. 
This aligns with the fourth selection criterion for ``warm" sources proposed by Kim \& Sanders (1998), aimed at including genuine ULIRGs while excluding a significant fraction of less luminous galaxies. The distribution range and mean value of $f60/f100$ color in our study fall intermediate to those of other ULIRG samples (e.g., Soifer \& Neugebauer 1991; Kim \& Sanders 1998; Hwang et al. 2007).

Soifer \& Neugebauer (1991) demonstrated an increasing mean $f60/f100$ color with rising infrared luminosity in the IRAS Bright Galaxy Survey. We plotted the mean colors in each luminosity bin using filled squares to explore any dependence of color on infrared luminosity. Our sample exhibits no significant infrared luminosity dependence of the ${\rm log}~ f60/f100$ color, up to $\rm 10^{12.8}~ L_{\sun}$; nevertheless, our colors are warmer than those typically found in less infrared-luminous galaxies, which tend to have cooler $f60/f100$ colors.

Additionally, we focus on the distribution of the $f25/f60$ color. 
Given the tendency of ULIRGs in interacting systems to undergo collisions and mergers, these processes are believed to be the primary mechanisms responsible for a significant portion of the infrared emission. 
In Veilleux et al. (2002), it was found that Seyfert galaxies, particularly Type I, warm ULIRGs ($f25/f60 > 0.2$), and more luminous sources ($\rm >10^{12.5} ~L_{\sun}$) showed a strong inclination toward advanced mergers with a single nucleus. 
In essence, the $f25/f60$ color effectively distinguishes the dominant energy source in ULIRGs (i.e., starburst and AGNs) and categorizes ULIRGs into "cool" and "warm" systems (e.g., Sanders et al. 1988b; Veilleux et al. 2002).

The top-right panel of Figure \ref{fig:f10} depicts the relationship between infrared luminosity and the color of $f25/f60$ for ULIRGs. The two vertical lines (${log} ~f25/f60 = -0.8$ and $-0.65$) represent the boundaries for the classification of normal galaxies, Seyfert 2 galaxies, and quasars as suggested by Neff \& Hutchings (1992). 
Among these ULIRGs, black-filled circles represent 70 sources with high or moderate flux qualities at 25 $\mu$m, while open circles with arrows represent others with flux upper limits at 25 $\mu$m. 
The AllWISE source catalog provides effective detections of upper limit sources at the $W4$ band, represented by green-filled circles.

The number ratio of objects is then $N{\rm(Galaxy)} : N{\rm (Sy2)} : N{\rm(Quasar)} = 204: 26 : 53$. 
Adopting $f25/f60= 0.2$ as the classification boundary between ``cool" and ``warm" ULIRGs (Sanders et al. 1988b), the number ratio for our ULIRG sample becomes $N{\rm (Cool)} : N{\rm (Warm)} = 225 : 58$. 
This ratio is comparable to that of the 1-Jy sample ($N{\rm (Cool)} : N{\rm (Warm)} = 90 : 25$, Kim \& Sanders 1998), and much higher than that of ULIRGs with high or moderate flux qualities ($N{\rm (Cool)} : N{\rm (Warm)} = 38 : 30$, Hwang et al. 2007). The differences may be related to the distinct selection criteria and the exclusion (or not) of sources with flux upper limits at 25 $\mu$m.

We conducted a cross-match between our ULIRG catalog and the latest release of the Million Quasars (Milliquas\footnote{https://heasarc.gsfc.nasa.gov/w3browse/all/milliquas.html}) v8 catalog (Flesch 2023). 
Milliquas comprises 907,144 type-I Quasars \& AGNs, along with 2814 blazars, 45,816 type-II objects, and 66,026 radio/X-ray associated quasar candidates with $p\rm_{quasar} = 99\%$ and is largely complete based on literature sources.
The cross-match resulted in 163 matches within a 3-arcsec radius, including 38 quasars (`Q' type), 99 type-I Seyferts/host-dominated AGNs (`A' type), 22 type-II narrow-line core-dominated (`N' type), 3 type-II Seyferts/host-dominated (`K' type), and 1 BL Lac-type object (`B' type). That is, more than half of the ULIRGs exhibit an active core at their centers.
On the face of it, the fraction of AGNs among them is approximately twice as high as that indicated by the $f25/f60$ classification.
The possible reason is that although the central black holes of these ULIRGs have been ignited, part of these ULIRGs perhaps do not produce enough energy in the nucleus to dominate the infrared emission of the whole galaxy. They have not yet transitioned from the ``cool'' to the ``warm'' systems in the $f25/f60$ diagnosis.

In the mid-infrared (MIR) waveband, spectroscopy presents a diagnostic avenue for distinguishing between AGN and starburst activities. This is due to the strong emission features at 6.2, 7.7, 8.6, and \mbox{11.3 $\mu$m} that are characteristic of starburst galaxies, which are attributed to aromatic carbonaceous material. In contrast, AGNs exhibit a strong continuum with low-contrast emission features in the nuclear region of the galaxy.
Genzel et al. (1998) and Lutz et al. (1998) proposed employing the line-to-continuum ($L/C$) ratio of the 7.7 $\mu$m band to distinguish between starburst- and AGN-dominated ULIRGs. 
Tran et al. (2001) suggested applying spectral decomposition to quantitatively characterize the relative contribution of star formation and AGN activity to the MIR emission of ULIRGs. 
This involved analyzing the spectra by overlaying a starburst and an AGN spectrum, both of which could be obscured to varying degrees.
Their studies revealed that approximately half of high-luminosity ULIRGs are dominated by starburst activity, while AGNs primarily influence the other half. 
This ratio aligns with the findings of our sample, as both sets exhibit a similar range in infrared luminosity, estimated to be around $\rm \sim 10^{12}-10^{13.1}~L_{\sun}$ according to Tran et al. (2001). 
Moreover, starbursts are the primary contributors to bolometric luminosity at the lower end of the spectrum, while AGNs take precedence at the higher end.

In general, the fraction of systems containing AGNs seems to escalate with infrared luminosity (e.g., Klaas 1989; Klass \& Elsasser 1993; Rowan-Robinson et al. 1991; Kim et al. 1998; Veilleux et al. 1999a,b). 
Our sample also indicates that the fraction of AGNs increases with infrared luminosity. 
The transition between star-dominated and AGN-dominated occurs at $\rm \sim 10^{12.4}~L_{\sun}$, with AGN fractions ranging from approximately $43\%-56\%$ when the infrared luminosity is below this threshold, jumping to around $\sim 76\%-92\%$ afterward.
Then, we constructed a [3.4]-[4.6]-[12] $\mu$m color-color diagram using magnitudes from the WISE Catalog for ULIRGs in the bottom-right panel of Figure \ref{fig:f10}. 
Contour data are adopted from Massaro et al. (2011), corresponding to WISE thermal sources detected in a 56 deg$^2$ region at high Galactic latitude. 
ULIRG sources are classified into three subclasses: galaxies (red), Type I (blue), and Type II (green). 
The former group includes 120 sources not in the Milliquas catalog. In contrast, the latter two groups, totaling 163 sources, are further divided based on the presence or absence of broad emission lines, as per Paliya et al. (2024) and Wu \& Shen (2022). Additional details can be found in Section 3.4.
Type I sources occupy overlapping regions predominantly populated by luminous quasars and Seyfert galaxies, while galaxy and Type II sources overlap regions dominated by ULIRGs and Liners. A small fraction of sources extends into regions occupied by starburst galaxies and obscured AGNs.

\subsection{Infrared to Radio Correlation}

\figurenum{11}
\begin{figure*}[tp]
\centering
\begin{minipage}[t]{0.48\textwidth}
\centering
\includegraphics[width=\textwidth]{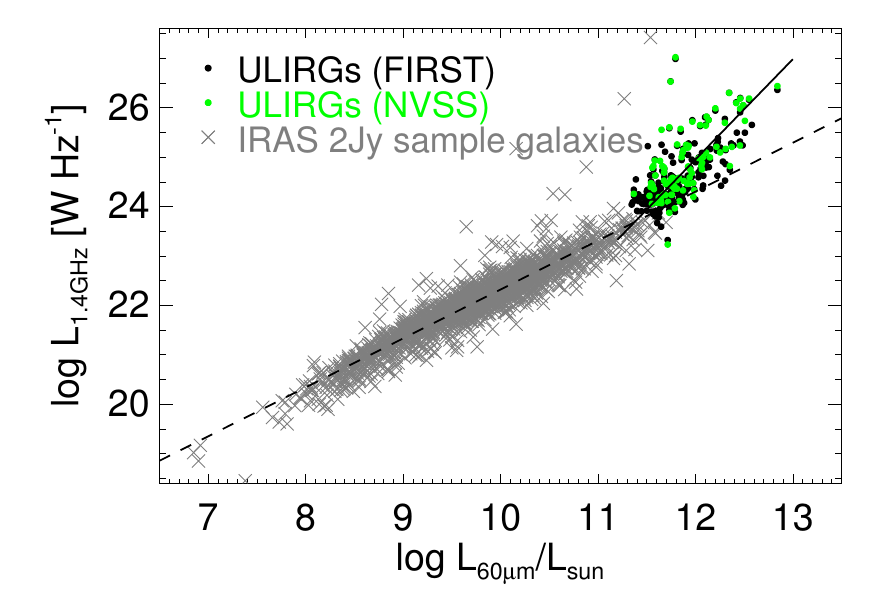}
\end{minipage}
\begin{minipage}[t]{0.48\textwidth}
\centering
\includegraphics[width=\textwidth]{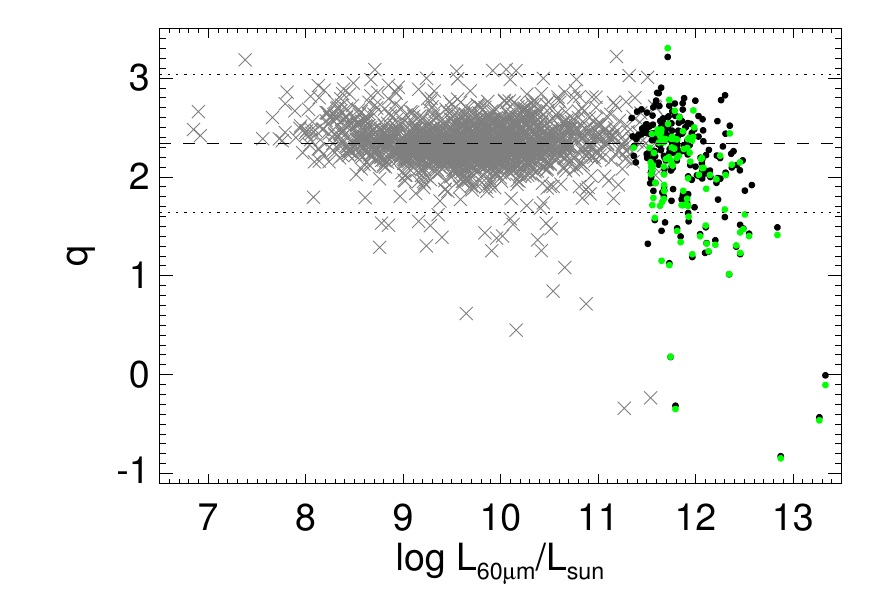}
\end{minipage}
\caption{Distribution of the 1.4 GHz radio luminosity (left panel) and q-value (right panel) of the ULIRGs plotted as a function of the 60 $\mu$m FIR luminosity. The sources in our ULIRG sample with detections of FIRST and NVSS are represented by black and green filled circles, and the IRAS 2 Jy sample galaxies are overplotted by gray crosses for contrast. In the left panel, the dashed line means the best-fitting line for an all-sky sample of infrared-detected galaxies from IRAS (Yun et al. 2001). In the right panel, the dashed line is at $ q = 2.34 $ which is the mean value obtained by Yun et al. (2001), the above and below dotted lines are limits for five times FIR excess and radio excess from the mean respectively. \label{fig:f11}}
\end{figure*}

It has been suggested that AGN activities can be distinguished from starbursts by using the FIR/radio correlation, a possible connection between the host galaxy and nuclear activity was involved (e.g., Condon et al. 1982; Rush, Malkan, \& Edelson 1996). 
Objects displaying a radio excess are primarily radio-loud AGNs, as suggested by Roy \& Norris 1997, possibly involving complex mechanisms such as jet emission. 
Objects displaying a radio excess are primarily radio-loud AGNs, as suggested by Roy \& Norris 1997, possibly involving complex mechanisms such as jet emission. 
Dust heating by low-mass stars (cirrus emission) might contribute solely to the observed FIR emission (e.g., Helou 1986; Lonsdale Persson \& Helou 1987; Fitt et al. 1988). Observational evidence also indicates that the well-known correlation between FIR and radio luminosity is observed only in star-forming ULIRGs (Yun et al. 2001).

To obtain the radio properties of our ULIRGs, we cross-identified them with the latest catalogs of the Faint Images of Radio Sky at Twenty centimeters (FIRST)\footnote{http://sundog.stsci.edu/first/catalogs/readme\_14dec17.html} (Becker et al. 1995) and the NRAO VLA Sky Survey (NVSS)\footnote{https://www.cv.nrao.edu/nvss/NVSSlist.shtml} (Condon et al. 1998) using a matching radius of 3.0 arcsec. In the preliminary sample, a total of 241 and 103 ULIRGs had counterparts in FIRST and NVSS, respectively, with flux peak densities at 20 cm (1.4 GHz) and high signal-to-noise ratios. In the end, 245 ULIRGs exhibited effective radio detections, with 98 of them observed in both surveys.

In Figure \ref{fig:f11}, we present the 1.4 GHz radio luminosity ($L_{\rm 1.4 GHz}$) and ``$q$'' parameter against the 60 $\mu$m FIR luminosity ($L_{\rm 60 \mu m}$) of our ULIRG sample.
The 60 $\mu$m and 1.4 GHz luminosities and $q$ parameter are calculated using the formulas (Condon et al. 1991; Yun et al. 2001):
\begin{equation}
{\rm log}~ L_{\rm 60 \mu m} ~ (L_{\sun}) = 6.014 + 2 ~{\rm log}~D_{\rm L} + {\rm log}~ f_{60\mu m},
\end{equation}
\begin{equation}
{\rm log}~ L_{\rm 1.4 GHz} ~ ({\rm W Hz^{-1}}) = 20.08 + 2 ~{\rm log}~D_{\rm L} + {\rm log}~ f_{\rm 1.4GHz},
\end{equation}
\begin{equation}
q\equiv {\rm log}~ (2.58~f_{\rm 60\mu m}+f_{100\mu m})/2.98 - {\rm log} ~f_{\rm 1.4GHz} ,
\label{eq:q}
\end{equation}
where $D_{\rm L}$ is the luminosity distance in Mpc, and $f_{\rm 60\mu m}$, $f_{\rm 100\mu m}$ and $f_{\rm 1.4GHz}$ are the observed flux densities at 60 $\mu$m, 100 $\mu$m and 1.4 GHz in units of Jy.

In the left panel, we present the linear relation of the 60 $\mu$m and 1.4 GHz luminosities for ULIRGs with a solid line. A formal fit to the observed radio-FIR luminosity correlation yields: 
\begin{equation}
{\rm log}~L_{\rm 1.4GHz}=(2.03\pm0.11)~{\rm log} ~L_{60\mu m}+(0.59\pm1.13).
\label{eq:radio_FIR}
\end{equation}
When combined with the IRAS 2 Jy galaxy sample to view the FIR and radio luminosity distribution comprehensively, as reported by Yun et al. (2001), most sources in star-forming ULIRGs follow the radio-FIR luminosity correlation with a unity slope (dashed line). However, a steepening relation exists at low luminosities ($L\rm_{60\mu m} \leq 10^9~L_{\sun}$), and a higher dispersion for luminosities above $\rm 10^{10.5}~L_{\sun}$ (also see Fig. 5a in Yun et al. (2001)). In other words, two notable features emerge: a downward trend at the low-luminosity end and an upturn at the high-luminosity end.

The deviations at low/high luminosities are more intuitively presented in the distribution of  $q$ values (see the right panel of Figure \ref{fig:f11}). $q$ is a measure of the difference between the logarithmic FIR/radio flux density (Equation \ref{eq:q}). 
It should be noted that the value of $q$ is overestimated for sources with flux upper limits at 100 $\mu m$. 
The dashed line represents $q = 2.34$, the mean value obtained by Yun et al. (2001), while the above and below dotted lines are limits for five times FIR excess and radio excess from the mean, respectively. 
Deviations from the linear relation can occur if the FIR or radio luminosity is not directly proportional to star-forming activity. 
It is essential to acknowledge a potential bias toward sources with higher FIR/radio flux density ratios in FIR-selected samples, with the selection effect being particularly severe at low luminosities (e.g., Condon \& Broderick 1986; Yun et al. 2001).

Furthermore, the IRAS 2 Jy sample spans over five decades in luminosity, with the majority of galaxies having $L_{\rm 60\mu m}$ between $10^8$ and $10^{11.5}~L_{\sun}$. In contrast, our ULIRGs are collected at the highest luminosity ($L\rm_{60\mu m} \gtrsim 10^{11.4}~L_{\sun}$). 
Upon categorizing the ULIRGs based on AGN activity into three subclasses and analyzing their distributions within the radio-FIR diagrams, it becomes apparent that AGN-free ULIRGs (represented by red filled circles) adhere to the Yun et al. radio-FIR relation. However, both Type I and Type II ULIRGs (represented by blue and green filled circles) exhibit a significant deviation from this pattern. The steeper radio-FIR relation (Equation \ref{eq:radio_FIR}) is predominantly attributed to AGNs. 
This result is consistent with that of Hou et al. (2009). Therefore, the radio-FIR relation originates from starburst-related non-thermal radiation, and the radio excess objects are due to AGN-related radio emission (Roy \& Norris 1997). 

\subsection{Type I AGNs in ULIRGs}

\figurenum{12}
\begin{figure*}[tp]
\centering
\begin{minipage}[t]{0.48\textwidth}
\centering
\includegraphics[width=\textwidth]{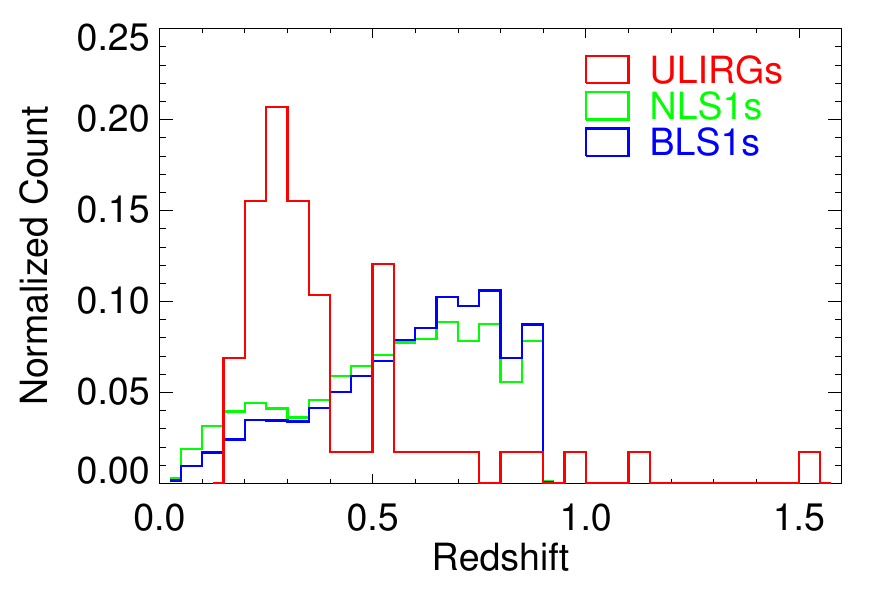}
\end{minipage}
\begin{minipage}[t]{0.48\textwidth}
\centering
\includegraphics[width=\textwidth]{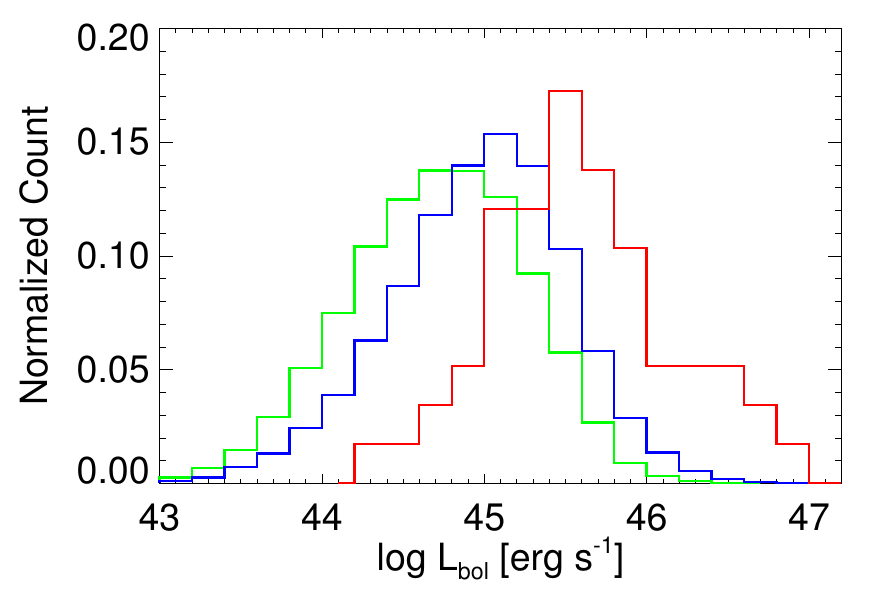}
\end{minipage}
\begin{minipage}[t]{0.48\textwidth}
\centering
\includegraphics[width=\textwidth]{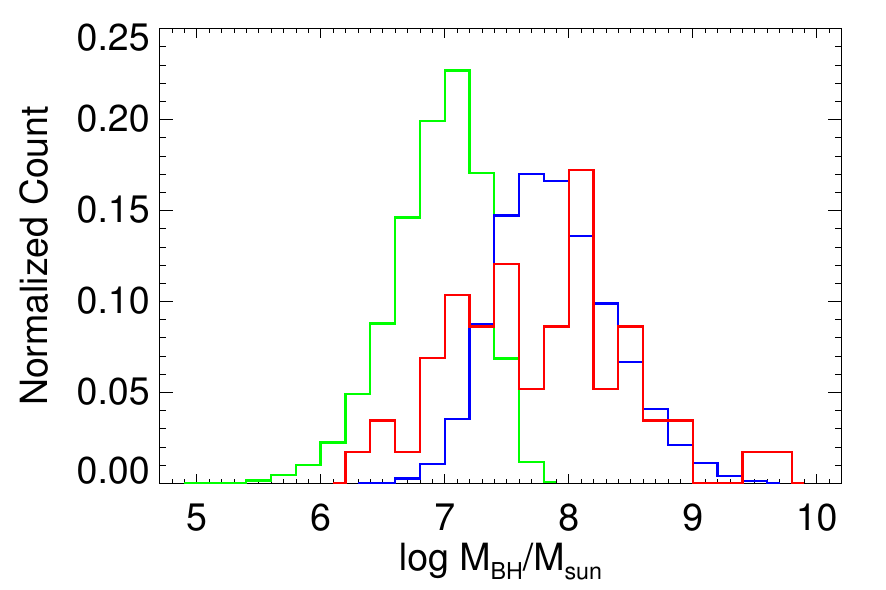}
\end{minipage}
\begin{minipage}[t]{0.48\textwidth}
\centering
\includegraphics[width=\textwidth]{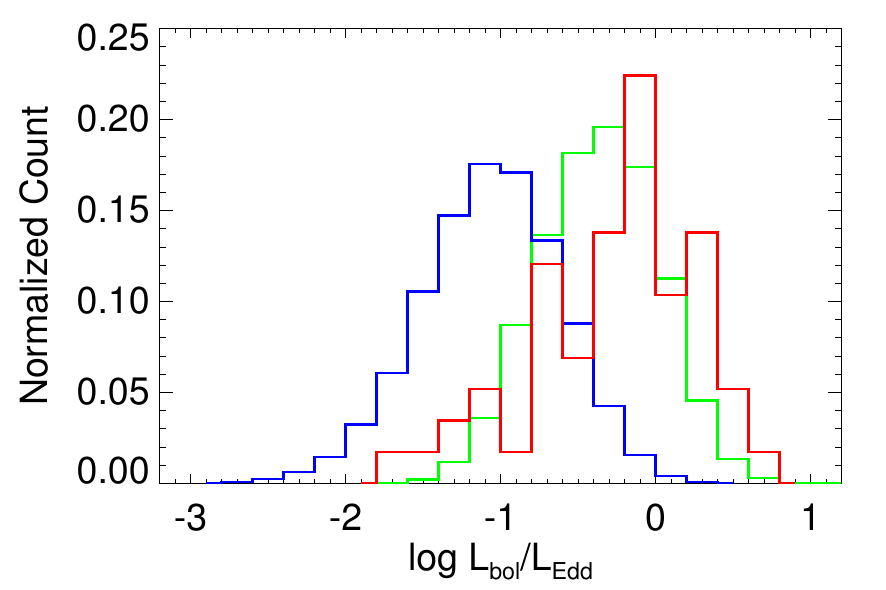}
\end{minipage}
\begin{minipage}[t]{0.48\textwidth}
\centering
\includegraphics[width=\textwidth]{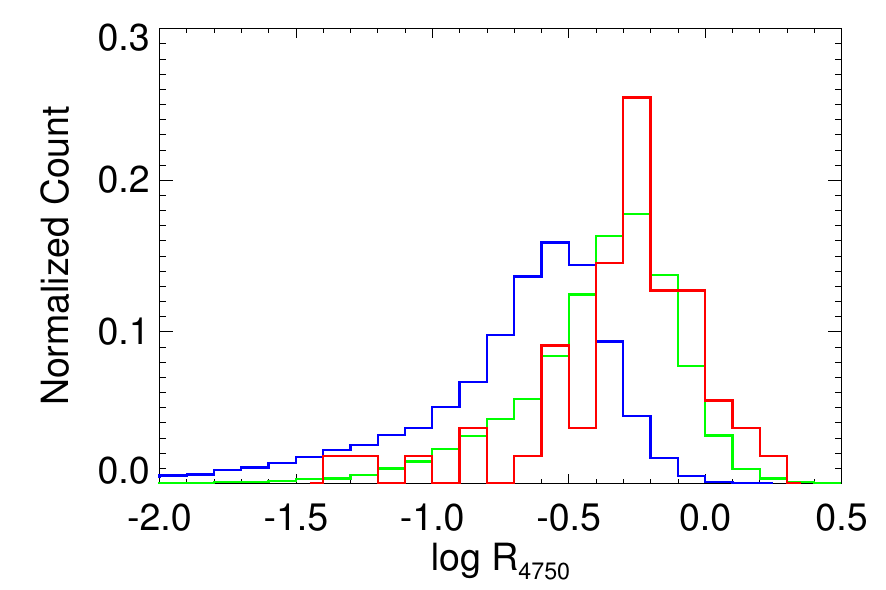}
\end{minipage}
\begin{minipage}[t]{0.48\textwidth}
\centering
\includegraphics[width=\textwidth]{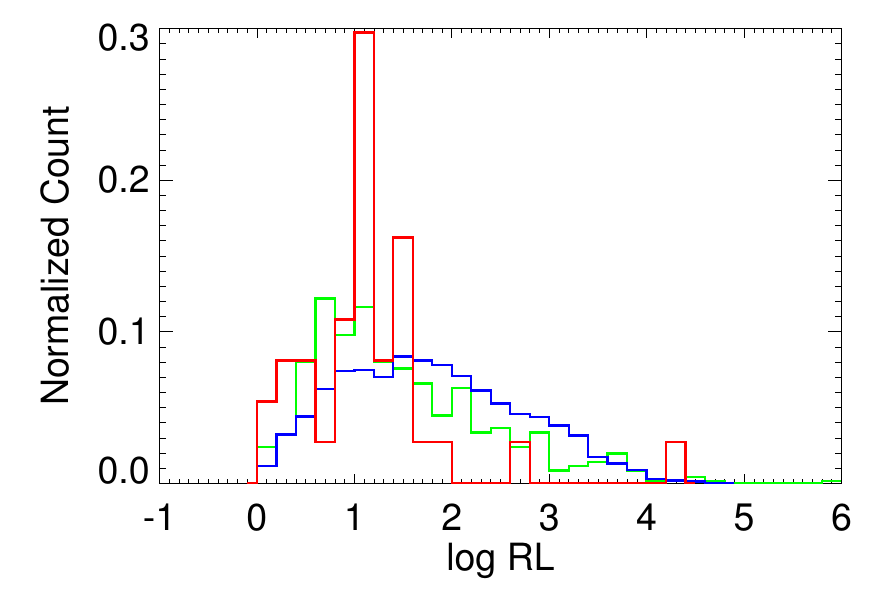}
\end{minipage}
\caption{Comparisons of various optical spectroscopic parameters measured/derived for Type I ULIRGs (red), NLS1 (green), and BLS1 (blue) sources. All measurements are obtained from Paliya et al. (2024).
\label{fig:f12}}
\end{figure*}
 
As mentioned in the third paragraph of Section 3.2, a cross-match with the Milliquas catalog identified 163 sources in the ULIRG sample displaying significant AGN activity at their nuclei. The AGN parameter properties are extracted from the following two articles.

Paliya et al. (2024) meticulously conducted spectral decomposition on over two million optical spectra of extragalactic sources at low redshifts ($z \le 0.8$ for SDSS-I/II, and $z \le 0.9$ for SDSS-III/IV) from the SDSS Data Release 17 (DR17). Their exhaustive work yielded a catalog containing 22,656 narrow-line Seyfert 1 (NLS1) galaxies and 52,273 broad-line Seyfert 1 (BLS1) galaxies. The optical spectra underwent meticulous replication to account for multi-component contributions from both AGN and host galaxies, involving simultaneous fitting of both emission line and continuum features.

Wu \& Shen (2022) presented a catalog of continuum and emission-line properties for 750,414 broad-line quasars included in the SDSS DR16 quasar catalog (DR16Q; Lyke et al. 2020), measured from optical spectroscopy. These quasars cover broad ranges in redshift ($0.1 \lesssim z \lesssim 6$) and luminosity ($44 \lesssim {\rm log~} L{\rm_{ bol} [erg~ s^{-1}]} \lesssim 48$). In contrast to Paliya et al. (2024), Wu \& Shen (2022) focused on high redshift, luminous quasars. The quasars' spectra were fitted with a global continuum+emission lines model using the public code PyQSOFIT (e.g., Guo et al. 2018), and the model does not include host galaxy emission in the spectral fits. Thus, the measurement reliability for the sources at low redshift with strong host galaxy starlight is yet to be verified.

Guided by redshift information, we selected the spectral continuum and emission-line properties of the three sources with redshifts exceeding 0.9 (namely F08105+2554, F10026+4949, and F13080+3237) from the catalog of the DR16Q properties. 
The measurements for the remaining 160 low-redshift sources were obtained from Paliya et al. (2024). 
Among these, the optical spectra of 55 sources showcase broad H$\alpha$ and/or H$\beta$  emission lines. As a result, our sample encompasses 58 Type I and 105 Type 2 AGNs within a total of 283 ULIRGs. 
In a prior investigation, Hou et al. (2009) identified 308 ULIRGs, with 62 sources classified as Type I. 
The percentage of Type I ULIRGs in their study was approximately 20\%, a figure consistent with our findings.

Figure \ref{fig:f12} presents various parameters and physical properties of Type I ULIRGs and Type I AGNs.  Leveraging measurements conducted by Paliya et al. (2024), we have redrawn the parameter distribution for Type I AGNs, specifically NLS1s and BLS1s,  to enable more effective comparative analysis.
In the top-left panel, the redshift distribution indicates an increase in the number of Type I AGNs with redshift, whereas Type I ULIRGs are more frequently observed within the redshift range of $z < 0.55$, peaking at $z=0.28$.

Furthermore, the distribution of black hole masses in Type I ULIRGs is similar to that in BLS1s, with the average value surpassing that of NLS1s by 0.83 dex.
Type I ULIRGs exhibit significantly larger bolometric luminosities compared to Type I AGNs. The average bolometric luminosities  for NLS1, BLS1s, and Type I ULIRGs gradually increases from $5.1\times10^{44}$ \mbox{erg s$^{-1}$} to $9.3\times10^{44}$ \mbox{erg s$^{-1}$}, and then to $4.4\times10^{45}$ \mbox{erg s$^{-1}$}. Consequently, Type I ULIRGs have exceptionally high accretion rates, with the mean value  ($\overline{L_{\rm bol}/L_{\rm Edd}} = 0.92$) being approximately 1.5 times higher than that of NLS1s  ($\overline{L_{\rm bol}/L_{\rm Edd}} = 0.63$).
When compared with previous studies, the black hole mass distributions in Type I ULIRGs from both Hou et al.'s and our study show largely identical profiles, with mean values of $\rm 7.6\times 10^7 ~M_{\sun}$ and $\rm5.8 \times 10^7~M_{\sun}$, respectively. These values are systematically smaller than those of DR16Q quasars (with a mean value of $\rm 8.3 \times 10^8~ M_{\sun}$). However, the average accretion rate for Hou et al.'s Type I ULIRGs is relatively low, comparable only to that of NLS1s but significantly larger than that of BLS1s and DR16Q quasars (with mean Eddington ratios of about 0.16 and 0.23, respectively).
In addition, the relative strength of $R_{4570}$ $\rm (= FeII \lambda \lambda 4434-4684 /H\beta)$ in NLS1s is generally about twice that in normal AGNs on average (e.g., Zhou et al. 2006; Rakshit et al. 2017; Paliya et al. 2024).   We found the median $R_{4570}$ parameter value of Type I ULIRGs to be even a quarter higher than NLS1s.
These findings align with the prevalent evolutionary model suggesting that ULIRGs are in a pre-QSO phase, and their central black holes are still in a growing phase.

Furthermore, the bottom-right panel of Figure \ref{fig:f12} illustrates a comparison of the radio-loudness ($RL$) parameters. The overall distributions reveal similarities between BLS1s and NLS1s, with median average $RL$ values of approximately 51 and 17, respectively. In contrast, Type 1 ULIRGs demonstrate relatively low $RL$ parameters, depicting a narrow distribution peaking at about 13. Remarkably, all sources are below 100, with only two exceptions.
The most noteworthy case is F16413+3954, distinguished by an exceptionally high $RL$, just below that of merely 10 sources in NLS1 and BLS1 catalogs. Its optical counterpart, SDSS J164258.80+394837.0, also known as 3C 345, has been detected in the Fermi-Large Area Telescope's all-sky gamma-ray survey (Abdo et al. 2009; Paliya et al. 2018). Recognized as a flat spectrum radio quasar in the constellation of Hercules, it is renowned for hosting a superluminal jet and displaying variability across almost all wave bands.
The second-highest radio-loudness source is F09220+2759, associated with the optical counterpart SDSS J092501.78+274607.9. This extremely red quasar is classified as a radio-detected Type 1 quasar (Ross et al. 2015). Notably, it exhibits weak broad H$\beta$ and anomalous [\ion{O}{3}] profiles in the optical spectrum (Boroson \& Lauer 2010).

\figurenum{13}
\begin{figure}[tp]
\centering
\includegraphics[width=0.48\textwidth]{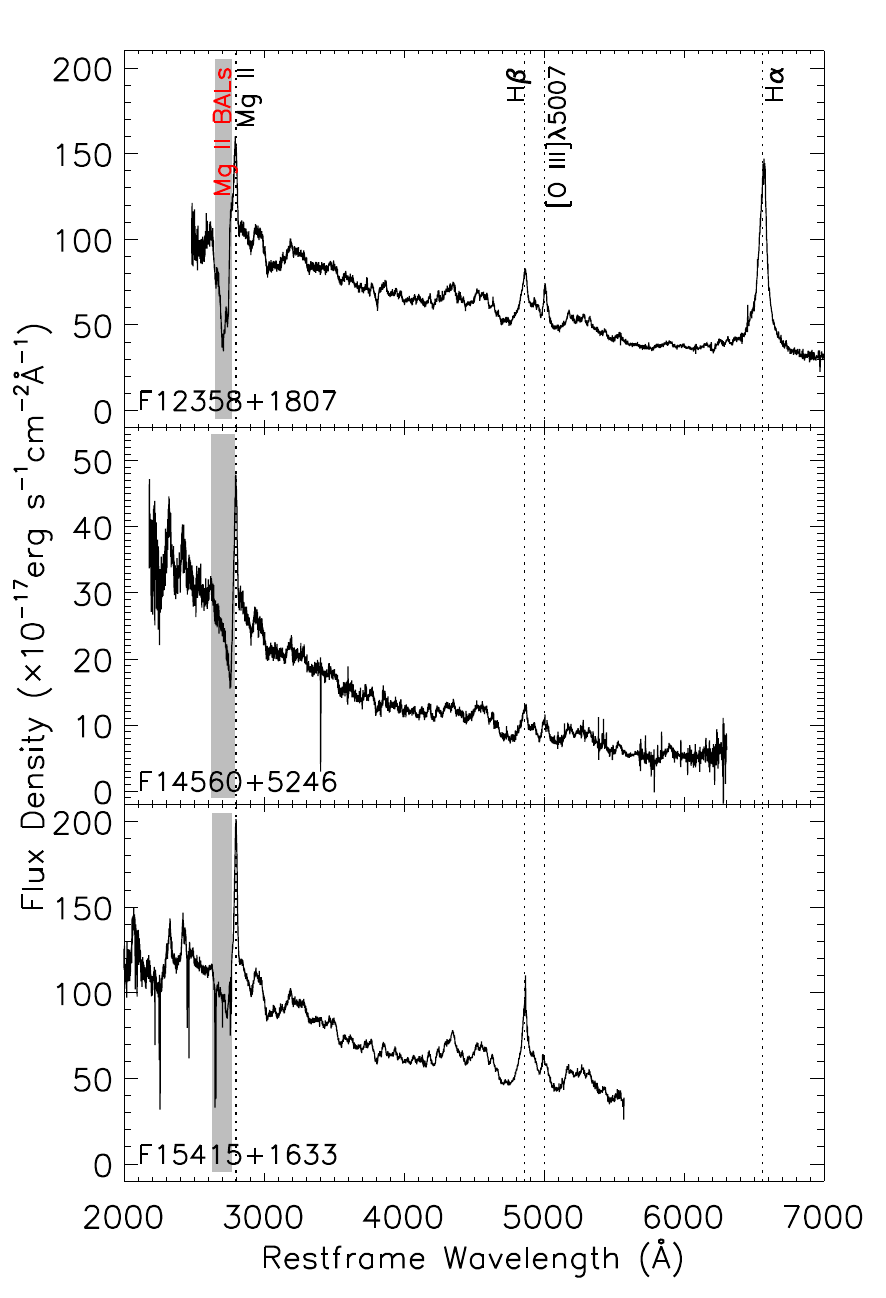}
\caption{SDSS spectra in restframe of 3 ULIRGs with \ion{Mg}{2} BAL features. The main emission lines are represented by dashed lines, and the gray shaded regions illustrate the \ion{Mg}{2} BAL troughs. 
\label{fig:f13}}
\end{figure}

Broad Absorption Lines (BALs) are frequently observed in quasars, characterized by specific spectral features such as a narrower H$\beta$ emission line, weaker [\ion{O}{3}], [\ion{Ne}{3}], and [\ion{Ne}{5}] emission lines, along with stronger optical \ion{Fe}{2} multiplets (Boroson 2002; Trump et al. 2006; Ganguly et al. 2007; Gibson et al. 2009; Zhang et al. 2010). 
The BAL fraction is significantly higher in quasars with elevated Eddington ratios and luminosity compared to those with lower values.
In Figure \ref{fig:f12}, we observe that Type I ULIRGs exhibit similar spectral lines and parameter characteristics, suggesting a potential connection between AGN outflows and Type I ULIRGs. Subsequently, we examined the SDSS spectra of the optical counterparts of Type I ULIRGs and identified three sources displaying \ion{Mg}{2} BAL features in their spectra (Figure \ref{fig:f13}).

Among these Type I ULIRGs, 18 sources have redshifts of $z \geqslant 0.4$, with this redshift cutoff chosen to ensure that \ion{Mg}{2} falls within the wavelength coverage of the SDSS spectrograph. Consequently, the fraction of \ion{Mg}{2} BALs in Type I ULIRGs is approximately 16.7\%. Typically, previous studies report a fraction of about 1\% for \ion{Mg}{2} BAL quasars (e.g., Turnshek et al. 1997; Trump et al. 2006; Zhang et al. 2010). Considering the bias of the optically selected sample, the fraction of \ion{Mg}{2} BAL quasars ranges from 2\% to 7\% in quasars (see more details in Section 3.3 of Zhang et al. 2010).
This high anomaly in the fraction of \ion{Mg}{2} BALs within Type I ULIRGs aligns with the evolutionary scenario of BAL quasars,  indicating an early stage of quasar evolution with a gas/dust richer nuclear environment (Voit et al. 1993; Hamann et al. 1993; Egami et al. 1996).

\section{Conclusion}

We present a novel dataset of ULIRGs identified through the cross-matching of IRAS sources in the FSC92 with the spectroscopic sample of the SDSS DR16. The combination of the reliability and extensive sky coverage of both surveys provides a uniformly sampled collection suitable for investigating the distinctive properties of ULIRGs, such as color, morphology, and AGN activity, using archived photometric and spectroscopic data.

A total of 283 ULIRGs have been identified, with 102 newly discovered within the SDSS DR16 coverage area. Notably, 120 false ULIRGs previously reported by Hwang et al. (2007) and Hou et al. (2009) have been excluded. In prior research, the application of the likelihood ratio method led to the misidentification of SDSS counterparts for IRAS sources and the acquisition of incorrect redshifts. Moreover, all infrared emissions from IRAS detections were erroneously attributed to their likelihood-selected optical counterpart, resulting in a significant overestimation of the infrared luminosities of these false sources and their erroneous classification as ULIRGs.
To address this, we employed cross-matching with the AllWISE source catalog and validated the reliability of IRAS-SDSS associations based on positional uncertainty ellipses. We also evaluated the infrared contribution from the SDSS counterpart for each IRAS source, yielding a high-confidence ULIRG sample.

A detailed examination of the SDSS $gri$ composite images revealed that 40\% (114 out of 283) of ULIRGs exhibit apparent interaction features. Approximately half of them display a single nucleus with tail features, while the rest show two nuclei with well-defined tidal tails or bridges, or closely situated nuclei, possibly overlapping. 
Notably, the higher redshifts of ULIRGs in this study, compared to previous research, have led to a slightly lower fraction of sources with interaction features, which decreases with increasing redshift  (92\% for $z<0.1$ and 11\% for $z \ge 0.4$), likely due to selection effects.	
We expect that the forthcoming CSST will provide high-resolution, high signal-to-noise ratio, and deep images, enabling a more detailed understanding of ULIRG morphology and interactions.

Our investigation into the infrared colors of these ULIRGs revealed that the fraction of AGNs, as defined by the $f_{25}/f_{60}$ classification, is 26\% (79/283). However, when matched with Milliquas, this fraction increases to 58\% (163/283), more than doubling the initial value. Interestingly, the fraction of AGNs appears to correlate with infrared luminosity.
Within the radio band, both Type I and Type II ULIRGs demonstrate an overabundance of radio emission attributable to AGN activity, resulting in a steeper radio-FIR relationship. 
Analysis of the [3.4]-[4.6]-[12] $\mu$m color-color diagram revealed that Type I ULIRGs predominantly occupy regions associated with luminous quasars and Seyfert galaxies. Conversely, regions dominated by ULIRGs and Liners overlap with those occupied by Type II ULIRGs and those lacking AGN activity. These intriguing patterns imply that while central black holes in certain ULIRGs have been ignited, they may not produce sufficient energy within the nucleus to dominate the entire galaxy's infrared emission.

Moreover, spectral decomposition conducted by Paliya et al. (2024) and Wu \& Shen (2022) identified 58 ULIRGs with significant H$\alpha$ and/or H$\beta$ broad emission lines. Approximately 20\% of these ULIRGs are thus classified as Type I. 
Comparative analysis revealed various parameter properties between Type I ULIRGs and Type I AGNs. Type I ULIRGs share similarities with NLS1s, particularly in terms of higher accretion rate $L_{\rm bol}/L_{\rm Edd}$ and larger relative strength of $R_{4570}$. However, their black hole masses align more closely with those of BLS1s and are still systematically smaller than those of DR16Q quasars, while their bolometric luminosities are even larger than those of NLS1s and BLS1s. Moreover, the fraction of \ion{Mg}{2} BALs in Type I ULIRGs is approximately 16.7\%,  over ten times that observed in quasars. These results are consistent with the current evolutionary model, indicating that ULIRGs are still in a pre-QSO phase and their central black holes are still growing.

ULIRGs have been proposed as the future quasar activity sites once the AGN, which is obscured by the massive gas and dust driven to the galactic center by the merger, has emerged. Additionally, they are viewed as local analogs of the SMGs discovered at large redshifts. Submillimeter observations, particularly sensitive to star-forming regions, dust distribution, molecular gas, and AGNs, offer superior resolution and more detailed insights. This distinctive class of galaxies is valuable and well-suited for submillimeter observations in the study of galaxy evolution.
Shanghai Normal University (ShNU) is currently collaborating with the California Institute of Technology (Caltech) and the University of Concepcion (UdeC) to promote the Leighton Chajnantor Telescope (LCT) project. This project involves relocating the 10.4-meter Caltech submillimeter telescope (CSO) from Mauna Kea in Hawaii to the Atacama Desert in Chile for terminal upgrades, with operations slated to commence in 2026. We believe that multiband and spectral submillimeter observations with LCT on this uniform and high-confidence sample will help improve the current paradigm of galaxy formation and evolution.

\begin{acknowledgments}

This work was supported in part by the National Natural Science Foundation of China (Grant No. 12173026 and 12141302), the National Key Research and Development Program of China (Grant No. 2022YFC2807303), the Shanghai Science and Technology Fund (Grant No.  23010503900).
SHZ acknowledges support from the Program for Professor of Special Appointment (Eastern Scholar) at Shanghai Institutions of Higher Learning and the Shuguang Program  (23SG39) of the Shanghai Education Development Foundation and Shanghai Municipal Education Commission. 
ZJL acknowledges the support from the Shanghai Science and Technology Foundation (Grant No. 20070502400) and the
science research grants from the China Manned Space Project.
HBX acknowledges support from the National Natural Science Foundation of China (Grant No. 12203034) and the Shanghai Science and Technology Fund  (Grant No. 22YF1431500).
This publication uses data products from the Infra-Red Astronomical Satellite, the Sloan Digital Sky Survey, the Wide-field Infrared Survey Explorer, the VLA Faint Images of the Radio Sky at Twenty-Centimeters, and the NRAO VLA Sky Survey.
Funding for the Sloan Digital Sky Survey IV has been provided by the Alfred P. Sloan Foundation, the U.S. Department of Energy Office of Science, and the Participating Institutions. SDSS-IV acknowledges support and resources from the Center for High-Performance Computing at the University of Utah. The SDSS website is www.sdss.org.
The Wide-field Infrared Survey Explorer is a joint project of the University of California, Los Angeles, and the Jet Propulsion Laboratory/California Institute of Technology, funded by the National Aeronautics and Space Administration.
The VLA is operated by the National Radio Astronomy Observatory, which is a facility of the National Science Foundation under a cooperative agreement by Associated Universities, Inc.
\end{acknowledgments}

\clearpage

\startlongtable
\centerwidetable
\begin{deluxetable*}{cccc ccccc cc cc c}
\tabletypesize{\scriptsize}
\tablewidth{0pt}
\tablenum{1}
\tablecaption{Parameters of the ULIRG Sample
\label{tab:t1} }
\tablehead{\colhead{ Name}  &\colhead{ M-P-F}& \colhead{redshift} &\colhead{${\rm log}~ L\rm_{IR}$}&\colhead{$F\rm_{12\mu m}$} &\colhead{$F\rm_{25\mu m}$} & \colhead{$F\rm_{60\mu m}$} & \colhead{$F\rm_{100\mu m}$} & \colhead{Quality} & \colhead{$N\rm_{WISE}$} & \colhead{Flux Ratio} &\colhead{$F\rm_{FIRST}$} &\colhead{$F\rm_{NVSS}$} & \colhead{AGN} \\
  & & &$L_{\sun}$&  Jy&Jy &Jy &Jy & & & \%& mJy & mJy &   \\
(1)&(2)&(3)&(4)&(5)&(6)&(7)&(8)&(9)&(10)&(11)&(12)&(13)&(14)}
\startdata
F00242+3344 &56537-06530-0561& 0.1734&  12.14& 0.0760& 0.1142&  0.6008&  1.3310&1132&  5& 83& &    18.97& \\
F00344-0130 &52261-00690-0216& 0.2936&  12.38& 0.0821& 0.1722&  0.3238&  0.6836&1131&  6& 82&     0.82& &II\\
F01093-1002 &52199-00659-0597& 0.1315&  12.00& 0.1140& 0.1364&  0.9330&  1.4040&1132&  5& 67&     4.89& & \\
F01329+1439 &51882-00426-0333& 0.2178&  12.17& 0.1870& 0.1861&  0.3769&  0.8811&1131& 11& 64& & & \\
F01478+1254 &51820-00429-0011& 0.1467&  12.02& 0.1273& 0.1827&  0.4800&  1.9110&1131&  3& 91& & & \\
F01501+0433 &55536-04403-0167& 0.4930&  12.86& 0.1004& 0.1310&  0.2516&  0.7125&1131& 13& 52&     0.68& &II\\
F01572+0009 &51871-00403-0550& 0.1635&  12.52& 0.1232& 0.5416&  2.2240&  2.1640&2332&  3& 91&    22.91&    25.38&I\\
F02394-0031 &52199-00706-0101& 0.2186&  12.06& 0.0991& 0.1215&  0.3058&  0.6354&1132&  7& 62&     0.95& & \\
F02417-0043 &51871-00409-0315& 0.2002&  12.14& 0.1053& 0.1595&  0.4822&  0.8260&1132&  6& 67&     9.46&     9.65&II\\
F02486-0714 &51901-00457-0324& 0.3267&  12.32& 0.0719& 0.1422&  0.2443&  0.3989&1131&  7& 68&     6.02&     6.70&I\\
F03209-0806 &51924-00460-0093& 0.1664&  12.28& 0.1003& 0.1343&  1.0050&  1.6920&1132&  2& 97&     5.55& &II\\
F03319-0800 &54476-01630-0045& 0.2870&  12.53& 0.0732& 0.1379&  0.4645&  1.0700&1131&  2& 94& & &II\\
F07384+4515 &56323-06375-0554& 0.2071&  12.07& 0.0951& 0.1181&  0.3009&  0.8828&1131&  5& 64&     1.77& & \\
F07407+2720 &52592-01059-0220& 0.2527&  12.10& 0.0729& 0.2620&  0.2539&  0.4688&1131&  7& 84&     0.72& &I\\
F07464+2841 &52592-01059-0507& 0.3370&  12.45& 0.1099& 0.2025&  0.2854&  0.5497&1131&  5& 73&    27.95&    26.20&II\\
F07548+4227 &57391-08291-0871& 0.2112&  12.13& 0.1104& 0.2195&  0.3263&  0.9755&1131&  8& 80&     1.81& &I\\
F07592+3736 &52253-00758-0259& 0.2385&  12.14& 0.1008& 0.1287&  0.2845&  0.6625&1132&  6& 64&     3.36& &II\\
F08025+4523 &56243-06377-0949& 0.3040&  12.56& 0.0783& 0.1445&  0.3482&  1.2260&1131&  2& 78&     2.07& & \\
F08030+5243 &53297-01781-0309& 0.0833&  12.08& 0.0954& 0.1836&  2.9880&  4.3940&1232&  2& 91&    15.36&    14.50&II\\
F08072+1622 &53713-02267-0561& 0.1855&  12.15& 0.1041& 0.1604&  0.2647&  1.8220&1131&  5& 64&     2.14& & \\
F08079+2822 &55858-04457-0604& 0.3358&  12.43& 0.0679& 0.1705&  0.2841&  0.5190&1132&  8& 50&     4.50& & \\
F08090+0849 &54081-02570-0571& 0.2836&  12.12& 0.0910& 0.1451&  0.2009&  0.3853&1131&  4& 66&     0.82& & \\
F08105+2554 &52709-01266-0219& 1.5099&  14.03& 0.1054& 0.1453&  0.2843&  0.5380&1132& 10& 71& & &I\\
F08162+2716 &55888-04461-0564& 0.2612&  12.52& 0.1327& 0.1359&  0.6348&  1.0710&1132&  2& 79&     5.94&     5.99&II\\
F08180+5612 &53386-01783-0354& 0.1588&  12.08& 0.0838& 0.1618&  0.6525&  1.3350&1131&  3& 93&    11.69&    11.66&II\\
F08201+2801 &52932-01267-0383& 0.1678&  12.30& 0.0850& 0.1616&  1.1710&  1.4290&1132&  4& 88&     4.74& & \\
F08209+2458 &52962-01585-0627& 0.2342&  12.20& 0.1175& 0.1713&  0.2795&  0.9380&1131&  9& 64&     0.94& & \\
F08219+1549 &53713-02272-0548& 0.2201&  12.15& 0.1293& 0.1780&  0.3602&  0.7955&1131&  2& 91&     3.11& &II\\
F08220+3842 &52615-00894-0135& 0.2058&  12.09& 0.0737& 0.1133&  0.3988&  0.7057&1132&  4& 70&     2.36& & \\
F08238+0752 &53084-01758-0187& 0.3114&  12.41& 0.1515& 0.1803&  0.2445&  0.7888&1131&  8& 68& & &I\\
F08252+4632 &51986-00548-0541& 0.2807&  12.35& 0.0672& 0.1385&  0.2481&  0.9269&1131&  3& 65&     1.29& & \\
F08266+3855 &52312-00827-0408& 0.1957&  12.09& 0.0570& 0.1325&  0.4636&  0.7513&1132&  4& 64&     2.51&     2.56& \\
F08274+1930 &53709-02275-0389& 0.1862&  12.11& 0.0722& 0.1259&  0.5817&  0.7599&1132&  2& 86& & & \\
F08280+5429 &55925-05156-0656& 0.2924&  12.27& 0.1067& 0.0839&  0.2516&  0.5526&1131&  5& 66&     5.37& &II\\
F08287+5254 &55212-03695-0976& 0.1694&  12.02& 0.0709& 0.0767&  0.6005&  0.7565&1132&  4& 57&     2.89& & \\
F08297+4728 &51981-00549-0409& 0.2996&  12.38& 0.1010& 0.1309&  0.2742&  0.7420&1131&  5& 76&     1.68& & \\
F08313+4855 &51873-00443-0019& 0.1745&  12.21& 0.0877& 0.0931&  0.3347&  2.4260&1131&  4& 69&     7.88& & \\
F08322+3609 &52668-01197-0613& 0.2012&  12.03& 0.0647& 0.1358&  0.3614&  0.6634&1131&  4& 76&     1.28& & \\
F08344+5105 &51873-00445-0488& 0.0969&  12.05& 0.0712& 0.1330&  2.1400&  2.6940&1232&  1&100&     7.53& & \\
F08407+2630 &53350-01932-0293& 0.2576&  12.10& 0.0938& 0.1119&  0.2015&  0.5487&1131& 10& 50& & &I\\
F08435+3141 &52991-01270-0423& 0.3867&  12.46& 0.0855& 0.0927&  0.2021&  0.4346&1131& 11& 54& & &I\\
F08449+2332 &53358-01931-0015& 0.1519&  12.09& 0.1289& 0.1539&  0.8674&  1.1950&1132&  2& 94&     3.24&     5.16& \\
F08449+2332 &53379-02085-0279& 0.1517&  12.09& 0.1289& 0.1539&  0.8674&  1.1950&1132&  2& 94&     3.24&     5.16& \\
F08504+2538 &53381-01933-0182& 0.2560&  12.38& 0.0829& 0.1053&  0.4872&  0.8258&1132&  5& 51&    10.46& &II\\
F08507+3636 &52643-00935-0179& 0.2604&  12.22& 0.0728& 0.1252&  0.2497&  0.7246&1131&  7& 51&    16.16&    32.45&II\\
F08519+2017 &53729-02283-0329& 0.3060&  12.78& 0.2838& 0.4335&  0.8911&  1.1580&3332&  2& 98&  1182.12&  1480.56&II\\
F08542+1920 &53711-02281-0618& 0.3313&  12.46& 0.1143& 0.1235&  0.3316&  0.5156&1132&  7& 77&     4.43& &I\\
F08542+5132 &55946-05155-0182& 0.3660&  12.84& 0.0968& 0.0618&  0.5983&  1.0920&1132&  2& 87&     7.11&     6.66&II\\
F08559+1053 &54085-02575-0125& 0.1480&  12.22& 0.1047& 0.1921&  1.1190&  1.9520&1232&  1&100&    12.72&    11.83&II\\
F08572+3915 &52703-01199-0041& 0.0582&  12.05& 0.3178& 1.7030&  7.4330&  4.5880&3332&  1&100&     4.89&     3.98& \\
F08589+5733 &56329-05150-0971& 0.2848&  12.51& 0.0718& 0.0828&  0.4911&  0.9441&1132&  5& 65&     7.28&     7.07&II\\
F08591+4341 &52294-00831-0377& 0.7029&  13.08& 0.0936& 0.1000&  0.2006&  0.4356&1131&  5& 67&     3.87& &I\\
F08591+5248 &51999-00553-0331& 0.1573&  12.21& 0.0999& 0.1617&  1.0100&  1.5260&1132&  2& 91&     5.96& & \\
F09005+0223 &51924-00471-0322& 0.3291&  12.56& 0.0672& 0.1037&  0.3626&  0.8260&1131&  4& 70&    22.50&    23.58&II\\
F09029+2430 &53401-02086-0149& 0.2316&  12.15& 0.0519& 0.0933&  0.3613&  0.5896&1131&  3& 71&     3.48&     3.37& \\
F09039+0503 &52649-01192-0115& 0.1250&  12.14& 0.1681& 0.2077&  1.4840&  2.0630&1132&  3& 86&     4.92&     5.76& \\
F09105+4108 &52668-01200-0073& 0.4408&  12.87& 0.1297& 0.3335&  0.5254&  0.4379&2331&  1&100&     8.30&    14.44&II\\
F09116+0334 &52252-00567-0200& 0.1453&  12.18& 0.0859& 0.1406&  1.0920&  1.8220&1132&  4& 83&     9.41&     9.87&II\\
F09117+3111 &56353-05809-0016& 0.3064&  12.23& 0.0903& 0.2109&  0.2764&  0.2509&1131&  4& 79&     5.45&     6.24&II\\
F09198+0323 &52254-00568-0267& 0.1741&  12.05& 0.1317& 0.1074&  0.2405&  1.7090&1131&  9& 51& & & \\
F09220+2759 &58130-09632-0133& 0.5316&  12.93& 0.0761& 0.2147&  0.3006&  0.5403&1131&  6& 75&    26.74&    26.00&I\\
F09246+0115 &51965-00475-0296& 0.1687&  12.07& 0.1024& 0.1401&  0.5049&  1.2880&1132&  4& 71&     2.66& & \\
F09320+6134 &51910-00486-0231& 0.0394&  12.02& 0.2499& 1.0340& 11.5400& 20.2300&3332&  1&100&   146.74&   144.78&II\\
F09322+0432 &52264-00569-0459& 0.1978&  12.13& 0.0868& 0.1714&  0.5835&  0.6059&1132&  6& 70& &     2.92&II\\
F09382+1449 &55978-05315-0118& 0.3838&  12.89& 0.1012& 0.2257&  0.6485&  0.9760&1132&  1&100&     7.64&     6.35&II\\
F09395+3939 &52725-01215-0082& 0.1944&  12.14& 0.0666& 0.1020&  0.4704&  1.0140&1132&  4& 70&     3.53& & \\
F09398+0013 &51602-00266-0235& 0.1459&  12.06& 0.0950& 0.1137&  0.5014&  2.2160&1131&  4& 83&     4.87& & \\
F09400+4338 &55631-04569-0576& 0.3634&  12.48& 0.0874& 0.1754&  0.2521&  0.5185&1131&  4& 84&     7.97& &II\\
F09418+1433 &54139-02582-0393& 0.2164&  12.19& 0.1941& 0.2951&  0.4059&  0.9099&1131&  3& 89&     5.44& & \\
F09438+4735 &56686-07299-0377& 0.5394&  13.01& 0.0576& 0.1325&  0.2476&  0.8972&1131& 10& 57&     2.84& &I\\
F09444+1019 &52757-01305-0509& 0.2020&  12.01& 0.0894& 0.1579&  0.2245&  0.9271&1131&  7& 52&     1.97& & \\
F09483+1443 &54139-02582-0583& 0.2333&  12.20& 0.1117& 0.1301&  0.2103&  1.1370&1131&  8& 61&     1.80& &I\\
F09496+5755 &56625-07085-0500& 0.3840&  12.70& 0.0619& 0.1006&  0.3720&  0.7202&1132&  3& 79&     6.48&     5.96& \\
F09510+2128 &56269-05785-0070& 0.2451&  12.29& 0.1569& 0.2289&  0.4152&  0.7829&1132&  4& 80&     2.99& &II\\
F09554+2718 &56270-06467-0602& 0.2499&  12.21& 0.1117& 0.1636&  0.3424&  0.5894&1131&  7& 64&     4.26&     3.39& \\
F09583+4714 &52339-00872-0280& 0.0859&  12.02& 0.0846& 0.4819&  2.6430&  3.0610&1332&  3& 87&    24.95& &II\\
F10015-0018 &51581-00269-0182& 0.2885&  12.52& 0.0821& 0.1430&  0.3203&  1.3330&1131&  5& 63&     2.65& &II\\
F10026+4347 &57401-08287-0228& 0.1783&  12.07& 0.1277& 0.1848&  0.5582&  0.8740&1232&  2& 92&     2.81& &I\\
F10026+4949 &56660-07282-0898& 1.1218&  13.72& 0.0858& 0.1772&  0.2655&  0.6190&1231&  3& 85& & &I\\
F10030+4126 &52672-01217-0021& 0.3285&  12.49& 0.0979& 0.0892&  0.3933&  0.5044&1132&  5& 59&     2.23& & \\
F10035+4852 &52339-00872-0408& 0.0651&  12.03& 0.0984& 0.2827&  4.5930&  6.2420&2332&  1&100&     8.58& & \\
F10035+2740 &53757-02347-0321& 0.1656&  12.30& 0.1371& 0.1698&  1.1440&  1.6300&1132&  5& 64&     5.44&     5.31&II\\
F10037+1112 &53053-01308-0483& 0.2736&  12.15& 0.0804& 0.2390&  0.2596&  0.3922&1131&  5& 68&     1.14& & \\
F10040+0932 &52751-01236-0534& 0.1709&  12.22& 0.1366& 0.2806&  0.8290&  1.4010&1132&  1&100&     2.58& & \\
F10077+1427 &54174-02588-0309& 0.1989&  12.08& 0.1221& 0.2729&  0.4673&  0.6248&1231&  2& 98&     8.82&    10.96&II\\
F10106+2227 &53739-02365-0389& 0.2739&  12.15& 0.0974& 0.1321&  0.2199&  0.4779&1131&  8& 70&     1.11& &I\\
F10107+4708 &52339-00872-0021& 0.2056&  12.33& 0.0911& 0.1183&  0.7034&  1.2520&1132&  1&100&     2.15& & \\
F10124+2742 &53757-02348-0020& 0.2103&  12.24& 0.1465& 0.1785&  0.5816&  0.8409&1132&  4& 73&     5.36&     5.03& \\
F10124+1631 &54174-02588-0551& 0.2402&  12.06& 0.1144& 0.2155&  0.2475&  0.5099&1131&  7& 62&     2.47& &II\\
F10139+0429 &55653-04801-0030& 0.2661&  12.51& 0.0955& 0.1746&  0.5305&  1.2110&1132&  1&100&     6.81& & \\
F10190+1322 &53062-01746-0194& 0.0763&  12.07& 0.0992& 0.3785&  3.3280&  5.5740&1232&  4& 71&    16.41&    15.79& \\
F10194+2427 &53734-02349-0266& 0.1882&  12.04& 0.1223& 0.2014&  0.3701&  0.9331&1132&  2& 87&     4.08& &II\\
F10211+2436 &53734-02349-0233& 0.2092&  12.06& 0.0678& 0.1852&  0.3724&  0.6185&1131&  4& 81& & & \\
F10234+3052 &58161-10460-0253& 0.3403&  12.47& 0.0592& 0.1572&  0.2588&  0.6713&1131&  8& 63&     1.82& &I\\
F10341+1312 &53112-01748-0194& 0.1740&  12.06& 0.0804& 0.1764&  0.5687&  0.8803&1132&  1&100&     3.16& & \\
F10345+3809 &53433-01998-0358& 0.2028&  12.05& 0.0646& 0.0787&  0.2743&  0.9466&1131&  4& 56& & & \\
F10369+4913 &52354-00875-0053& 0.1755&  12.06& 0.0835& 0.1110&  0.6420&  0.6584&1132&  1&100&     7.25&     7.35& \\
F10372+4801 &52620-00962-0487& 0.4861&  12.81& 0.0771& 0.1488&  0.2190&  0.6733&1131&  8& 67&     4.28& &II\\
F10378+1108 &53090-01600-0061& 0.1363&  12.34& 0.1141& 0.2351&  2.2810&  1.8160&1232&  4& 87&     8.55&     8.17& \\
F10425+2224 &54058-02477-0010& 0.2113&  12.18& 0.0573& 0.1004&  0.2321&  1.4030&1131&  6& 64&     2.56& & \\
F10445+4205 &53047-01361-0036& 0.1990&  12.14& 0.0863& 0.0958&  0.4289&  0.9897&1131&  4& 69&     1.00& & \\
F10482+1909 &54175-02482-0430& 0.2192&  12.42& 0.0664& 0.1561&  0.6850&  1.4400&1132&  3& 73&     3.08& & \\
F10494+4424 &53053-01434-0052& 0.0921&  12.25& 0.1171& 0.1648&  3.5270&  5.4120&1232&  1&100&    20.74&    20.39&II\\
F10531+5531 &52373-00907-0167& 0.2564&  12.01& 0.0673& 0.0857&  0.1994&  0.3728&1131&  9& 58& & &I\\
F10558+3845 &53474-02007-0610& 0.2081&  12.23& 0.0516& 0.0857&  0.6240&  0.7535&1132&  4& 71&     2.25& & \\
F10565+2448 &56354-06418-0766& 0.0428&  12.06& 0.2170& 1.1380& 12.1200& 15.1300&3332&  2& 99&    51.80&    51.61& \\
F10594+3818 &53469-01988-0235& 0.1578&  12.31& 0.0853& 0.1525&  1.2930&  1.8950&1132&  2& 88&     9.60&     9.18& \\
F11028+3442 &53466-02034-0230& 0.5083&  12.85& 0.1025& 0.1471&  0.2375&  0.6289&1231&  6& 86& & &I\\
F11028+3130 &53472-01990-0525& 0.1988&  12.43& 0.0863& 0.1182&  1.0210&  1.4440&1132&  1&100&     2.32& & \\
F11057+4053 &53003-01439-0263& 0.1657&  12.08& 0.1003& 0.1304&  0.6830&  1.0170&1132&  1&100&     3.90& & \\
F11066+4242 &53003-01439-0377& 0.2317&  12.11& 0.1075& 0.1234&  0.2469&  0.7610&1131&  3& 85&     1.73& &I\\
F11070+4249 &55652-04620-0744& 0.2614&  12.34& 0.0663& 0.1754&  0.4693&  0.5827&1132&  3& 90&    16.96&    15.20&II\\
F11087+5351 &52649-01012-0229& 0.1429&  12.03& 0.0833& 0.1100&  0.8109&  1.2820&1132&  1&100&    30.97&    29.45&II\\
F11119+3257 &56367-06438-0524& 0.1876&  12.51& 0.1672& 0.3484&  1.5880&  1.5230&3332&  3& 90&   107.77&   106.80&II\\
F11134+0225 &52636-00511-0315& 0.2109&  12.01& 0.1685& 0.1889&  0.3992&  0.3291&1131&  6& 72&     0.57& &I\\
F11162+6020 &52398-00951-0410& 0.2643&  12.55& 0.0792& 0.1178&  0.6751&  1.1100&1132&  3& 77&    56.22& &II\\
F11163+3207 &53431-01979-0373& 0.2618&  12.22& 0.0797& 0.1414&  0.2967&  0.6030&1131&  4& 82&    34.38&    35.90&II\\
F11188+1138 &53062-01605-0093& 0.1849&  12.36& 0.1228& 0.2403&  0.9126&  1.7200&1132&  4& 76&     6.53& &II\\
F11206+3639 &53446-02037-0099& 0.2425&  12.42& 0.0975& 0.0872&  0.4852&  1.3410&1131&  5& 65&    13.54& &II\\
F11213+6556 &51942-00491-0088& 0.2641&  12.28& 0.0941& 0.1241&  0.2404&  0.9077&1131& 11& 50& & & \\
F11215+1058 &52781-01223-0325& 0.1988&  12.08& 0.0816& 0.1375&  0.3249&  0.9979&1131&  6& 59&     1.38& & \\
F11273+4610 &56390-06647-0298& 0.2811&  12.73& 0.0920& 0.0923&  0.8939&  1.4410&1132&  1&100&     6.19& & \\
F11277+2016 &54169-02498-0590& 0.1930&  12.02& 0.1110& 0.1365&  0.3215&  0.8742&1131&  6& 53& & & \\
F11370+4647 &53050-01442-0025& 0.1735&  12.06& 0.0815& 0.1037&  0.3227&  1.5680&1131&  3& 89&     4.95&     4.08& \\
F11387+4116 &53062-01445-0242& 0.1488&  12.16& 0.2003& 0.1376&  1.0250&  1.5130&1132&  2& 95&     5.39& & \\
F11388+0204 &51994-00514-0285& 0.3821&  12.54& 0.1078& 0.1318&  0.2230&  0.6244&1131&  6& 50& & &I\\
F11394+0108 &51584-00283-0411& 0.2450&  12.22& 0.1665& 0.2659&  0.4147&  0.5166&1132&  5& 80&    17.68&    18.79&II\\
F11417+1151 &54894-03245-0248& 0.2702&  12.33& 0.1542& 0.2205&  0.3674&  0.6861&1132&  5& 91&     3.01& &I\\
F11506+1331 &53144-01610-0331& 0.1273&  12.39& 0.0995& 0.2949&  2.5830&  3.3230&1132&  4& 85&    13.16&    13.24& \\
F11553-0259 &52370-00330-0012& 0.2145&  12.06& 0.1170& 0.2981&  0.3325&  0.6266&1132&  8& 69&     3.08& &I\\
F11579+0848 &53089-01623-0321& 0.2481&  12.10& 0.0914& 0.4126&  0.2540&  0.5270&1131&  9& 55&     1.66& & \\
F11588+2052 &56067-05973-0482& 0.5301&  13.13& 0.1316& 0.1983&  0.2791&  1.3740&1131& 10& 60&     5.35& &II\\
F11595+1144 &53147-01611-0295& 0.1937&  12.34& 0.1077& 0.1573&  0.9452&  1.1500&1132&  3& 74&     5.77&     6.87& \\
F12018+1941 &54476-02609-0301& 0.1679&  12.45& 0.1068& 0.3735&  1.7610&  1.7760&1232&  2& 95&     5.55& &II\\
F12043+5215 &52370-00882-0065& 0.3992&  12.51& 0.0993& 0.0945&  0.2220&  0.4313&1131&  7& 70&     1.32& &I\\
F12047+0233 &52024-00517-0142& 0.2217&  12.14& 0.1149& 0.3462&  0.4121&  0.5774&1132&  4& 69&     1.74& & \\
F12112+0305 &52282-00518-0467& 0.0730&  12.38& 0.1100& 0.5093&  8.5030&  9.9760&1332&  3& 90& &    22.81& \\
F12126+0943 &52672-01230-0221& 0.2632&  12.09& 0.0932& 0.2684&  0.2143&  0.4491&1131&  5& 78& & & \\
F12232+5532 &52721-01020-0390& 0.2324&  12.36& 0.2413& 0.0954&  0.6008&  0.9410&1132&  3& 78&     4.96&     3.47& \\
F12266+3240 &53818-02013-0629& 0.1730&  12.03& 0.0873& 0.1533&  0.5715&  0.7624&1132&  2& 89&     2.97&     2.72& \\
F12297+5222 &52379-00885-0307& 0.3907&  12.61& 0.0739& 0.1188&  0.2624&  0.6350&1131&  7& 66&     4.12&     3.57& \\
F12358+1807 &54234-02599-0503& 0.4522&  12.67& 0.1359& 0.1279&  0.2621&  0.3896&1131&  6& 89&     6.07& &I\\
F12375+3721 &53495-02010-0621& 0.2402&  12.12& 0.0534& 0.0890&  0.2478&  0.6759&1131&  6& 65&     6.13&     7.21&II\\
F12410+2344 &54485-02648-0414& 0.2120&  12.03& 0.0902& 0.1115&  0.3392&  0.5528&1132&  5& 74&     2.63& &I\\
F12433+6540 &52316-00601-0479& 0.3197&  12.36& 0.0594& 0.0695&  0.2081&  0.6485&1131& 10& 61& & &I\\
F12442+4550 &53063-01373-0327& 0.1957&  12.28& 0.0670& 0.1374&  0.7700&  1.0400&1132&  1&100&     2.77& &II\\
F12447+3721 &53772-01989-0282& 0.1582&  12.14& 0.1155& 0.1215&  1.0430&  0.8434&1132&  1&100&     2.15&     2.55& \\
F12465+4458 &53063-01373-0197& 0.2286&  12.09& 0.0669& 0.0977&  0.2094&  0.8256&1131&  6& 71& & &II\\
F12471+4759 &53089-01455-0597& 0.3043&  12.55& 0.1179& 0.0672&  0.3568&  1.1680&1131&  7& 58&     2.53&     3.04&I\\
F12484+6245 &56444-06970-0269& 0.4608&  12.66& 0.0617& 0.0566&  0.1875&  0.5126&1131&  3& 83&    10.84&    10.40&II\\
F12487+0235 &52026-00523-0312& 0.2533&  12.38& 0.0888& 0.1270&  0.4717&  0.8864&1132&  6& 63&     2.22& & \\
F12489+6619 &51988-00495-0207& 0.2817&  12.17& 0.0604& 0.0712&  0.1739&  0.5768&1131&  3& 88& & &I\\
F12510+2603 &54505-02661-0518& 0.4836&  12.86& 0.1253& 0.1444&  0.2168&  0.8561&1131& 11& 63&     2.89& &II\\
F12514+1027 &55982-05415-0618& 0.3190&  12.70& 0.0632& 0.1904&  0.7124&  0.7553&1232&  4& 82&     7.91&     8.35&II\\
F12532-0322 &51694-00338-0287& 0.1687&  12.06& 0.1279& 0.1518&  0.6741&  0.8086&1132&  2& 90&     3.74& &II\\
F12551+0825 &54504-01794-0331& 0.2724&  12.49& 0.0898& 0.1191&  0.5876&  0.7789&1131&  5& 82&     6.58&     6.29&I\\
F12559+2257 &56066-05990-0641& 0.2083&  12.30& 0.1521& 0.0794&  0.5616&  1.3340&1132&  5& 69&     3.79&     3.87& \\
F13080+3237 &53819-02029-0602& 0.9959&  13.69& 0.1744& 0.2710&  0.4198&  0.5760&1232&  1&100&  1506.33&  1618.18&I\\
F13161+0927 &53851-01798-0109& 0.2821&  12.22& 0.0955& 0.1573&  0.2159&  0.5929&1131&  7& 60&     0.99& &II\\
F13190+4050 &53112-01462-0140& 0.1850&  12.08& 0.1039& 0.1119&  0.4423&  1.0110&1132&  5& 78&     2.62& &II\\
F13209+6353 &52056-00603-0240& 0.2000&  12.21& 0.0582& 0.0687&  0.5896&  0.9138&1132&  2& 90&     6.66&     6.29&II\\
F13218+0552 &52375-00852-0367& 0.2028&  12.41& 0.2621& 0.4032&  1.1740&  0.7134&3332&  1&100&     4.26& &II\\
F13231+6235 &52327-00784-0541& 0.2376&  12.31& 0.0436& 0.0643&  0.3584&  1.1900&1131&  3& 73&     2.86& & \\
F13342+3932 &53472-02005-0441& 0.1791&  12.37& 0.1192& 0.2528&  1.1140&  1.6340&1132&  3& 97&     6.94&     6.11&I\\
F13372+2614 &56089-05999-0984& 0.3446&  12.48& 0.0855& 0.1219&  0.2343&  0.7195&1131&  6& 62&     8.22&     8.42&II\\
F13403-0038 &51671-00299-0098& 0.3256&  12.39& 0.1110& 0.2365&  0.2094&  0.6905&1131& 11& 66& & &I\\
F13428+5608 &52764-01321-0638& 0.0373&  12.16& 0.2352& 2.2820& 21.7400& 21.3800&3332&  1&100&   132.02&   134.58&II\\
F13443+0802 &54152-01803-0567& 0.1350&  12.17& 0.1187& 0.2351&  1.2990&  1.9350&1132&  2& 63&    10.09& &II\\
F13446+3727 &53858-02101-0622& 0.2147&  12.06& 0.1069& 0.0761&  0.3377&  0.5906&1132&  7& 59&     2.02& &II\\
F13451+1232 &53142-01701-0483& 0.1205&  12.18& 0.1433& 0.6695&  1.9160&  2.0600&1232&  2& 95&  4859.88&  5251.91&II\\
F13464+1828 &56038-05444-0671& 0.1795&  12.11& 0.0677& 0.0803&  0.4818&  1.2120&1132&  6& 61&     4.53&     4.72& \\
F13469+5833 &52668-01158-0429& 0.1575&  12.29& 0.0517& 0.0719&  1.2670&  1.7340&1132&  2& 89&     3.02& & \\
F13485+3739 &53476-02033-0373& 0.2478&  12.27& 0.1096& 0.0859&  0.3043&  0.9397&1131&  4& 73&     1.84& & \\
F13515+0317 &52026-00530-0616& 0.2786&  12.22& 0.1139& 0.1471&  0.2298&  0.5912&1131&  6& 67&     2.53& &II\\
F13531+2005 &56045-05868-0682& 0.3143&  12.53& 0.1264& 0.1434&  0.4244&  0.7346&1131&  8& 79&     4.07& & \\
F13539+2920 &53854-02116-0051& 0.1088&  12.12& 0.0872& 0.1224&  1.8320&  2.7290&1232&  3& 90&    10.70&    11.09&II\\
F13583+2343 &54529-02784-0436& 0.2160&  12.05& 0.1082& 0.1498&  0.3149&  0.6056&1131&  4& 72&     1.97& & \\
F14014+3718 &53115-01642-0490& 0.2105&  12.10& 0.1432& 0.1364&  0.3811&  0.7066&1132&  6& 71&     1.96& & \\
F14026+4341 &53115-01467-0216& 0.3229&  12.71& 0.1178& 0.2854&  0.6217&  0.9936&2332&  3& 86&     1.46& &I\\
F14041+0117 &51616-00302-0403& 0.2363&  12.48& 0.1648& 0.3190&  0.8956&  0.8293&1231&  4& 84&     9.19&    13.93&II\\
F14060+2919 &53794-02126-0307& 0.1167&  12.13& 0.0959& 0.1442&  1.6110&  2.4170&1232&  1&100&     7.86&     9.11& \\
F14079+4025 &56045-05169-0304& 0.4546&  12.83& 0.1018& 0.1375&  0.3363&  0.6586&1131&  5& 71&    49.62&    49.34& \\
F14082+0205 &51993-00532-0064& 0.2011&  12.07& 0.0711& 0.1076&  0.3062&  0.9522&1131&  9& 55&     2.09& &I\\
F14088+0212 &51993-00532-0080& 0.2023&  12.05& 0.1076& 0.2348&  0.3842&  0.6646&1132&  8& 61&     1.02& & \\
F14143+1244 &56011-05458-0342& 0.3278&  12.56& 0.1377& 0.2589&  0.4287&  0.6601&1131& 11& 72&     5.46& &II\\
F14167+4247 &53108-01394-0044& 0.4203&  12.70& 0.0976& 0.0925&  0.3073&  0.5710&1231&  5& 77&     1.25& &I\\
F14170+4545 &52728-01287-0193& 0.1502&  12.02& 0.0498& 0.0809&  0.7000&  1.2010&1132&  2& 86&     5.86& & \\
F14202+2615 &53819-02131-0280& 0.1587&  12.36& 0.1834& 0.1522&  1.4920&  1.9860&1232&  2& 98&     8.31& & \\
F14204+4533 &52728-01287-0144& 0.1673&  12.01& 0.0831& 0.1138&  0.5255&  0.9385&1132&  1&100&     7.87&     7.26& \\
F14225+1044 &56000-05460-0199& 0.4801&  12.92& 0.0486& 0.2529&  0.4063&  0.5923&1132&  3& 89&     3.76& & \\
F14229+1636 &54506-02760-0220& 0.1850&  12.12& 0.1020& 0.2495&  0.5385&  0.9784&1132&  4& 85&     2.44& & \\
F14248-0045 &51613-00305-0079& 0.1623&  12.18& 0.1165& 0.3196&  0.9107&  1.2810&1132&  2& 96&     3.51& & \\
F14290+1712 &54534-02761-0235& 0.2102&  12.12& 0.0514& 0.1023&  0.2859&  1.0430&1131&  5& 68&     1.76& & \\
F14305+0546 &55682-04780-0710& 0.2982&  12.51& 0.0934& 0.1563&  0.3058&  1.1890&1131&  2& 89&     9.79&     9.19&II\\
F14312+2825 &53876-02134-0502& 0.1748&  12.22& 0.1492& 0.1035&  0.7790&  1.4110&1232&  4& 72&     5.63&     5.79&II\\
F14315+2955 &53917-02133-0019& 0.5264&  12.82& 0.1269& 0.1161&  0.2020&  0.5392&1131&  9& 66& & &I\\
F14318-0252 &52409-00919-0201& 0.1865&  12.11& 0.0825& 0.1827&  0.3912&  1.2390&1131&  3& 81& & & \\
F14330+0141 &52024-00536-0234& 0.2325&  12.16& 0.1319& 0.2227&  0.3511&  0.6480&1131&  6& 54&    15.08& &II\\
F14379+5420 &52668-01162-0095& 0.2685&  12.22& 0.0645& 0.0528&  0.3328&  0.4276&1132&  4& 70&     1.93& & \\
F14390+6209 &52368-00607-0625& 0.2755&  12.31& 0.0595& 0.0910&  0.3153&  0.6831&1131&  4& 77&     2.86& &I\\
F14394+5332 &52781-01327-0619& 0.1050&  12.08& 0.0721& 0.3464&  1.9540&  2.3950&1332&  1&100&    39.55&    41.60&II\\
F14401+6316 &56447-06983-0612& 0.2805&  12.28& 0.0657& 0.0828&  0.2666&  0.6651&1131&  3& 82&     7.93& &II\\
F14413+3730 &53115-01382-0587& 0.2598&  12.20& 0.0634& 0.0679&  0.3107&  0.5328&1132&  1&100&     1.76& &II\\
F14424+1050 &56011-05475-0494& 0.4097&  12.68& 0.0690& 0.1533&  0.2765&  0.6615&1131&  9& 54&     4.63& &II\\
F14455+3507 &53498-01646-0578& 0.6624&  13.10& 0.0663& 0.1049&  0.2173&  0.5833&1131&  4& 80&    14.44&    15.26&I\\
F14481+4454 &58171-08492-0889& 0.6618&  13.03& 0.0764& 0.0851&  0.1897&  0.5005&1231&  7& 86&    11.06&    11.26&II\\
F14488+3521 &53116-01383-0083& 0.2057&  12.30& 0.0681& 0.0928&  0.6389&  1.2120&1232&  5& 73&     5.26&     5.88&II\\
F14501+0639 &54560-01830-0380& 0.2672&  12.22& 0.0493& 0.1422&  0.2917&  0.5550&1131&  6& 63&   291.60&   289.40&II\\
F14503+6006 &56444-06982-0393& 0.5748&  12.91& 0.0411& 0.0788&  0.2263&  0.4721&1231&  3& 83&    18.03&    17.51&II\\
F14513-0235 &52380-00921-0139& 0.2094&  12.10& 0.0851& 0.1160&  0.2545&  1.0510&1131&  6& 56&     1.19& & \\
F14538+1730 &54535-02764-0521& 0.1033&  12.02& 0.1012& 0.1458&  1.4420&  3.0070&1132&  4& 82&    12.68&    11.50&II\\
F14541+3813 &58224-10748-0819& 0.2833&  12.01& 0.0815& 0.1295&  0.1625&  0.2905&1131&  9& 73& & &I\\
F14541+4906 &52736-01048-0505& 0.2467&  12.31& 0.0825& 0.1999&  0.5303&  0.5428&1332&  2& 85&     6.35&     6.64&II\\
F14556+1328 &56014-05478-0866& 0.3341&  12.48& 0.0525& 0.0596&  0.2577&  0.7606&1131& 10& 53&     8.76&     7.97&II\\
F14560+5246 &56390-06719-0072& 0.6380&  13.02& 0.0532& 0.0659&  0.1560&  0.6510&1131&  8& 52&     2.07& &I\\
F14589+2340 &54509-02149-0006& 0.2578&  12.08& 0.0647& 0.0845&  0.1954&  0.5151&1231&  7& 75&     3.97& &I\\
F15001+1433 &54535-02765-0287& 0.1622&  12.46& 0.1220& 0.1674&  1.8710&  2.0430&1232&  2& 98&    16.74&    14.85& \\
F15002+4945 &52767-01329-0103& 0.3368&  12.52& 0.0533& 0.0738&  0.3952&  0.5099&1232&  5& 80&     3.09& & \\
F15004+0351 &52055-00589-0197& 0.2182&  12.13& 0.0876& 0.0824&  0.2327&  1.0920&1131&  9& 53&     2.10& & \\
F15043+5754 &52056-00610-0060& 0.1506&  12.16& 0.1221& 0.0718&  1.0190&  1.5020&1232&  2& 90&     2.64&     2.31& \\
F15069+1808 &54556-02792-0089& 0.1705&  12.12& 0.0914& 0.1192&  0.6851&  1.0390&1232&  3& 91&    11.87&    11.40&I\\
F15099+1521 &54242-02766-0224& 0.5007&  12.82& 0.0484& 0.0691&  0.1885&  0.7123&1131& 10& 56& & &I\\
F15101+5624 &57894-08416-0020& 0.2584&  12.15& 0.0467& 0.0644&  0.2450&  0.5577&1132&  2& 59& & & \\
F15104+3431 &55691-04720-0744& 0.8565&  13.26& 0.0576& 0.0541&  0.1893&  0.4003&1132&  6& 51& & &I\\
F15111+2458 &53820-02155-0534& 0.2200&  12.19& 0.0556& 0.1732&  0.5165&  0.5627&1332&  2& 90&     3.18& &II\\
F15206+3342 &53118-01387-0458& 0.1253&  12.18& 0.0812& 0.3468&  1.7670&  1.8870&2332&  1&100&     9.96& & \\
F15206+3631 &53470-01400-0254& 0.1527&  12.05& 0.0629& 0.0935&  0.7706&  1.0640&1232&  2& 91&     4.82&     4.61&II\\
F15221+5619 &54525-02883-0414& 0.2438&  12.02& 0.0497& 0.0706&  0.2060&  0.4726&1231&  5& 88&     2.04& &II\\
F15225+2350 &53878-02161-0373& 0.1387&  12.15& 0.0686& 0.1798&  1.3000&  1.4830&1332&  2& 89&     5.53&     6.38& \\
F15239+4331 &52468-01051-0123& 0.3561&  12.68& 0.0481& 0.0564&  0.4869&  0.6974&1132&  1&100&     6.85& & \\
F15261+5502 &53437-00614-0213& 0.2292&  12.22& 0.0478& 0.0592&  0.4616&  0.6577&1232&  1&100&     1.50& & \\
F15320+0325 &52026-00593-0145& 0.2057&  12.06& 0.0715& 0.0979&  0.3285&  0.7635&1132&  8& 65&     2.95&     3.85&II\\
F15327+2340 &53823-02163-0058& 0.0184&  12.22& 0.4837& 7.9070&103.8000&112.4000&3332&  1&100&   316.13&   318.80& \\
F15390+3913 &53171-01680-0189& 0.2394&  12.02& 0.0544& 0.0786&  0.1983&  0.5398&1131&  8& 54&     1.24& & \\
F15396+1715 &54563-02795-0589& 0.2570&  12.12& 0.0638& 0.0931&  0.2498&  0.4837&1131&  6& 56&     1.59& & \\
F15415+1633 &55337-03932-0246& 0.8483&  13.36& 0.0781& 0.1170&  0.2315&  0.5459&1231&  6& 87&    12.45&    14.87&I\\
F15432+3502 &58223-10740-0291& 0.5166&  12.81& 0.0509& 0.0771&  0.1787&  0.6219&1131&  8& 65&    15.93&    17.68&I\\
F15437+4647 &52731-01168-0227& 0.2284&  12.05& 0.0645& 0.1076&  0.2639&  0.5625&1231&  1&100&     3.67&     3.31&II\\
F15438+0438 &55679-04804-0808& 0.2362&  12.47& 0.0649& 0.1082&  0.7833&  1.0230&1132&  3& 81&     6.54& &II\\
F15439+4855 &57893-08429-0896& 0.3997&  12.58& 0.0743& 0.1082&  0.2828&  0.4475&1332&  1&100&     2.11& &I\\
F15453+1131 &55737-04886-0146& 0.2890&  12.52& 0.0519& 0.1126&  0.5145&  0.8263&1132&  5& 54&     7.51& &II\\
F15453+4316 &56101-06042-0339& 0.2973&  12.41& 0.0392& 0.0567&  0.3457&  0.6949&1131&  2& 90&    12.36&    10.51& \\
F15458+0041 &51691-00342-0514& 0.2524&  12.05& 0.0585& 0.1265&  0.1892&  0.5158&1231&  9& 66&     1.98& &II\\
F15478+5014 &52401-00796-0022& 0.1977&  12.11& 0.0520& 0.0705&  0.4178&  0.9203&1132&  3& 86&     2.44& & \\
F15496+0331 &54592-02951-0294& 0.1937&  12.06& 0.0837& 0.1436&  0.3519&  0.9330&1231&  7& 74&    29.37& &II\\
F15529+4545 &52753-01169-0382& 0.5164&  12.81& 0.0694& 0.0548&  0.2276&  0.4863&1132&  8& 67&     1.32& &I\\
F15531+2513 &53786-01850-0565& 0.1846&  12.02& 0.0601& 0.0888&  0.3025&  1.1240&1131&  8& 53& & & \\
F15554+5322 &56799-07562-0403& 0.5628&  12.88& 0.0532& 0.0868&  0.1820&  0.5547&1131&  3& 86&     8.78&     7.35&II\\
F15577+3816 &52761-01055-0312& 0.2181&  12.14& 0.0635& 0.0829&  0.4467&  0.5607&1132&  2& 89&     2.39& &II\\
F15583+4002 &52516-01054-0072& 0.2189&  12.06& 0.0901& 0.1150&  0.3149&  0.6076&1231&  3& 91&     2.02& & \\
F16006+0955 &54582-02526-0302& 0.1520&  12.05& 0.1176& 0.1049&  0.4826&  1.8380&1131&  3& 89&     3.89& & \\
F16019+0828 &53858-01729-0457& 0.2283&  12.21& 0.0860& 0.0982&  0.1959&  1.3080&1131&  6& 50& & & \\
F16048+1342 &54568-02524-0064& 0.2090&  12.11& 0.0570& 0.0738&  0.3835&  0.7919&1132&  3& 57& & &I\\
F16075+2838 &53496-01578-0005& 0.1696&  12.16& 0.0894& 0.1064&  0.8405&  1.0060&1232&  2& 90&     4.78& &II\\
F16098+2624 &52824-01393-0113& 0.1843&  12.08& 0.0705& 0.0939&  0.3054&  1.4110&1131&  7& 52&     1.61& & \\
F16109+1309 &54569-02527-0172& 0.1996&  12.09& 0.0805& 0.0792&  0.3166&  1.0410&1131&  3& 83&     1.84& & \\
F16122+1531 &53918-02198-0475& 0.3078&  12.42& 0.0541& 0.1490&  0.2671&  0.8163&1231&  7& 81&    20.27&    23.13&II\\
F16126+1953 &53795-02206-0198& 0.2529&  12.27& 0.0658& 0.0980&  0.3239&  0.8370&1131&  5& 59&     2.93& &II\\
F16133+2107 &53566-01853-0227& 0.0906&  12.00& 0.0827& 0.1621&  2.1640&  2.9280&1332&  1&100&    11.24& &II\\
F16172+4432 &57900-08525-0882& 0.3351&  12.41& 0.0545& 0.1131&  0.2757&  0.4972&1232&  3& 90&     3.09&     2.84&I\\
F16184+5012 &54526-02884-0145& 0.2834&  12.25& 0.0638& 0.0790&  0.2685&  0.5284&1132&  5& 74&     2.35& &II\\
F16300+1558 &53558-02207-0123& 0.2418&  12.77& 0.0733& 0.1240&  1.4830&  1.9900&1132&  1&100&     5.93&     7.10& \\
F16310+2254 &53177-01572-0054& 0.2217&  12.06& 0.1049& 0.0710&  0.2278&  0.7851&1231&  7& 71&     1.50& & \\
F16403+2537 &53167-01423-0175& 0.1602&  12.04& 0.0842& 0.0584&  0.3877&  1.6780&1131& 10& 58&     4.40& & \\
F16413+3954 &56090-06038-0144& 0.5934&  13.39& 0.1068& 0.2918&  0.5999&  1.3840&3331&  1&100&  6598.19&  6930.92&I\\
F16533+6216 &51695-00351-0329& 0.1060&  12.02& 0.0588& 0.1689&  1.4800&  2.5030&1322&  2& 94&    14.43&    16.26& \\
F17040+6048 &51703-00353-0377& 0.3715&  12.60& 0.0700& 0.1276&  0.2959&  0.6971&1311&  1&100& & &I\\
F17051+3824 &52079-00633-0003& 0.1683&  12.23& 0.0621& 0.0788&  0.9865&  1.2060&1232&  3& 69&     4.30& &II\\
F17081+3300 &52426-00973-0039& 0.2786&  12.20& 0.0722& 0.0863&  0.1956&  0.6187&1132&  7& 66& & &II\\
F17135+4153 &58041-08545-0839& 0.7274&  13.34& 0.0641& 0.0526&  0.2781&  0.8838&1132&  2& 66& & & \\
F17175+6603 &51691-00350-0442& 0.2926&  12.25& 0.0647& 0.0478&  0.2075&  0.5919&1231&  7& 78& & &I\\
F17234+6228 &51694-00352-0082& 0.2407&  12.14& 0.0435& 0.0680&  0.2587&  0.6977&1131&  2& 95&     1.93& &I\\
F20460-0035 &52914-01115-0317& 0.4327&  12.82& 0.1084& 0.1045&  0.3060&  0.8955&1131&  4& 71&     3.65&     3.46& \\
F20461+0043 &52524-01022-0525& 0.3487&  12.82& 0.0903& 0.1166&  0.3049&  2.0140&1131&  5& 50&     2.28& &II\\
F20522-0120 &52443-00983-0044& 0.1735&  12.22& 0.0527& 0.1041&  0.3907&  2.4570&1131& 10& 54& & & \\
F21292-0115 &55450-04194-0246& 0.3875&  12.61& 0.0720& 0.1340&  0.2586&  0.6916&1131&  6& 76&     4.41& &II\\
F21581-0053 &55481-04199-0482& 0.3566&  12.66& 0.1530& 0.1350&  0.3502&  0.9194&1131&  7& 57&    28.63&    29.00&II\\
F22032+0038 &55481-04199-0750& 0.2845&  12.40& 0.0802& 0.1667&  0.2154&  1.1570&1131& 11& 67&     3.52&     4.52&I\\
F22098-0748 &52203-00719-0323& 0.1427&  12.00& 0.0980& 0.1693&  0.7600&  1.2240&1132&  3& 83&     4.56&     3.94&II\\
F22288+1258 &56187-05043-0287& 0.1852&  12.08& 0.1500& 0.1311&  0.3195&  1.3300&1131&  4& 60& & &II\\
F22505+1704 &56244-05033-0996& 0.1786&  12.30& 0.0994& 0.1415&  0.5785&  2.3600&1131&  5& 75& & &II\\
F22559+0733 &53710-02310-0458& 0.3955&  12.93& 0.1211& 0.2242&  0.6066&  1.1360&1231&  7& 69&    27.70&    33.04&I\\
F23060+0505 &55882-04414-0184& 0.1728&  12.26& 0.1978& 0.4258&  1.1520&  0.8329&3332&  3& 85&     5.50&     6.52&II\\
F23233+2817 &56537-06587-0910& 0.1135&  12.01& 0.1287& 0.2756&  1.2620&  2.1070&1332&  4& 84& &    34.76 & II
\enddata
\tablecomments{(1) Source name in the IRAS FSC92 catalog. (2) Mjd, plate, and fiberid of the SDSS optical counterpart's spectrum.
(3) Spectroscopic redshift from the SDSS pipeline. (4) Total infrared luminosity.  (5-9)  Observed flux densities at 12 $\mu m$, 25 $\mu m$, 60 $\mu m$, 100 $\mu m$, and their qualities from the IRAS survey.  (10) Number of WISE counterparts within the IRAS source's positional uncertainty ellipse. (11) Flux ratio at the W4 band of the optical counterpart to all WISE counterparts. (12-13) Radio flux densities at 1.4 GHz from the FIRST and NVSS surveys. (14) Type I or Type II AGNs.
}
\end{deluxetable*}

\end{CJK}

\begin{thebibliography}{ }
\bibitem[Abdo et al.(2009)]{2009ApJ...707L.142A} Abdo, A.~A., Ackermann, M., Ajello, M., et al.\ 2009, \apjl, 707, L142. doi:10.1088/0004-637X/707/2/L142
\bibitem[Ahumada et al.(2020)]{2020ApJS..249....3A} Ahumada, R., Prieto, C.~A., Almeida, A., et al.\ 2020, \apjs, 249, 3. doi:10.3847/1538-4365/ab929e
\bibitem[Barger et al.(1998)]{1998Natur.394..248B} Barger, A.~J., Cowie, L.~L., Sanders, D.~B., et al.\ 1998, \nat, 394, 248. doi:10.1038/28338
\bibitem[Becker et al.(1995)]{1995ApJ...450..559B} Becker, R.~H., White, R.~L., \& Helfand, D.~J.\ 1995, \apj, 450, 559. doi:10.1086/176166
\bibitem[Boroson(2002)]{2002ApJ...565...78B} Boroson, T.~A.\ 2002, \apj, 565, 78. doi:10.1086/324486
\bibitem[Boroson \& Lauer(2010)]{2010AJ....140..390B} Boroson, T.~A. \& Lauer, T.~R.\ 2010, \aj, 140, 390. doi:10.1088/0004-6256/140/2/390
\bibitem[Bridge et al.(2013)]{2013ApJ...769...91B} Bridge, C.~R., Blain, A., Borys, C.~J.~K., et al.\ 2013, \apj, 769, 91. doi:10.1088/0004-637X/769/2/91
\bibitem[Bushouse et al.(2002)]{2002ApJS..138....1B} Bushouse, H.~A., Borne, K.~D., Colina, L., et al.\ 2002, \apjs, 138, 1. doi:10.1086/324019
\bibitem[Cao et al.(2006)]{2006ChJAA...6..197C} Cao, C., Wu, H., Wang, J.-L., et al.\ 2006, \cjaa, 6, 197. doi:10.1088/1009-9271/6/2/7
\bibitem[Cao et al.(2018)]{2018MNRAS.480.2178C} Cao, Y., Gong, Y., Meng, X.-M., et al.\ 2018, \mnras, 480, 2178. doi:10.1093/mnras/sty1980
\bibitem[Calzetti et al.(2000)]{2000ApJ...533..682C} Calzetti, D., Armus, L., Bohlin, R.~C., et al.\ 2000, \apj, 533, 682. doi:10.1086/308692
\bibitem[Canalizo \& Stockton(2001)]{2001ApJ...555..719C} Canalizo, G. \& Stockton, A.\ 2001, \apj, 555, 719. doi:10.1086/321520
\bibitem[Casey et al.(2014)]{2014ApJ...796...95C} Casey, C.~M., Scoville, N.~Z., Sanders, D.~B., et al.\ 2014, \apj, 796, 95. doi:10.1088/0004-637X/796/2/95
\bibitem[Clements et al.(1996)]{1996MNRAS.279..477C} Clements, D.~L., Sutherland, W.~J., McMahon, R.~G., et al.\ 1996, \mnras, 279, 477. doi:10.1093/mnras/279.2.477
\bibitem[Chen \& Liu(2024)]{2024NewA..10502083C} Chen, P.~S. \& Liu, J.~Y.\ 2024, \na, 105, 102083. doi:10.1016/j.newast.2023.102083
\bibitem[Cohen et al.(1987)]{1987AJ.....93.1199C} Cohen, M., Schwartz, D.~E., Chokshi, A., et al.\ 1987, \aj, 93, 1199. doi:10.1086/114401
\bibitem[Condon et al.(1982)]{1982ApJ...252..102C} Condon, J.~J., Condon, M.~A., Gisler, G., et al.\ 1982, \apj, 252, 102. doi:10.1086/159538
\bibitem[Condon et al.(1991)]{1991ApJ...378...65C} Condon, J.~J., Huang, Z.-P., Yin, Q.~F., et al.\ 1991, \apj, 378, 65. doi:10.1086/170407
\bibitem[Condon et al.(1998)]{1998AJ....115.1693C} Condon, J.~J., Cotton, W.~D., Greisen, E.~W., et al.\ 1998, \aj, 115, 1693. doi:10.1086/300337
\bibitem[Condon \& Broderick(1986)]{1986AJ.....92...94C} Condon, J.~J. \& Broderick, J.~J.\ 1986, \aj, 92, 94. doi:10.1086/114139
\bibitem[Cutri et al.(2013)]{2013wise.rept....1C} Cutri, R.~M., Wright, E.~L., Conrow, T., et al.\ 2013, Explanatory Supplement to the AllWISE Data Release Products, by R. M. Cutri et al.
\bibitem[Cui et al.(2001)]{2001AJ....122...63C} Cui, J., Xia, X.-Y., Deng, Z.-G., et al.\ 2001, \aj, 122, 63. doi:10.1086/321127
\bibitem[Dawson et al.(2016)]{2016AJ....151...44D} Dawson, K.~S., Kneib, J.-P., Percival, W.~J., et al.\ 2016, \aj, 151, 44. doi:10.3847/0004-6256/151/2/44
\bibitem[Efstathiou \& Rowan-Robinson(2003)]{2003MNRAS.343..322E} Efstathiou, A. \& Rowan-Robinson, M.\ 2003, \mnras, 343, 322. doi:10.1046/j.1365-8711.2003.06679.x
\bibitem[Efstathiou et al.(2021)]{2021MNRAS.503L..11E} Efstathiou, A., Ma{\l}ek, K., Burgarella, D., et al.\ 2021, \mnras, 503, L11. doi:10.1093/mnrasl/slaa206
\bibitem[Efstathiou et al.(2022)]{2022MNRAS.512.5183E} Efstathiou, A., Farrah, D., Afonso, J., et al.\ 2022, \mnras, 512, 5183. doi:10.1093/mnras/stab3642
\bibitem[Egami et al.(1996)]{1996AJ....112...73E} Egami, E., Iwamuro, F., Maihara, T., et al.\ 1996, \aj, 112, 73. doi:10.1086/117989
\bibitem[Eisenhardt et al.(2012)]{2012ApJ...755..173E} Eisenhardt, P.~R.~M., Wu, J., Tsai, C.-W., et al.\ 2012, \apj, 755, 173. doi:10.1088/0004-637X/755/2/173
\bibitem[Farrah et al.(2001)]{2001MNRAS.326.1333F} Farrah, D., Rowan-Robinson, M., Oliver, S., et al.\ 2001, \mnras, 326, 1333. doi:10.1111/j.1365-2966.2001.04721.x
\bibitem[Farrah et al.(2003)]{2003MNRAS.343..585F} Farrah, D., Afonso, J., Efstathiou, A., et al.\ 2003, \mnras, 343, 585. doi:10.1046/j.1365-8711.2003.06696.x
\bibitem[Farrah et al.(2017)]{2017ApJ...844..106F} Farrah, D., Petty, S., Connolly, B., et al.\ 2017, \apj, 844, 106. doi:10.3847/1538-4357/aa78f2
\bibitem[Flesch(2023)]{2023arXiv230801505F} Flesch, E.~W.\ 2023, arXiv:2308.01505. doi:10.48550/arXiv.2308.01505
\bibitem[Ganguly et al.(2007)]{2007ApJ...665..990G} Ganguly, R., Brotherton, M.~S., Cales, S., et al.\ 2007, \apj, 665, 990. doi:10.1086/519759
\bibitem[Genzel et al.(1998)]{1998ApJ...498..579G} Genzel, R., Lutz, D., Sturm, E., et al.\ 1998, \apj, 498, 579. doi:10.1086/305576
\bibitem[Gibson et al.(2009)]{2009ApJ...692..758G} Gibson, R.~R., Jiang, L., Brandt, W.~N., et al.\ 2009, \apj, 692, 758. doi:10.1088/0004-637X/692/1/758
\bibitem[Gong et al.(2019)]{2019ApJ...883..203G} Gong, Y., Liu, X., Cao, Y., et al.\ 2019, \apj, 883, 203. doi:10.3847/1538-4357/ab391e
\bibitem[Guo et al.(2018)]{} Guo, H., Shen, Y., \& Wang, S. 2018, PyQSOFit: Python code to fit the spectrum of quasars, Astrophysics Source Code Library, record ascl:1809.008. http://ascl.net/1809.008
\bibitem[Hamann et al.(1993)]{1993ApJ...415..541H} Hamann, F., Korista, K.~T., \& Morris, S.~L.\ 1993, \apj, 415, 541. doi:10.1086/173185
\bibitem[Helou et al.(1988)]{1988ApJS...68..151H} Helou, G., Khan, I.~R., Malek, L., et al.\ 1988, \apjs, 68, 151. doi:10.1086/191285
\bibitem[Hou et al.(2009)]{2009ApJ...704..789H} Hou, L.~G., Wu, X.-B., \& Han, J.~L.\ 2009, \apj, 704, 789. doi:10.1088/0004-637X/704/1/789
\bibitem[Houck et al.(1984)]{1984ApJ...278L..63H} Houck, J.~R., Soifer, B.~T., Neugebauer, G., et al.\ 1984, \apjl, 278, L63. doi:10.1086/184224
\bibitem[Hughes et al.(1998)]{1998Natur.394..241H} Hughes, D.~H., Serjeant, S., Dunlop, J., et al.\ 1998, \nat, 394, 241. doi:10.1038/28328
\bibitem[Hwang et al.(2007)]{2007MNRAS.375..115H} Hwang, H.~S., Serjeant, S., Lee, M.~G., et al.\ 2007, \mnras, 375, 115. doi:10.1111/j.1365-2966.2006.11279.x
\bibitem[Jones et al.(2014)]{2014MNRAS.443..146J} Jones, S.~F., Blain, A.~W., Stern, D., et al.\ 2014, \mnras, 443, 146. doi:10.1093/mnras/stu1157
\bibitem[Kim \& Sanders(1998)]{1998ApJS..119...41K} Kim, D.-C. \& Sanders, D.~B.\ 1998, \apjs, 119, 41. doi:10.1086/313148
\bibitem[Kim et al.(1998)]{1998ApJ...508..627K} Kim, D.-C., Veilleux, S., \& Sanders, D.~B.\ 1998, \apj, 508, 627. doi:10.1086/306409
\bibitem[Klaas(1989)]{1989Ap&SS.157..245K} Klaas, U.\ 1989, \apss, 157, 245. doi:10.1007/BF00637336
\bibitem[Klaas \& Elsaesser(1993)]{1993A&A...280...76K} Klaas, U. \& Elsaesser, H.\ 1993, \aap, 280, 76
\bibitem[Koss et al.(2018)]{2018Natur.563..214K} Koss, M.~J., Blecha, L., Bernhard, P., et al.\ 2018, \nat, 563, 214. doi:10.1038/s41586-018-0652-7
\bibitem[Krawczyk et al.(2013)]{2013ApJS..206....4K} Krawczyk, C.~M., Richards, G.~T., Mehta, S.~S., et al.\ 2013, \apjs, 206, 4. doi:10.1088/0067-0049/206/1/4
\bibitem[Leech et al.(1988)]{1988MNRAS.231..977L} Leech, K.~J., Lawrence, A., Rowan-Robinson, M., et al.\ 1988, \mnras, 231, 977. doi:10.1093/mnras/231.4.977
\bibitem[Leech et al.(1994)]{1994MNRAS.267..253L} Leech, K.~J., Rowan-Robinson, M., Lawrence, A., et al.\ 1994, \mnras, 267, 253. doi:10.1093/mnras/267.2.253
\bibitem[Lonsdale et al.(2006)]{2006asup.book..285L} Lonsdale, C.~J., Farrah, D., \& Smith, H.~E.\ 2006, Astrophysics Update 2, 285. doi:10.1007/3-540-30313-8\_9
\bibitem[Lonsdale Persson \& Helou(1987)]{1987ApJ...314..513L} Lonsdale Persson, C.~J. \& Helou, G.\ 1987, \apj, 314, 513. doi:10.1086/165082
\bibitem[Lutz et al.(1998)]{1998ApJ...505L.103L} Lutz, D., Spoon, H.~W.~W., Rigopoulou, D., et al.\ 1998, \apjl, 505, L103. doi:10.1086/311614
\bibitem[Massaro et al.(2011)]{2011ApJ...740L..48M} Massaro, F., D'Abrusco, R., Ajello, M., et al.\ 2011, \apjl, 740, L48. doi:10.1088/2041-8205/740/2/L48
\bibitem[Melnick \& Mirabel(1990)]{1990A&A...231L..19M} Melnick, J. \& Mirabel, I.~F.\ 1990, \aap, 231, L19
\bibitem[Moshir et al.(1992)]{1992ifss.book.....M} Moshir, M., Kopman, G., \& Conrow, T.~A.~O.\ 1992, Pasadena: Infrared Processing and Analysis Center, California Institute of Technology, 1992, edited by Moshir, M.; Kopman, G.; Conrow, T. a.o.
\bibitem[Murphy et al.(1996)]{1996AJ....111.1025M} Murphy, T.~W., Armus, L., Matthews, K., et al.\ 1996, \aj, 111, 1025. doi:10.1086/117849
\bibitem[Neff \& Hutchings(1992)]{1992AJ....103.1746N} Neff, S.~G. \& Hutchings, J.~B.\ 1992, \aj, 103, 1746. doi:10.1086/116192
\bibitem[Neugebauer et al.(1984)]{1984ApJ...278L...1N} Neugebauer, G., Habing, H.~J., van Duinen, R., et al.\ 1984, \apjl, 278, L1. doi:10.1086/184209
\bibitem[Paliya et al.(2018)]{2018ApJ...853L...2P} Paliya, V.~S., Ajello, M., Rakshit, S., et al.\ 2018, \apjl, 853, L2. doi:10.3847/2041-8213/aaa5ab
\bibitem[Paliya et al.(2024)]{2024MNRAS.527.7055P} Paliya, V.~S., Stalin, C.~S., Dom{\'\i}nguez, A., et al.\ 2024, \mnras, 527, 7055. doi:10.1093/mnras/stad3650
\bibitem[P{\^a}ris et al.(2014)]{2014A&A...563A..54P} P{\^a}ris, I., Petitjean, P., Aubourg, {\'E}., et al.\ 2014, \aap, 563, A54. doi:10.1051/0004-6361/201322691
\bibitem[P{\^a}ris et al.(2017)]{2017A&A...597A..79P} P{\^a}ris, I., Petitjean, P., Ross, N.~P., et al.\ 2017, \aap, 597, A79. doi:10.1051/0004-6361/201527999
\bibitem[P{\^a}ris et al.(2018)]{2018A&A...613A..51P} P{\^a}ris, I., Petitjean, P., Aubourg, {\'E}., et al.\ 2018, \aap, 613, A51. doi:10.1051/0004-6361/201732445
\bibitem[P{\'e}rez-Torres et al.(2021)]{2021A&ARv..29....2P} P{\'e}rez-Torres, M., Mattila, S., Alonso-Herrero, A., et al.\ 2021, \aapr, 29, 2. doi:10.1007/s00159-020-00128-x
\bibitem[Perna et al.(2021)]{2021A&A...646A.101P} Perna, M., Arribas, S., Pereira Santaella, M., et al.\ 2021, \aap, 646, A101. doi:10.1051/0004-6361/202039702
\bibitem[Polletta et al.(2007)]{2007ApJ...663...81P} Polletta, M., Tajer, M., Maraschi, L., et al.\ 2007, \apj, 663, 81. doi:10.1086/518113
\bibitem[Rakshit et al.(2017)]{2017ApJS..229...39R} Rakshit, S., Stalin, C.~S., Chand, H., et al.\ 2017, \apjs, 229, 39. doi:10.3847/1538-4365/aa6971
\bibitem[Ross et al.(2015)]{2015MNRAS.453.3932R} Ross, N.~P., Hamann, F., Zakamska, N.~L., et al.\ 2015, \mnras, 453, 3932. doi:10.1093/mnras/stv1710
\bibitem[Rowan-Robinson et al.(1991)]{1991Natur.351..719R} Rowan-Robinson, M., Broadhurst, T., Lawrence, A., et al.\ 1991, \nat, 351, 719. doi:10.1038/351719a0
\bibitem[Rowan-Robinson et al.(2018)]{2018A&A...619A.169R} Rowan-Robinson, M., Wang, L., Farrah, D., et al.\ 2018, \aap, 619, A169. doi:10.1051/0004-6361/201832671
\bibitem[Roy \& Norris(1997)]{1997MNRAS.289..824R} Roy, A.~L. \& Norris, R.~P.\ 1997, \mnras, 289, 824. doi:10.1093/mnras/289.4.824
\bibitem[Rush et al.(1996)]{1996ApJ...473..130R} Rush, B., Malkan, M.~A., \& Edelson, R.~A.\ 1996, \apj, 473, 130. doi:10.1086/178132
\bibitem[Sanders et al.(2003)]{2003AJ....126.1607S} Sanders, D.~B., Mazzarella, J.~M., Kim, D.-C., et al.\ 2003, \aj, 126, 1607. doi:10.1086/376841
\bibitem[Sanders et al.(1988)]{1988ApJ...325...74S} Sanders, D.~B., Soifer, B.~T., Elias, J.~H., et al.\ 1988a, \apj, 325, 74. doi:10.1086/165983
\bibitem[Sanders et al.(1988)]{1988ApJ...328L..35S} Sanders, D.~B., Soifer, B.~T., Elias, J.~H., et al.\ 1988b, \apjl, 328, L35. doi:10.1086/185155
\bibitem[Sanders \& Mirabel(1996)]{1996ARA&A..34..749S} Sanders, D.~B. \& Mirabel, I.~F.\ 1996, \araa, 34, 749. doi:10.1146/annurev.astro.34.1.749
\bibitem[Soifer et al.(1984)]{1984ApJ...278L..71S} Soifer, B.~T., Rowan-Robinson, M., Houck, J.~R., et al.\ 1984, \apjl, 278, L71. doi:10.1086/184226
\bibitem[Soifer \& Neugebauer(1991)]{1991AJ....101..354S} Soifer, B.~T. \& Neugebauer, G.\ 1991, \aj, 101, 354. doi:10.1086/115691
\bibitem[Sutherland \& Saunders(1992)]{1992MNRAS.259..413S} Sutherland, W. \& Saunders, W.\ 1992, \mnras, 259, 413. doi:10.1093/mnras/259.3.413
\bibitem[Tran et al.(2001)]{2001ApJ...552..527T} Tran, Q.~D., Lutz, D., Genzel, R., et al.\ 2001, \apj, 552, 527. doi:10.1086/320543
\bibitem[Trump et al.(2006)]{2006ApJS..165....1T} Trump, J.~R., Hall, P.~B., Reichard, T.~A., et al.\ 2006, \apjs, 165, 1. doi:10.1086/503834
\bibitem[Turnshek et al.(1997)]{1997ApJ...476...40T} Turnshek, D.~A., Monier, E.~M., Sirola, C.~J., et al.\ 1997, \apj, 476, 40. doi:10.1086/303610
\bibitem[Veilleux et al.(1995)]{1995ApJS...98..171V} Veilleux, S., Kim, D.-C., Sanders, D.~B., et al.\ 1995, \apjs, 98, 171. doi:10.1086/192158
\bibitem[Veilleux et al.(1999)]{1999ApJ...522..113V} Veilleux, S., Kim, D.-C., \& Sanders, D.~B.\ 1999a, \apj, 522, 113. doi:10.1086/307634
\bibitem[Veilleux et al.(1999)]{1999ApJ...522..139V} Veilleux, S., Sanders, D.~B., \& Kim, D.-C.\ 1999b, \apj, 522, 139. doi:10.1086/307635
\bibitem[Veilleux et al.(2002)]{2002ApJS..143..315V} Veilleux, S., Kim, D.-C., \& Sanders, D.~B.\ 2002, \apjs, 143, 315. doi:10.1086/343844
\bibitem[Voit et al.(1993)]{1993ApJ...413...95V} Voit, G.~M., Weymann, R.~J., \& Korista, K.~T.\ 1993, \apj, 413, 95. doi:10.1086/172980
\bibitem[Wright et al.(2010)]{2010AJ....140.1868W} Wright, E.~L., Eisenhardt, P.~R.~M., Mainzer, A.~K., et al.\ 2010, \aj, 140, 1868. doi:10.1088/0004-6256/140/6/1868
\bibitem[Wu \& Shen(2022)]{2022ApJS..263...42W} Wu, Q. \& Shen, Y.\ 2022, \apjs, 263, 42. doi:10.3847/1538-4365/ac9ead
\bibitem[Xia et al.(2012)]{2012ApJ...750...92X} Xia, X.~Y., Gao, Y., Hao, C.-N., et al.\ 2012, \apj, 750, 92. doi:10.1088/0004-637X/750/2/92
\bibitem[Yun et al.(2001)]{2001ApJ...554..803Y} Yun, M.~S., Reddy, N.~A., \& Condon, J.~J.\ 2001, \apj, 554, 803. doi:10.1086/323145
\bibitem[Zhan(2011)]{2011SSPMA..41.1441Z} Zhan, H.\ 2011, Scientia Sinica Physica, Mechanica \& Astronomica, 41, 1441. doi:10.1360/132011-961
\bibitem[Zhan(2018)]{2018cosp...42E3821Z} Zhan, H.\ 2018, in 42nd COSPAR Scientific Assembly, 42, E1.16
\bibitem[Zhan(2021)]{zhan2021} Zhan, H. \ 2021, Chinese Science Bulletin, 66, 11. doi:10.1360/TB-2021-0016
\bibitem[Zhang et al.(2010)]{2010ApJ...714..367Z} Zhang, S., Wang, T.-G., Wang, H., et al.\ 2010, \apj, 714, 367. doi:10.1088/0004-637X/714/1/367
\bibitem[Zheng et al.(1999)]{1999A&A...349..735Z} Zheng, Z., Wu, H., Mao, S., et al.\ 1999, \aap, 349, 735
\bibitem[Zhou et al.(2006)]{2006ApJS..166..128Z} Zhou, H., Wang, T., Yuan, W., et al.\ 2006, \apjs, 166, 128. doi:10.1086/504869
\bibitem[Zou et al.(1991)]{1991MNRAS.252..593Z} Zou, Z., Xia, X., Deng, Z., et al.\ 1991, \mnras, 252, 593. doi:10.1093/mnras/252.4.593
\end{thebibliography}
\end{document}